\documentclass[a4paper,11pt]{article}
\usepackage{cite}
\usepackage{latexsym,amssymb,enumerate,amsmath,epsfig,amsthm,dsfont,bm}
\usepackage[margin=1in]{geometry}
\usepackage{setspace,color}
\usepackage{tikz}
\usepackage{url}
\usetikzlibrary{positioning,arrows.meta,spy}
\usetikzlibrary{decorations.text,calc,arrows.meta}
\usepackage{floatrow}
\usepackage{graphicx,subfigure}
\usepackage[linesnumbered,ruled,vlined]{algorithm2e}
\usepackage{algcompatible}
\usepackage{epstopdf}
\usepackage[multidot]{grffile}
\usepackage{comment}
\usepackage{multirow}

\makeatletter
\newcommand\figcaption{\def\@captype{figure}\caption}
\newcommand\tabcaption{\def\@captype{table}\caption}
\makeatother

\theoremstyle{plain}
\newtheorem{Th}{Theorem}[section]

\newtheorem{Prop}[Th]{Proposition}

\theoremstyle{definition}

\newtheorem{Rem}[Th]{Remark}
\newtheorem{?}[Th]{Problem}

\definecolor{arrowblue}{RGB}{98,145,224}

\title{Silhouette Vectorization by Affine Scale-space }
\author{Yuchen He, Sung Ha Kang, Jean-Michel Morel}
\date{}
\begin{document}
	\maketitle
	
	\begin{abstract}
		Silhouettes or 2D planar shapes are extremely important in human communication, which involves many logos, graphics symbols and fonts in vector form. Many more shapes can be extracted from image by binarization or segmentation, thus in raster form that requires a vectorization.
		There is a need for disposing of a mathematically well defined and justified shape vectorization process, which in addition provides a minimal set of control points with geometric meaning. In this paper we propose a silhouette vectorization method which extracts the outline of a 2D shape from a raster binary image, and converts it to a combination of cubic B\'{e}zier polygons and perfect circles.   Starting from the boundary curvature extrema computed at sub-pixel level, we identify a set of control points based on the affine scale-space induced by the outline.  These control points capture similarity invariant geometric features of the given silhouette and give precise locations of the shape's corners.of the given silhouette. Then, piecewise B\'{e}zier cubics are computed  by least-square fitting combined with an adaptive splitting to guarantee a predefined accuracy.  When there are no curvature extrema identified, either the outline is recognized as a circle using the isoperimetric inequality, or a pair of the most distant outline points are chosen to initiate the fitting.

		Given their construction, most of our control points are geometrically stable
		under affine transformations. By comparing with other feature detectors, we show that our method can be used as a reliable feature point detector for silhouettes. Compared to state-of-the-art image vectorization software, our algorithm demonstrates superior  reduction on the number of control points, while maintaining high accuracy.  
		
	\end{abstract}	
	
	\section{Introduction}
	A silhouette is a subset of the plane that  was traditionally obtained by copying on paper the shadow projected on a wall by a person placed in front of a point light source\footnote{\url{https://en.wikipedia.org/wiki/Silhouette}}.
	In digital images, silhouettes of objects can be obtained by a mere luminance threshold (e.g. Otsu's algorithm \cite{ipol.2016.158}) as soon as the object is darker or brighter than its surroundings. Then the silhouette appears as one of the connected components of an upper or lower set of the image. More generally, the study of shape promoted by Mathematical Morphology \cite{matheron1975random} calls 2D shapes any such connected component.  We will call   our object of study here  2D shape or silhouette.  Silhouettes are essential for the human perception of shapes, and the distribution of  corners along its outlines are closely linked to the neurological models of the visual system~\cite{attneave1954some}.  The geometric features captured by the vectorization are important in feature identification~\cite{nadal1990complementary}, remote sensing~\cite{kirsanov2010contour}, and others~\cite{yang2001bezier,tombre2000vectorization,zou2001cartoon}. As proved in \cite{ambrosio2001connected}, if a  closed subset of the plane has finite perimeter, then it can be described by its essential boundary, which is a countable set of Jordan curves with finite length.  In image processing upper level sets can be extracted by a mere thresholding, in which case they are a finite union of pixels, bounded by a finite number of Jordan curves made of vertical and horizontal segments. Using a  parametric interpolation such as the bilinear, one can also extract the boundary of a level set as a union of pieces of hyperbolae \cite{cao2008theory}. Following \cite{monasse2000scale} an image can therefore be decomposed in a tree of connected shapes ordered by inclusion, and each of these shapes (or silhouettes) can be described by its raster or by its boundary, which is a finite number of Jordan curves described as polygons or concatenations of pieces of hyperbolae.
	
	Thus, there is a standard way to lead back image analysis to the analysis of 2D shapes, and eventually to the analysis of its outline, described by a  finite set of nested Jordan curves that also are level lines of the image. This is not the only way to extract shapes from images. Any segmentation algorithm divides an image into connected regions. For example,   many software vectorization software\footnote{See (e.g.) \url{https://en.wikipedia.org/wiki/Adobe_Illustrator} or Vector Magic} proceed by a mere color quantization which reduces the image to a piecewise constant image and therefore to a union of disjoint 2D shapes. The boundaries of these shapes can then be encoded in Scalable Vector Graphics (SVG) format\footnote{\url{https://fr.wikipedia.org/wiki/Scalable_Vector_Graphics}}. 
	
	A crucial point of such vector representation is that it is  scalable, and therefore used for all 2D shapes that, like logos or fonts, require printing at many sizes.

	Common silhouette vectorization methods from its outline consist of two steps: identification of control points and approximation of curves connecting the control points.  In a founding work,  Montanari~\cite{montanari1970note} introduced a polygonal approximation of outlines of rasterized silhouettes. After the discrete boundary is traced, the sub-pixel locations of the polygonal vertices are determined by minimizing a global length energy with an $L^\infty$ loss to the initial outline.  In more recent literature, B\'{e}zier curves have become widely adopted to replace  polygonal lines~\cite{mortenson1999mathematics}. Most developments on silhouette outline vectorization use piecewise B\'{e}zier curves, or B\'{e}zier polygons~\cite{ramer1972iterative,cinque1998shape,montero2012skeleton,pal2007cubic,yang2001bezier}. Ramer~\cite{ramer1972iterative} proposed an iterative splitting scheme for identifying a set of control points on a polygonal line $\mathcal{C}$ such that the B\'{e}zier polygon $\widehat{\mathcal{C}}$ defined by these vertices approximates $\mathcal{C}$. The Hausdorff distance between $\widehat{\mathcal{C}}$  and $\mathcal{C}$ is constrained to stay below a predefined threshold, and the number of control points is suboptimal.  More recently, Safraz~\cite{sarfraz2010vectorizing} proposed an outline vectorization
	algorithm that splits the outline at corners which are identified without computing curvatures~\cite{chetverikov2003simple}, then new control points are introduced to improve curve fitting.  
	The control points produced by some of these works may correspond to curvature extrema of the outline, but this happens by algorithmic convergence rather than by design. It is well-known that the direct computation of curvature is not reliable~\cite{alvarez1994formalization}.  The above mentioned methods reflect the challenges of estimating the  outline's curvature on shapes extracted from raster images. 
	
	
	In this paper, we propose a mathematically founded  outline vectorization algorithm. It identifies (i) curvature extrema of the outline computed at sub-pixel level by  (ii) backpropagating control points detected as curvature extrema at coarser scale in the affine scale-space,  then (iii) computing piecewise least-square cubic B\'{e}zier joining these control points while fitting the initial outline with a predefined accuracy.

	The main contribution of this paper is to propose a new approach using the sub-pixel curvature extrema and affine scale-space for silhouette vectorization.  We shall illustrate by comparisons how the proposed method can give an accurate vectorization with a generally smaller number of more meaningful control points.
	

	
	We organize the paper as follows. In section \ref{sec_overview}, an overview of proposed algorithm is presented.    In Section~\ref{sec_1}, we review the level line extraction and sub-pixel curvature computation~\cite{ciomaga2017image}. In Section~\ref{sec_2}, we introduce the affine scale-space induced by the smooth bilinear outline and define the candidate control points. In Section~\ref{sec_3}, we describe an adaptive piecewise least-square B\'{e}zier polygon fitting, where the set of candidate points is modified to achieve a compact representation and to guarantee a predefined accuracy.  The overall algorithm is summarized in Section~\ref{sec_algOutline}. We include various numerical results and comparison with other vectorization methods in Section~\ref{sec_5}, and conclude the paper in Section~\ref{sec_6}.

\section{Overview of the Proposed Method}\label{sec_overview}
	
On a rectangular domain $\Omega=[0,H]\times[0,W] \subset\mathbb{R}^2$ with $H>0$ and $W>0$, a \textit{silhouette} is a compact subset $\mathcal{S}\subset\Omega$ whose topological boundary  $\partial \mathcal{S}$,  the \textit{outline}, is a piecewise smooth curve.  Suppose $\mathcal{S}$ is represented by a \textit{raster binary image} $I:\Omega\cap \mathbb{N}^2\to\{0,255\}$, that is, the set of black pixels
\begin{align}
\overline{\mathcal{S}}=\{(i,j)\in\Omega\cap\mathbb{N}^2~|~I(i,j)=0\}\label{eq_raster_sil}
\end{align}
approximates $\mathcal{S}$. We assume that  $\mathcal{S}\cap\partial\Omega=\varnothing$. Our objective is to find a cubic B\'{e}zier polygon close to $\partial \mathcal{S}$ in the Hausdorff distance. The proposed silhouette vectorization method has three main steps:
\begin{enumerate}
\item Estimate the curvature extrema of  $\partial \mathcal{S}$ across different scales  in sub-pixel level.

\item Based on the affine scale-space induced by $\partial\mathcal{S}$, identify salient curvature extrema which are robust against pixelization and noise as the candidate control points.

\item Fit the outline using a B\'{e}zier polygon from the candidate control points which are adaptively modified to achieve a compact representation while guaranteeing a desired accuracy.

		
\end{enumerate}
	
In the following sections, we give the details of the proposed method.
\begin{figure}
	\centering
	\includegraphics[scale=0.4]{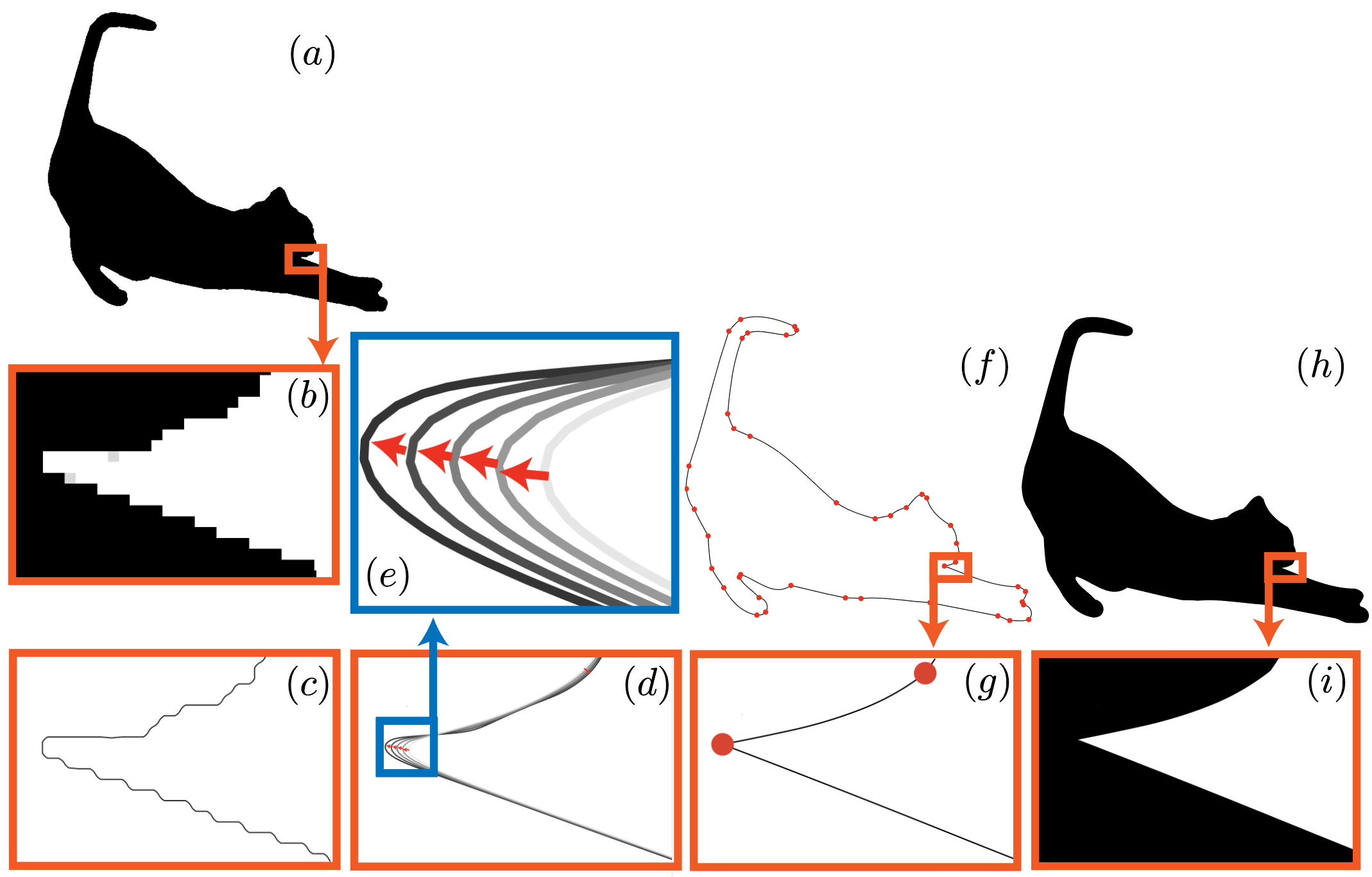}
	\caption{Flowchart of the proposed method. (a) A raster image of a cat's silhouette. (b) Zoom-in of (a). (c) Extracted bilinear outline of (a). (d) Inversely tracing the curvature extrema along the affine shortening flow. (e) Zoom-in of (d). (f) The vectorized outline of (a) with control points marked as red dots. (g) Zoom-in of (f).  (h) Vectorized silhouette of (a) by the proposed method. (i) Zoom-in of (h). Notice the difference between the given image (a) and our result (h), as well as the zoom of (b) and (i). }\label{fig_flow}
\end{figure}

	\section{Sub-pixel Curvature Extrema Localization} \label{sec_1} 


	 Following the work of~\cite{ciomaga2017image}, we consider the  bilinear interpolation $u:\Omega\to[0,255]$ for the raster image $I$ whose continuous function form is
	\begin{align}
	u(x,y) = axy+bx+cy+d\;,\quad(x,y)\in\Omega\;,
	\end{align}  
	where $a,b,c,d$ are scalar functions depending on $(\lfloor x\rfloor, \lfloor y\rfloor)$, and 
	\begin{align}
	u(i+1/2,j+1/2) = I(i,j)\;,\quad (i,j)\in \Omega\cap\mathbb{N}^2\;.
	\end{align}
	Here $\lfloor r\rfloor$ is the floor function giving the greatest integer smaller than the real number $r$. For any $\lambda\in(0,255)$, the level line of $u$ corresponding to $\lambda$ is defined as $\mathcal{C}_\lambda=\{(x,y)\in\Omega~|~u(x,y) = \lambda\}$. Since $I$ is binary, the Hausdorff distance between any $\mathcal{C}_\lambda$ and the raster silhouette $\overline{\mathcal{S}}$~\eqref{eq_raster_sil} is bounded above by $\sqrt{2}$. Hence $\mathcal{C}_\lambda$ for an arbitrary level $\lambda\in(0,255)$ approximates the discrete outline as a piecewise $C^2$ Jordan curve except for at finitely many points, e.g., saddle points~\cite{caselles2009geometric}. In the following, we focus on a single level line $\mathcal{C}_{\lambda^*}$ for some $\lambda^*\in(0,255)$ extracted by the level line extraction algorithm detailed in~\cite{caselles2010}.
	
   Specifically,  $\mathcal{C}_{\lambda^*}$ is either piecewise linear line (horizontal or vertical) or a part of a hyperbola  whose asymptotes are adjacent edges of a single pixel.  See Figure~\ref{fig_flow}~(c). Due to pixelization,  $\mathcal{C}_{\lambda^*}$ shows strong staircase effects~\cite{cao2003geometric}, and such oscillatory behavior is effectively reduced by the affine shortening flow~\cite{cao2003geometric,sapiro1993affine}. For any planar curve $\mathcal{C}$, we smooth it via solving the PDE
	\begin{align}
	\frac{\partial \mathcal{C}(s,t)}{\partial t} = \kappa^{1/3}(s,t)\mathbf{N}(s,t)\;,\quad\mathcal{C}(s,0) = \mathcal{C}(s)\;,\quad s\in [0,\text{Length}(\mathcal{C}(\cdot,t))]\label{eq_affine}
	\end{align}
	to some short time $T>0$. Here each curve $\mathcal{C}(\cdot,t)$  is arc-length parametrized for any $t$,  $\kappa$ denotes the curvature, and $\mathbf{N}$ is the inward normal at $\mathcal{C}(s,t)$. The flow~\eqref{eq_affine} is affine intrinsic, that is, its solution is invariant under affine transformations; hence it preserves
	the geometric properties of the original curve during the evolution.
	
	\begin{figure}
		\centering
		\begin{tabular}{cc}
			(a)&(b)\\
			\includegraphics[scale=0.25]{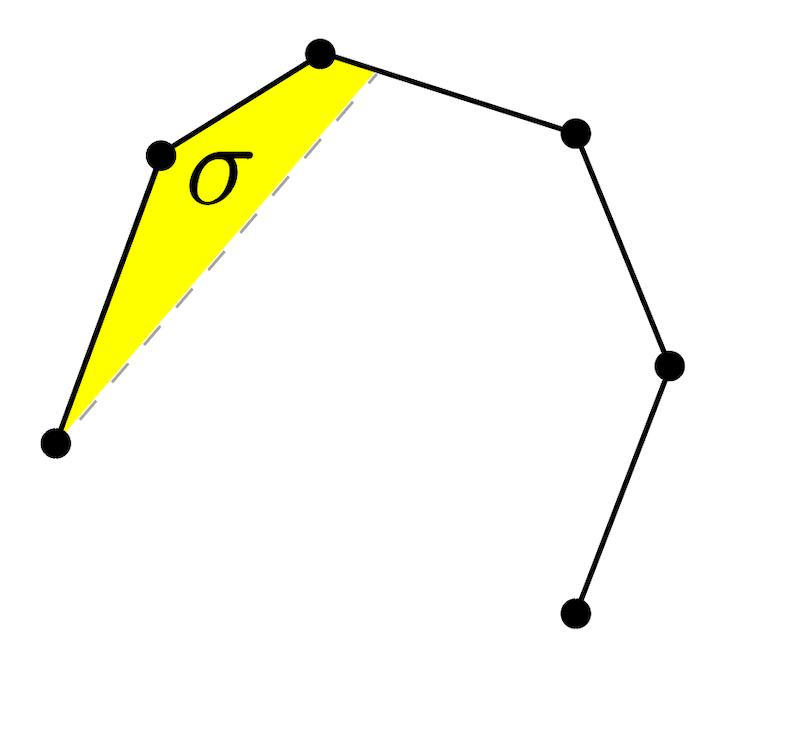}&
			\includegraphics[scale=0.25]{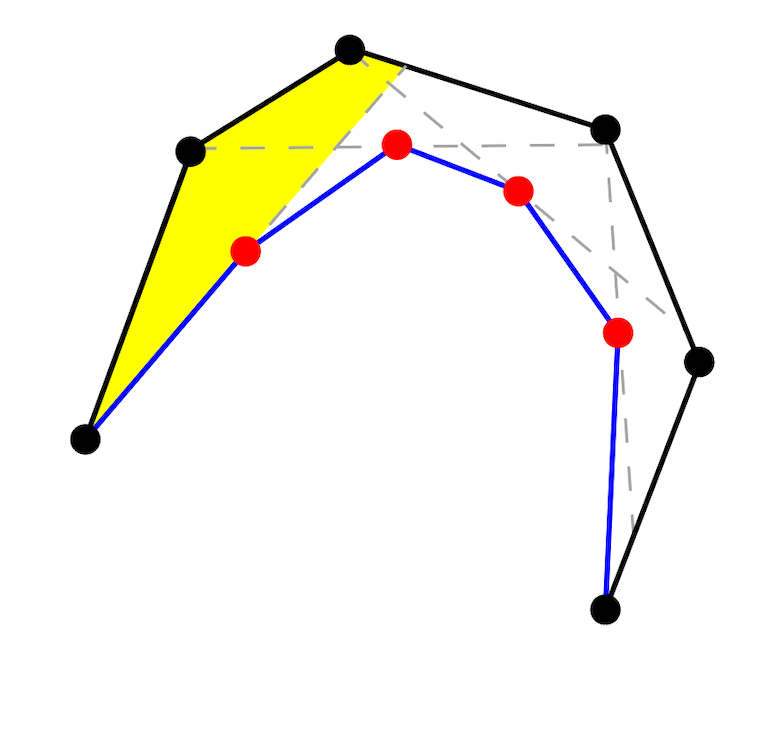}
		\end{tabular}
		\caption{Illustration of the geometric scheme for affine shortening. (a) A convex component of a discrete curve and a $\sigma$-chord (dashed line). (b) The result of discrete $\sigma$-affine erosion is a polygonal line (blue lines) whose vertices are middle points (red dots) of the $\sigma$-chords.}\label{fig_sigma_chord}
	\end{figure}

	To solve \eqref{eq_affine}, we apply the fully consistent geometric scheme \cite{moisan1998affine} which  is independent of grid discretization. The idea is that by iterating the discrete affine erosion with a sufficiently small parameter, the convergent morphological operator becomes equivalent to the differential operator in~\eqref{eq_affine}.  Given a discrete curve partitioned by its inflection points,  the $\sigma$-affine erosion of each covex component is a polygonal line whose vertices are the middle points of $\sigma$-chords. See Figure~\ref{fig_sigma_chord}. A $\sigma$-chord is a segment joining two points on the curve such that the area enclosed by the segment and the curve is $\sigma$.  After the erosion, we glue the evolved components at the inflection points, resample the resulted curve by arc-length, and iterate the procedure above to reach the desired scale $T$.  For a sufficiently smooth convex curve,  applying the $\sigma$-affine erosion is equivalent to solving~\eqref{eq_affine} till time $\omega\sigma^{2/3}$ for some absolute constant $\omega>0$~\cite{moisan1998affine}.

Denoting the smooth bilinear outline  obtained above by $\Sigma_{\lambda^*}$, the curvature at any of its vertices can be computed without dependence on the grid discretization.  The following discussion applies for each connected component. 
Suppose $\Sigma_{\lambda^*}=\{P_i(x_i,y_i)\}_{i=0}^N$  with $P_0=P_N$ and is oriented same as  $\mathcal{C}_{\lambda^*}$.  Following~\cite{ciomaga2017image}, the discrete curvature at point $P_i$ is computed by
	\begin{align}
	\kappa(P_i) = \frac{-2\,\text{det}(P_iP_{i-1}~~ P_iP_{i+1})}{||P_{i-1}P_i||\,||P_{i}P_{i+1}||\,||P_{i-1}P_{i+1}||}\;,~i=0,1,2,\dots,N-1\label{eq_curvature}
	\end{align}
	where 
	\begin{align}
	\text{det}(P_iP_{i-1}~~ P_iP_{i+1}):=\text{det}\begin{bmatrix}
	x_{i-1}-x_i&x_{i+1}-x_i\\
	y_{i-1}-y_i&y_{i+1}-y_i
	\end{bmatrix}\;,
	\end{align}
	$P_{-1}=P_{N-1}$, and $||\cdot||$ denotes the Euclidean $2$-norm.  It computes the discrete curvature of $\Sigma_{\lambda^*}$ at $P_i$ as the curvature of the circumcircle that passes through three consecutive points $P_{i-1}$, $P_i$, and $P_{i+1}$.  The discrete curvature values can be obtained at arbitrary resolution based on the sampling frequency applied to the bilinear outline $\mathcal{C}_{\lambda^*}$, hence it is called ``curvature microscope". 
	
	To identify the curvature extrema, we process the data $\{\kappa(P_i)\}_{i=0}^{N-1}$ by repeatedly applying the filter $(1/18,4/18,8/18,4/18,1/18)$ with periodic boundary condition for $20$ times to suppress the noise. 
Based on the filtered data $\{\widetilde{\kappa}(P_i)\}_{i=0}^{N-1}$,  $P_i$ is a curvature extramum if
	\begin{align}
	|\widetilde{\kappa}(P_i)|>|\widetilde{\kappa}(P_j)|\;,~\text{for}~j=i\pm 1, i\pm 2\;.~\label{eq_criterion}
	\end{align}
	In practice, to further stabilize the identification, we also require that a curvature extremum should have $|\widetilde{\kappa}(P_i)|>\delta$ for some small value $\delta>0$. In this paper, we take $\delta =0.001$.
	
	\begin{Rem}
		Our method is also applicable when the input is a raster gray-scale image where the intensity variation concentrates around the topological boundary of the underlying silhouette.  The higher the image gradient across the silhouette's boundary, the more stable the position of the extracted outline with respect to the choice of levels.
	\end{Rem}

	\section{Affine Scale-space Control Points Identification} 
	\label{sec_2}
	
	The curvature extrema form a good set of control points, since they capture the geometrical changes in the outline.  We propose to refine the control points via affine scale-space approach, which is detailed in this section.

	\subsection{Backward Tracing via Inverse Affine Shortening Flow}
	The concept of scale-space, first introduced by Witkin~\cite{witkin1987scale},  provides a formalism for multiscale analysis of signals. Later developments~\cite{alvarez1992axiomes,babaud1986uniqueness,perona1990scale,koenderink1984structure} established the axiomatic properties for defining a scale-space. In~\cite{sapiro1993affine}, Sapiro et al. proved that  the solution of the affine shortening flow~\eqref{eq_affine} defines an affine scale-space, where the scale is given by the time parameter, and the solution at any scale is affine invariant, i.e., it commutes with planar special affine transforms.  A critical property satisfied by the affine  scale-space is causality: no new information is created when passing from fine to coarse scales. In particular,  the following is proved in \cite{sapiro1993affine}:
	\begin{Prop}\label{prop1}
		In the affine invariant scale-space of a planar curve, the number of vertices, that is, the extrema of Euclidean curvature, is a nonincreasing function of time.
	\end{Prop}
	More precisely,  every curvature extremum  on the curve at a coarser scale  is the continuation of \textit{at least} one of the extrema at a finer scale. The lack of one-to-one correspondence is due to the possibility of multiple extrema (e.g. two maxima and one minimum) merging to a single one during the evolution.
	 
	In this paper, we propose a new approach for defining the control points as the curvature extrema on $\Sigma_{\lambda^*}$ which persist across different scales in its affine scale-space.  By inversely tracing curvature extrema from the coarser scales to the finer scales, the  proposed control points are more robust to noise and help to capture prominent corners of the silhouette.

	Given a sequence of discrete times $t_0=0<t_1<\cdots<t_K$ for some positive integer $K$,   we obtain the curve $\mathcal{C}(\cdot,t_{n})$ at scale $t_n$ by the affine shortening flow \eqref{eq_affine} for $n=0,1,\dots, K$.
	For any $1\leq n\leq K$, by a first order Taylor expansion, the affine shortening flow~\eqref{eq_affine}  is approximated as
	\begin{align}
	\frac{\mathcal{C}(s,t_{n})-\mathcal{C}(s,t_{n-1})}{t_{n}-t_{n-1}}=(\kappa^n(s))^{1/3}\mathbf{N}^n(s)+\mathbf{r}(s)\;,\label{eq_implicit1}
	\end{align}
	where $\kappa^n$ and $\mathbf{N}^n$ denote the curvature and normal at the scale $t_n$, and $\mathbf{r}$ is a remainder such that $||\mathbf{r}(s)|| = O(t_n-t_{n-1})$. Rearranging~\eqref{eq_implicit1} gives
	\begin{align}
	\mathcal{C}(s,t_{n-1})=\mathcal{C}(s,t_n)-(t_n-t_{n-1})(\kappa^n(s))^{1/3}\mathbf{N}^n(s)+(t_n-t_{n-1})\mathbf{r}(s)\;.\label{eq_implicit2}
	\end{align}
	This expression shows that,  if $t_n-t_{n-1}$ is sufficiently small,  by following  the opposite direction of the affine shortening flow at $\mathcal{C}(s,t_n)$, that is, 
	\begin{align}
	-\text{sign}(\kappa^n(s))\mathbf{N}^n(s)\;,\label{eq_inverse_direction}
	\end{align} 
	we can find $\mathcal{C}(s,t_{n-1})$ nearby.  Here $\text{sign}(r)$ denotes the sign function which gives $+1$ if $r>0$, $-1$ if $r<0$ and $0$ if $r=0$. This gives
	a well-defined map from the curve at a coarser scale $t_n$ to a finer scale $t_{n-1}$ via the inverse affine shortening flow.
	
	Starting from $K$, for any curvature extremum $X^K$ on $\mathcal{C}_K=\mathcal{C}(\cdot,t_K)$, we set up the following constrained optimization problem to find  a curvature extremum $X^{K-1}$ on $\mathcal{C}_{K-1}$ at scale $t_{K-1}$:
	\begin{equation}
	\max_{X\in\mathcal{C}_{K-1}} \frac{\langle X-X^K, -\text{sign}(\kappa^K)\mathbf{N}^K\rangle}{||X-X^K||}\;, \;\; 
	 \text{s.t.}~ 
	\begin{cases}
	\displaystyle{\frac{\langle X-X^K, -\text{sign}(\kappa^n)\mathbf{N}^n\rangle}{||X-X^K||}>\alpha} \\
	||X-X^K|| < D\\
	X \text{~is a curvature extremum on~}\mathcal{C}_{K-1}
	\end{cases}\;,\label{eq_opt}
	\end{equation} 
	where $D>0$ is a positive parameter that controls the closeness between $X$ and $X^K$, and $\alpha$ enforces that the direction of $X-X^K$ is similar to that of the inverse affine shortening flow.  The problem~\eqref{eq_opt} looks for the curvature extremum on $\mathcal{C}_{K-1}$ in the $D$-neighborhood of $X^K$ that is the nearest to the line passing $X^K$ in the direction of the inverse affine shortening flow.
	When $D$ and $\alpha$ are properly chosen, if~\eqref{eq_opt} has one solution, we define it to be $X^{K-1}$. If~\eqref{eq_opt} has multiple solutions, we choose the one that has the shortest distance from $X^K$ to be $X^{K-1}$. In case there are  multiple solutions having the same shortest distance from $X^K$, we can arbitrarily select one to be $X^{K-1}$. However, in practice, if~\eqref{eq_opt} has a solution, it is almost always unique.
	
	We repeat the optimization~\eqref{eq_opt} for $K-1$, $K-2$, etc. Either the solutions always exist until the scale $t_0$, or there exists some $m\geq 1$, such that~\eqref{eq_opt} at $t_m$ does not have any solution. In the first case, we call $X^K$ a \textit{complete} point, and in the second case, we call it \textit{incomplete}.
	For each curvature extremum $X^K$ on $\mathcal{C}_K$, we construct a sequence of points $\mathcal{L}(X^K)$ that contains the solutions of~\eqref{eq_opt} for $K,K-1,K-2$, etc., starting at  $X^K$ in a scale-decreasing order. If  $X^K$ is complete, then $\mathcal{L}(X^K)$ has exactly $K+1$ elements, and we call the sequence complete; otherwise, the size of $\mathcal{L}(X^K)$ is strictly smaller than $K+1$, and we call the sequence incomplete.

	We define the last elements of the complete sequences as the candidate control points, and denote them as $\{O_i(t_K)\}_{i=1}^{M(t_K)}$. This set of points is ordered following the orientation of $\Sigma_{\lambda^*}$. Here the parameter $t_K$ in the parenthesis indicates that the candidate control points are associated with the curvature extrema identified at the scale $t_K$. When the scale $t_K$ is fixed, we simply write $\{O_i\}_{i=1}^{M}$.

	\subsection{Degenerate Case}\label{sec_degenerate}
	In the discussion above, if $M(t_K)=0$, i.e., if there are no candidate control points identified on $\Sigma_{\lambda^*}$ associated with the curvature extrema at scale $t_K$, then we call it a \textit{degenerate case}. This situation occurs when the underlying silhouette is a disk, or has a smoothly varying boundary, provided that the image has sufficiently high resolution. This paper is not considering open curves. If we did,  the  absence of curvature  extrema only means that the curve has a monotone curvature, hence is a spiral.
	
	If $\mathcal{S}$ is indeed a disk, the  vectorization only requires  information about its center and radius.
	We use the isoperimetric inequality to determine if $\Sigma_{\lambda^*}$ represents a circle: for any closed plane curve with area $A$ and perimeter $L$, we have
	\begin{align}
	4\pi A \leq L^2\;,\label{eq_iso}
	\end{align}
	and the equality holds if and only if the curve is a circle. In practice, we decide that $\Sigma_{\lambda^*}$ is a circle only if the corresponding ratio $1-4\pi A/L^2<0.005$. By this criterion, if $\Sigma_{\lambda^*}$ is classified as a circle, then its center and radius are easily computed by arbitrarily three distinct points on $\Sigma_{\lambda^*}$. For numerical stability, we take three outline points that are equidistant from each other.
	
	When $1-4\pi A/L^2\geq 0.005$ for a degenerate case, we insert a pair of most distant points on $\Sigma_{\lambda^*}$ to be the candidate control points.  An efficient approach for finding these points is to combine a convex hull algorithm, e.g., the monotone chain method~\cite{andrew1979another}, which takes $O(N\log N)$ time,  with the rotating calipers~\cite{preparata2012computational}, which takes $O(N)$ time. Here $N$ is the number of vertices of the polygonal line $\Sigma_{\lambda^*}$.

	\section{Adaptive Cubic B\'{e}zier Polygon Approximation}
	\label{sec_3}
	
	Starting from the control points identified by the affine scale-space,  $H:=\{O_i\}_i^M$, we  adjust $H$ by deleting non-salient sub-pixel curvature extrema and inserting new control points for guaranteeing a predefined accuracy. This adaptive approach yields a cubic B\'{e}zier polygon $\mathcal{B}(H)$ whose vertices are points in the updated $H$ and edges are cubic B\'{e}zier curves computed by least-square fittings. 
	
	\subsection{B\'{e}zier Fitting with Chord-length Parametrization }\label{sec_cubicfitting}
	A cubic B\'{e}zier curve is specified by four points $B_0,B_1,B_2$, and $B_3$. Its parametric form is
	\begin{align}
	B(s) = (1-s)^3B_0+3(1-s)^2sB_1+3(1-s)s^2B_2+s^3B_3\;,~s\in[0,1]\;.\label{eq_cubic}
	\end{align}
	Specifically, it has the following properties: (i) $B_0$ and $B_3$ are the two endpoints for $B(s)$; and (ii) $B_1-B_0$ is the right tangent of $B(s)$ at $B_0$, and $B_2-B_3$ is the left tangent at $B_3$. To approximate a polygonal line segment $\Sigma=\{P_0,P_1,\dots, P_N\}$, we find a cubic B\'{e}zier curve that is determined by $B_0=P_0$, $B_1$, $B_2$, and $B_3=P_N$ such that the squared fitting error
	\begin{align}
	\widetilde{S}= \sum_{i=0}^N\left(P_i-((1-\widetilde{s_i})^3B_0+3(1-\widetilde{s_i})^2\widetilde{s_i}B_1+3(1-\widetilde{s_i})\widetilde{s_i}^2B_2+\widetilde{s_i}^3B_3)\right)^2\;~\label{eq_min2}
	\end{align}
	is minimized.
%
	Here  $\widetilde{s_i}=(\sum_{k=1}^{i}||P_{k}-P_{k-1}||)/(\sum_{k=1}^N||P_{k}-P_{k-1}||)$ is the chord-length parameter for $P_i$, $i=0,1,\dots,N$. We note that \eqref{eq_min2}  is used to initialize  an iterative algorithm in~\cite{plass1983curve} for a more accurate B\'{e}zier fitting. The benefit of  this approximating setup is that we have closed-form formulae for the minimizing $B_1$ and $B_2$ as follows:
	\begin{align}
	B_1=(A_2C_1-A_{12}C_2)/(A_1A_2-A_{12}^2)\;, \;\;\;
	B_2=(A_1C_2-A_{12}C_1)/(A_1A_2-A_{12}^2)\;,\label{eq_B2}
	\end{align}
	where
	\begin{equation*}
	A_1 = 9\sum_{i=1}^N\widetilde{t_i}^2(1-\widetilde{t_i})^4\;, \;\;\;
	A_2 = 9\sum_{i=1}^N\widetilde{t_i}^4(1-\widetilde{t_i})^2\;, \;\;\;
	A_{12}=9\sum_{i=1}^N\widetilde{t_i}^3(1-\widetilde{t_i})^3\;,
	\end{equation*}
	and 
	\begin{equation*}
	C_1 = \sum_{i=1}^N3\widetilde{s_i}(1-\widetilde{s_i})^2[P_i-(1-\widetilde{s_i})^3P_0-\widetilde{s_i}^3P_3]\;, \;\;\;
	C_2 = \sum_{i=1}^N3\widetilde{s_i}^2(1-\widetilde{s_i})[P_i-(1-\widetilde{s_i})^3P_0-\widetilde{s_i}^3P_3]\;.
	\end{equation*}
	Hence we gain a better computational efficiency. A similar strategy is also taken in~\cite{montero2012skeleton} to find a B\'{e}zier cubic to smooth discrete outlines. 
	
	\subsection{Control Point Update: Deletion of Sub-pixel Extrema}\label{sec_deletion}
	
	Recall that the candidate control points $H=\{O_i\}_{i=1}^{M}$ in Section~\ref{sec_2} are curvature extrema at sub-pixel level. Hence it is possible that some of them do not reflect salient corners of the silhouette. To remove spurious sub-pixel extrema from $H$, we propose to compare the left tangent and right tangent at each candidate control point, which are obtained via the least-square cubic B\'{e}zier fitting discussed above.
	
	We take advantage of the second property of cubic B\'{e}zier curves mentioned in Section~\ref{sec_cubicfitting}. For $i=1,\dots, M$, we fit a cubic B\'{e}zier to the polygonal line segment whose set of vertices is
	\begin{align}
	\{O_{i}=P_{j(i)},P_{j(i)+1},\dots, P_{j(i+1)}=O_{i+1}\}\;,
	\end{align}
	where we take $O_{M+1}=O_1$, and obtain the estimated defining points $B_{i,1}$ and $B_{i,2}$ for the B\'{e}zier curve. The left and right tangent at $O_i$ are computed as 
	\begin{align}
	T_{i}^-= B_{i-1,2}-O_{i}\;,\;\;\;T_{i}^+= B_{i,1}-O_{i}\;,\label{eq_tangent}
	\end{align}
	respectively, where $B_{-1,2} = B_{M,2}$. These tangent vectors are associated with all the points between neighboring candidate control points. Therefore, the angle formed by $T_{i}^- $ and $T_{i}^+$ measures the sharpness of $\Sigma_{\lambda^*}$ at $O_i$ from a more global perspective. We delete $O_i$ from the set of candidate control points $H$ if
	\begin{align}
	\frac{\langle T_i^+, T_i^+\rangle}{||T_i^+||\,||T_i^-||}+1<\varepsilon\;,\label{eq_tangent_cond}
	\end{align}
	for some small parameter $\varepsilon>0$, which is equivalent to the condition  that the angle between $T_i^+$ and $T_i^-$ is close to $\pi$.  The set $H$ is updated with the remaining control points.
	 
	It is possible that all the candidate control points $\{O_i\}_{i=1}^{M}$ are removed after this procedure, thus we encounter a degenerate case. If the underlying outline is a circle, we compute the center and radius; if it is not, we take the most distant pair of outline points to update $H$.
	\subsection{Control Point Update: Insertion  for Accuracy}\label{sec_insertion}
	The candidate control points  in $H$ split the outline $\Sigma_{\lambda^*}$ into polygonal line segments, each of which is approximated by a cubic B\'{e}zier using least square fitting as described in Section~\ref{sec_cubicfitting}.  We obtain a B\'{e}zier polygon that approximates $\Sigma_{\lambda^*}$, denoted by $\mathcal{B}(H)$.    A natural measure for the error of approximating $\Sigma_{\lambda^*}$ using the B\'{e}zier polygon $\mathcal{B}(H)$ is 
	\begin{align}
	e = \max_{P_i\in \Sigma_{\lambda^*}}\text{dist}(P_i,\mathcal{B}(H))\;,\label{eq_error}
	\end{align}
	where $\text{dist}(P_i, \mathcal{B}(H))=\inf_{P\in\mathcal{B}(H)}||P_i-P||$ is the distance from $P_i$ to the curve $\mathcal{B}(H)$. It is desirable that the user can specify the threshold for the error, $\tau_e>0$. To guarantee that $e\leq \tau_e$, we  apply the splitting strategy~\cite{ramer1972iterative} which  inserts $P_{\text{new}}\in\Sigma_{\lambda^*}$ to $H$ as a new control point if 
	\begin{align}
	\text{dist} (P_{\text{new}},\mathcal{B}(H))>\tau_e\;,~\label{eq_condition}
	\end{align}
	and among those points on $\Sigma_{\lambda^*}$ satisfying~\eqref{eq_condition}, $\text{dist} (P_{\text{new}},\mathcal{B}(H))$ is the largest. After the insertion, we fit $\Sigma_{\lambda^*}$ using a B\'{e}zier polygon based on the new set of control points in $H$. If the error of the newly fitted B\'{e}zier polygon is still greater than $\tau_e$,  we insert another point based on the same criterion. This series of insertions terminates once the condition $e\leq \tau_e$ is met.  
	
	Finally, $\mathcal{B}(H)$ with the updated set of control points $H$ gives a B\'{e}zier polygon that approximates the outline $\partial \mathcal{S}$, and with its interior filled with black, we obtain the vectorized silhouette for $\mathcal{S}$ from the raster image $I$. 

\begin{Rem}
For a further reduction on the size of $H$, we may consider an optional step to merge neighboring B\'{e}zier cubics if the union of the underlying polygonal line segments can be approximated by a single B\'{e}zier cubic via~\eqref{eq_min2} with an error below $\tau_e$. We can regard the insertion in Section~\ref{sec_insertion} as controlling the data fidelity, and the simplification described here as minimizing the complexity of an estimator. Alternatively iterating these procedures provides a numerical scheme for a constrained optimizing problem
\begin{equation*}
\min_{H\subseteq\Sigma_{\lambda^*}}|H|\;,\quad\text{\text{s.t.}}~\max_{P_i\in \Sigma_{\lambda^*}}\text{dist}(P_i,\mathcal{B}(H))\leq \tau_e\;,
\end{equation*}
where $|H|$ denotes the number of elements in $H$. For any $\tau_e\geq 0$, this problem always has a solution, yet the uniqueness largely depends on the geometric structure of $\Sigma_{\lambda^*}$.
\end{Rem}

\section{Pseudo-code for the Proposed Method}\label{sec_algOutline}
	
In this section, assuming that the outline of the given silhouette has only one connected component, we summarize the proposed algorithm for silhouette vectorization in three steps:

\begin{enumerate}
		\item \textit{Extraction of the smooth sub-pixel outline $\Sigma_{\lambda^*}$} Extract the level line corresponding to $\lambda^*$  from the bilinear interpolation of the image, then discretize it as a polygon with uniform sub-pixel sampling. To reduce the staircase effects, smooth the polygon via affine shortening at a scale specified by the smoothness parameter $\sigma_0$. In this paper, we take $\lambda^*=127.5$.
		
		\item \textit{Identification of candidate control points.}  Fix an increment $\Delta\sigma>0$ and a positive integer $K>0$.  Evolve $\Sigma_{\lambda^*}$ via the affine shortening using the scales $\sigma^*=k\Delta \sigma$,  $k=1,2,\dots,K$. Starting from each curvature extremum at  scale $K\Delta\sigma$, trace the curvature extrema  at smaller scales along the inverse affine shortening flow as described by~\eqref{eq_opt}. By doing so, each curvature extremum at scale $K\Delta \sigma$ induces a sequence of traced curvature extrema across different scales, which are arranged in a scale-decreasing order. The final elements of the complete sequences are defined as the candidate control points, denoted by $H=\{O_i\}_{i=1}^{M}$. In this paper, we fix $\Delta\sigma=0.5$ and $K =4$, so that $\sigma^*=2$.
		
		In case $M=0$, process the degenerate case as described in Section~\ref{sec_degenerate}.
		
		\item \textit{Refinement of the control points.}  Remove any candidate control point $O_i$ from $H$ whose left tangent and right tangent~\eqref{eq_tangent} form an angle close to $\pi$~\eqref{eq_condition}. If all the candidate control points are removed, follow the instruction in Section~\ref{sec_degenerate} to address the degenerate case.	Then, insert new control points into $H$ by the splitting strategy~\cite{ramer1972iterative} until the approximation error $e$~\eqref{eq_error} is bounded by a user-specified threshold $\tau_e>0$. 
		
		\item \textit{(Optional) Merging neighboring B\'{e}zier cubics} For $i=1,2,\dots,M$, delete $O_i$ from $H$ if the polygonal line segment bounded by the left and right neighboring control points of $O_i$ in $H$ can be fitted by a single B\'{e}zier cubic with error smaller than $\tau_e$.
	\end{enumerate}
	
As for the output, if $\Sigma_{\lambda^*}$ is not a circle, we write the points in $H$ together with the estimated defining points~\eqref{eq_B2} for each segment into a SVG format with the specification of drawing cubic B\'{e}zier curves. For the visualization purpose, if a fitted B\'{e}zier curve has a maximal absolute curvature smaller than $0.001$, we assign a straight line. Note that this value $0.001$ is consistent with the value for $\delta$ in Section~\ref{sec_1}. If $\Sigma_{\lambda^*}$ is a circle, we write its estimated center and radius into the SVG with the specification of drawing a circle. The pseudo-codes are presented in Algorithm~\ref{iterative} (with sub-procedures described in Algorithm~\ref{degenerate} and Algorithm~\ref{merge}), which can be parallelized to apply to outlines with multiple connected components.

%

	\begin{algorithm}
		\KwIn{ $\Sigma^0_{\lambda^*}$: a polygonal Jordan curve sampled from the level line of the bilinear interpolated image $u$ corresponding to the level $\lambda^*$. 
		$\tau_e$: approximation error threshold. $\sigma_0$: smoothness parameter. Fixed parameters: $\Delta\sigma = 0.5$ , $K = 4$.}
		\KwOut{Cubic B\'{e}zier polygon $\mathcal{B}(H)$ specified by the vertex set $H$, or a perfect circle.}

		Apply the affine shortening to smooth $\Sigma_{\lambda^*}^0$ up to scale $\sigma_0$, which yields the sub-pixel smooth outline $\Sigma_{\lambda^*}=\{P_i\}_{i=0}^N$.
		
		\For{$k=1,2,\dots, K$}{Evolve $\Sigma_{\lambda^*}$ up to scale $k\Delta\sigma$ via the affine shortening, denoted by $\Sigma_{\lambda^*}^k=\{P_i^k\}_{i=0}^{N^k}$. 
			
			Compute curvature $\kappa_i^k$ of $\Sigma_{\lambda^*}^k$ at $P_i^k$ according to~\eqref{eq_curvature}.
			
			Denoise the data $\{\kappa_i^k\}_{i=1}^{N^k}$ by moving average, based on which curvature extrema $\{X_i^{k}\}_{i=1}^{S^k}$ are located by~\eqref{eq_criterion}.}
		
		Initialize $H=\varnothing$.
		
		\If{$S^K\geq  1$}{
			\For{$i=1,\dots, S^K$}{
			 	Set $X_i^{(K)}=X_{i}^K$.
				
				\For{$k=K,K-1,\dots,1$}{
					Solve the problem~\eqref{eq_opt} associated with $X_i^{(k)}$
					
					\If{\eqref{eq_opt} has a solution }{
						Denote the solution by $X_i^{(k-1)}$.
						\If{k=1}{Insert $X_i^{(0)}$ into $H$.}
					}
					\Else{
						\textbf{break}
					}	
				}	
			}
		}
		\	\Else{
			Run Algorithm~\ref{degenerate}.
		}
		
		\For{$i=1,\dots, M$}{
			Fit the line segment $\{O_i=P_{j(i)},P_{j(i)+1},\dots, P_{j(i+1)}=O_{i+1}\}\subset\Sigma_{\lambda^*}$ using least square cubic B\'{e}zier~\eqref{eq_min2}.
			
			Obtain the right tangent $T_i^+$ at $O_i$ and left tangent $T_{i+1}^-$ at $O_{i+1}$ according to~\eqref{eq_tangent}.
		}
		
		\For{$i=1,\dots, M$}{
			\If{Condition~\eqref{eq_tangent_cond} holds}{
				Remove $O_i$ from $H$.}
		}
		
		\If{$H=\varnothing$}{
			Run Algorithm~\ref{degenerate}.
		}
		
		Compute approximation error $e$ of the B\'{e}zier polygon $\mathcal{B}(H)$ via~\eqref{eq_error}.
		
		\While{$e>\tau_e$}{
			Insert into $H$ a point $P_{\text{new}}\in\Sigma_{\lambda^*}$ furthest from $\mathcal{B}(H)$ that satisfies~\eqref{eq_condition}.
			
			Recompute the error $e$ of $\mathcal{B}(H)$.
			
		}
	
	\textit{(Optional)} Run Algorithm~\ref{merge} to further shrink the size of $H$.

		\caption{{\bf Shape Vectorization by Affine Scale-space} \label{iterative}}
	\end{algorithm}

		\begin{algorithm}
	
				\If{$\Sigma_{\lambda^*}$ is a circle}{
		Take three equidistant points on $\Sigma_{\lambda^*}$ to compute the center $O$ and radius $r$.
		
		\Return $O, r$.
	}
	\Else{
		Find the most distant pair of points on $\Sigma_{\lambda^*}$: $O_1$, $O_2$. Set $H=\{O_1,O_2\}$.
	}

		\caption{{\bf Sub-procedure for the Degenerate Case} \label{degenerate}}
\end{algorithm}

		\begin{algorithm}
		
	Suppose $H=\{O_i\}_{i=1}^M$.
	
	Define $M'=M$, $P^-=O_{M'}$, $P^0= O_{1}$, and $P^+= O_{2}$.
	
	\For{$i=1,\dots,M$}{
	Fit the the polygonal line segment bounded by $P^-$ and $P^+$ using least square cubic B\'{e}zier~\eqref{eq_min2}, and denote the fitting error by $e$.
	
	\If{$e<\tau_e$}{
	 $P^-\gets O_{i}$, $P^0\gets O_{\text{mod}(i+1,M')}$, $P^+\gets O_{\text{mod}(i+2,M')}$.
	}
	\Else{
		$M'\gets M'-1$
		
		$P^-\gets O_{\text{mod}(i-1,M')}$, $P^0\gets O_{\text{mod}(i+1,M')}$, $P^+\gets O_{\text{mod}(i+2,M')}$.
	}
}
	\caption{{\bf Simple B\'{e}zier Cubic Merging} \label{merge}}
\end{algorithm}
	
	\section{Numerical Results}\label{sec_5}
	In this section, we present some numerical experiments to demonstrate the performance of our proposed algorithm. After obtaining the SVGs from~\cite{SVG}, we rasterize them as PNG images, which are used as inputs in the following experiments. The inputs are either binary or gray-scale. We choose the level line corresponding to $\lambda^*=127.5$ to approximate the outlines throughout the experiments. By default, we set the error threshold $\tau_e = 1$, so that the vectorized outline is guaranteed to have sub-pixel level of accuracy; and the smoothness parameter $\sigma_0=1$. For the parameters in~\eqref{eq_opt}, we fix $D=10$ and $\alpha = 0.9$. The silhouettes used in the following experiments are collectively displayed in Table~\ref{tab_dataset}. If without any specifications, we apply the proposed method without the optional merging step.
	\subsection*{General Performance}
\begin{figure}
		\centering
		\begin{tabular}{cc}
			(a)&(b)\\
			\includegraphics[scale=1]{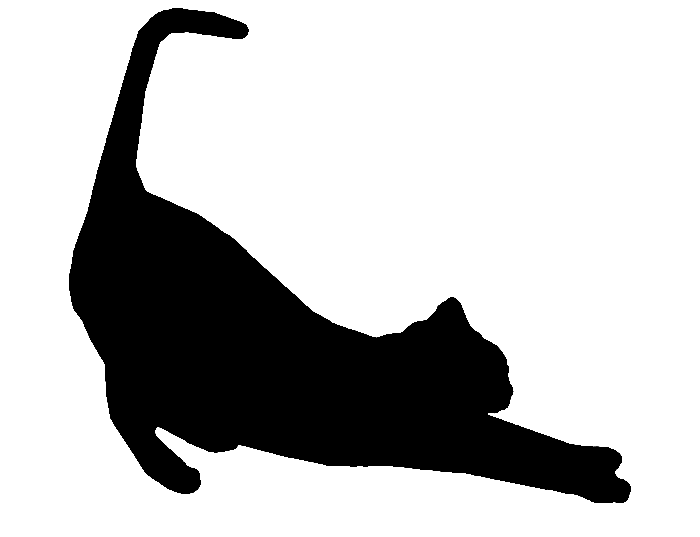}&
			\includegraphics[scale=0.24]{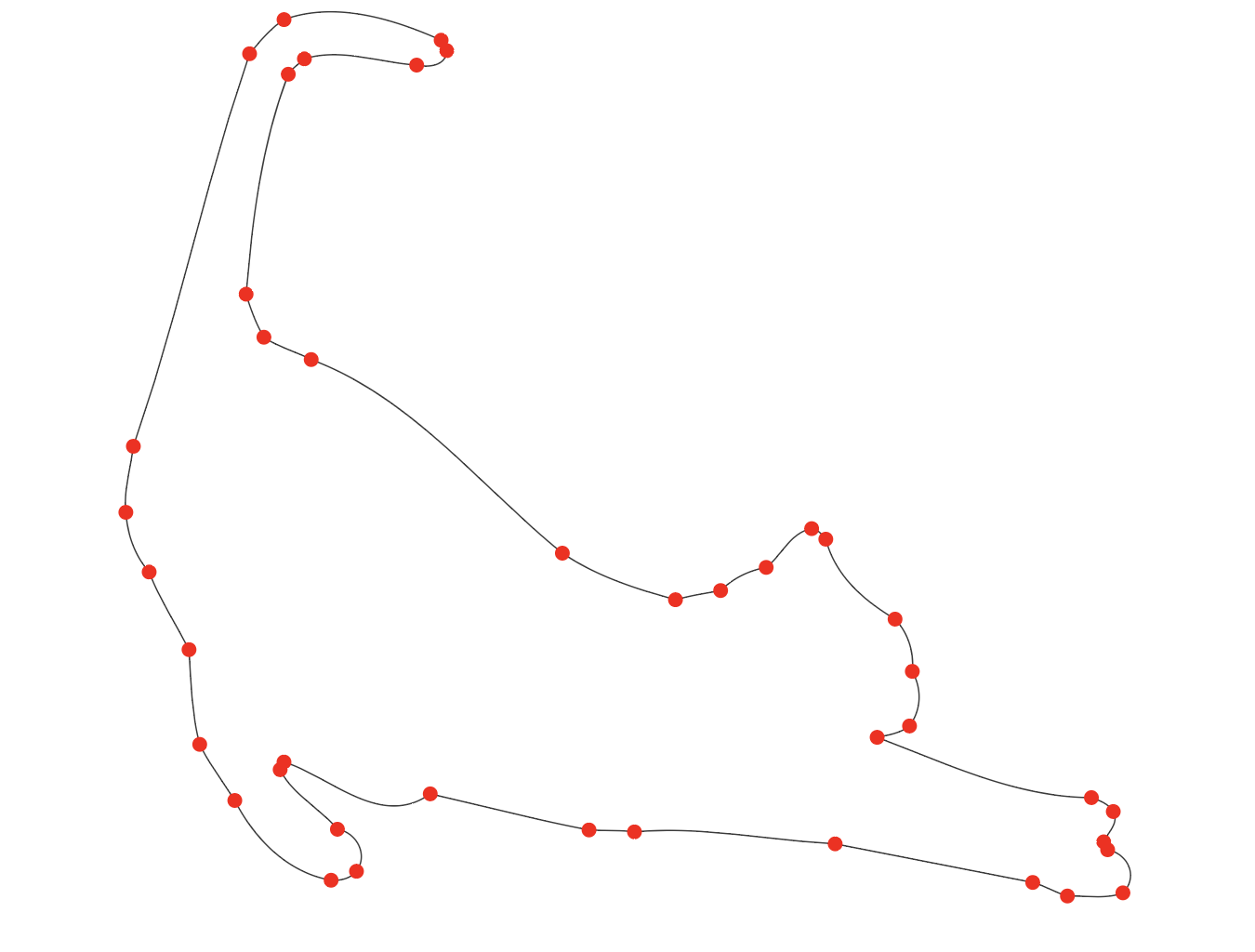}\\
			(c)&(d)\\
			\includegraphics[scale=0.5]{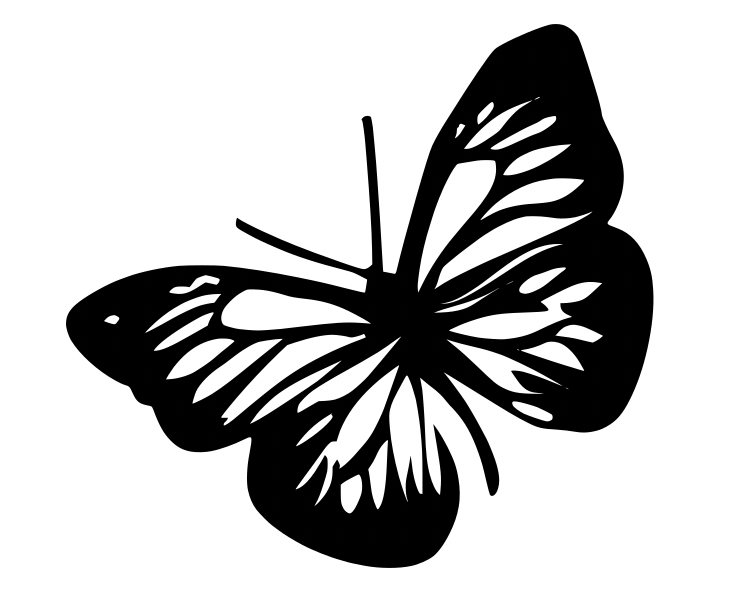}&
			\includegraphics[scale=0.27]{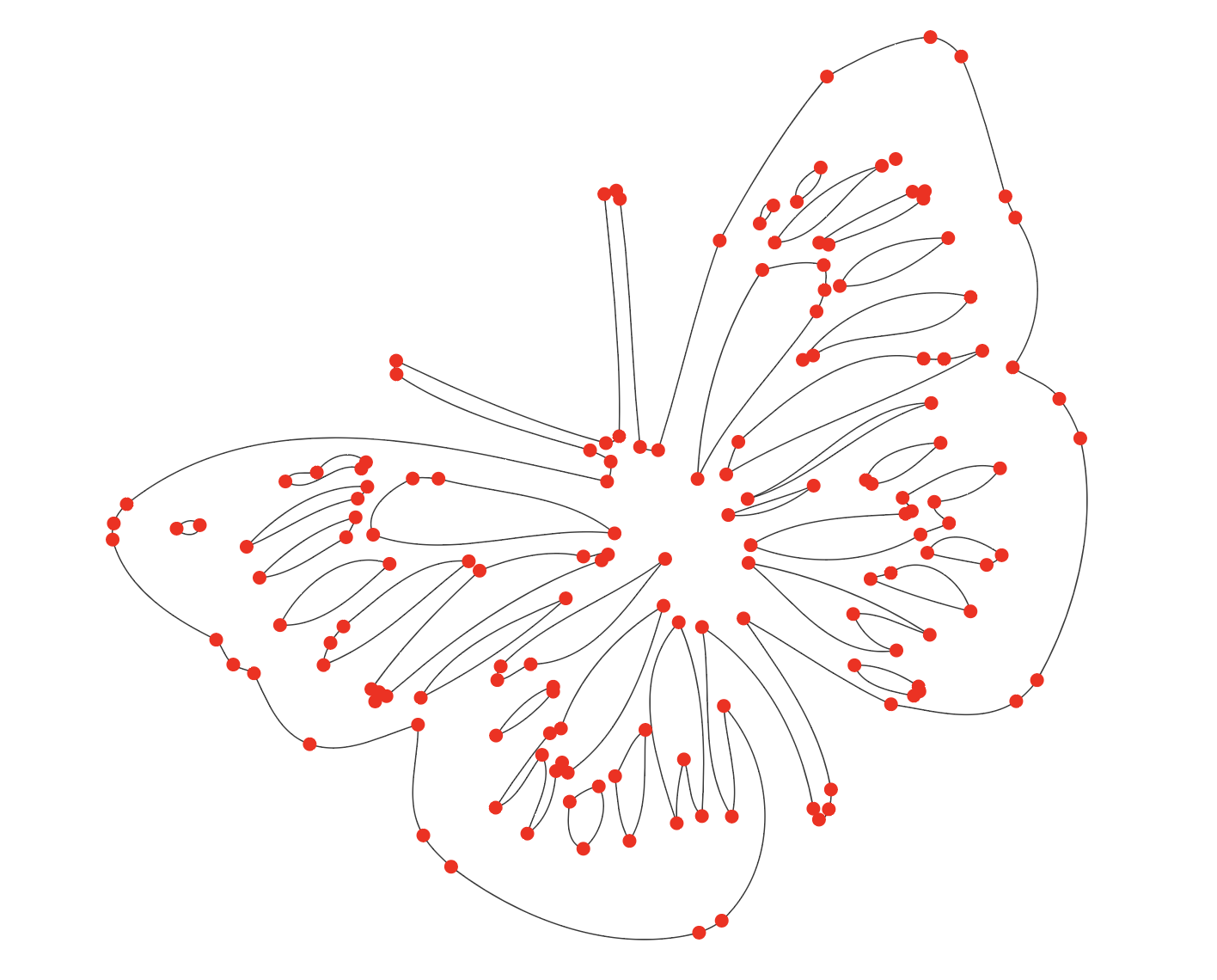}\\
			(e)&(f)\\
			\includegraphics[scale=0.41]{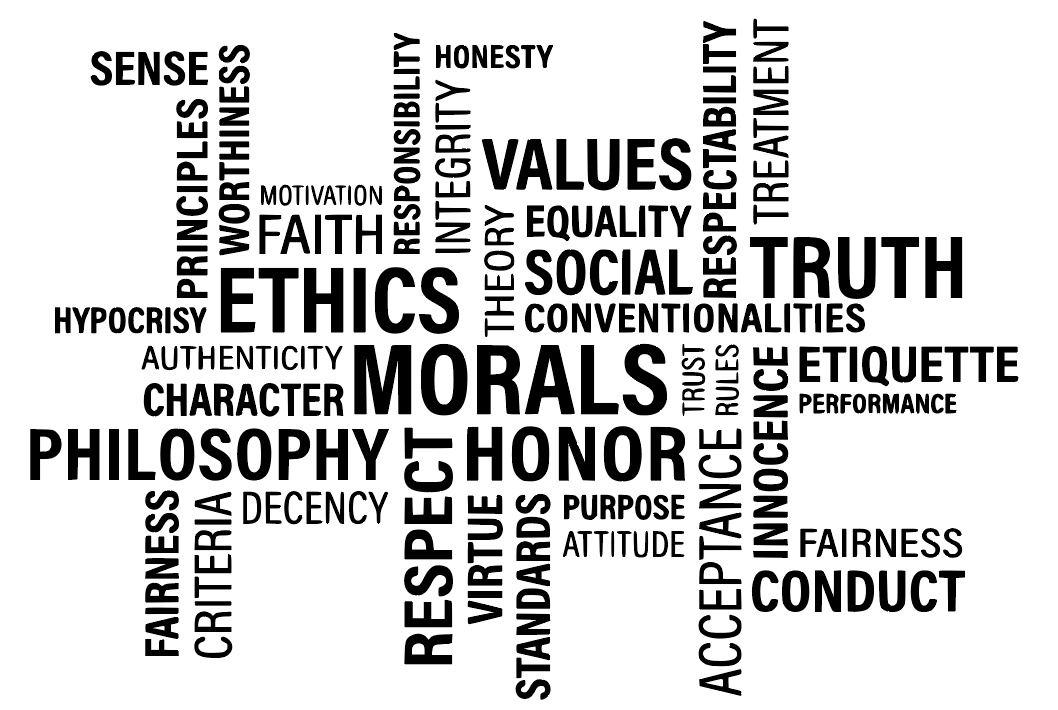}&
			\includegraphics[scale=0.175]{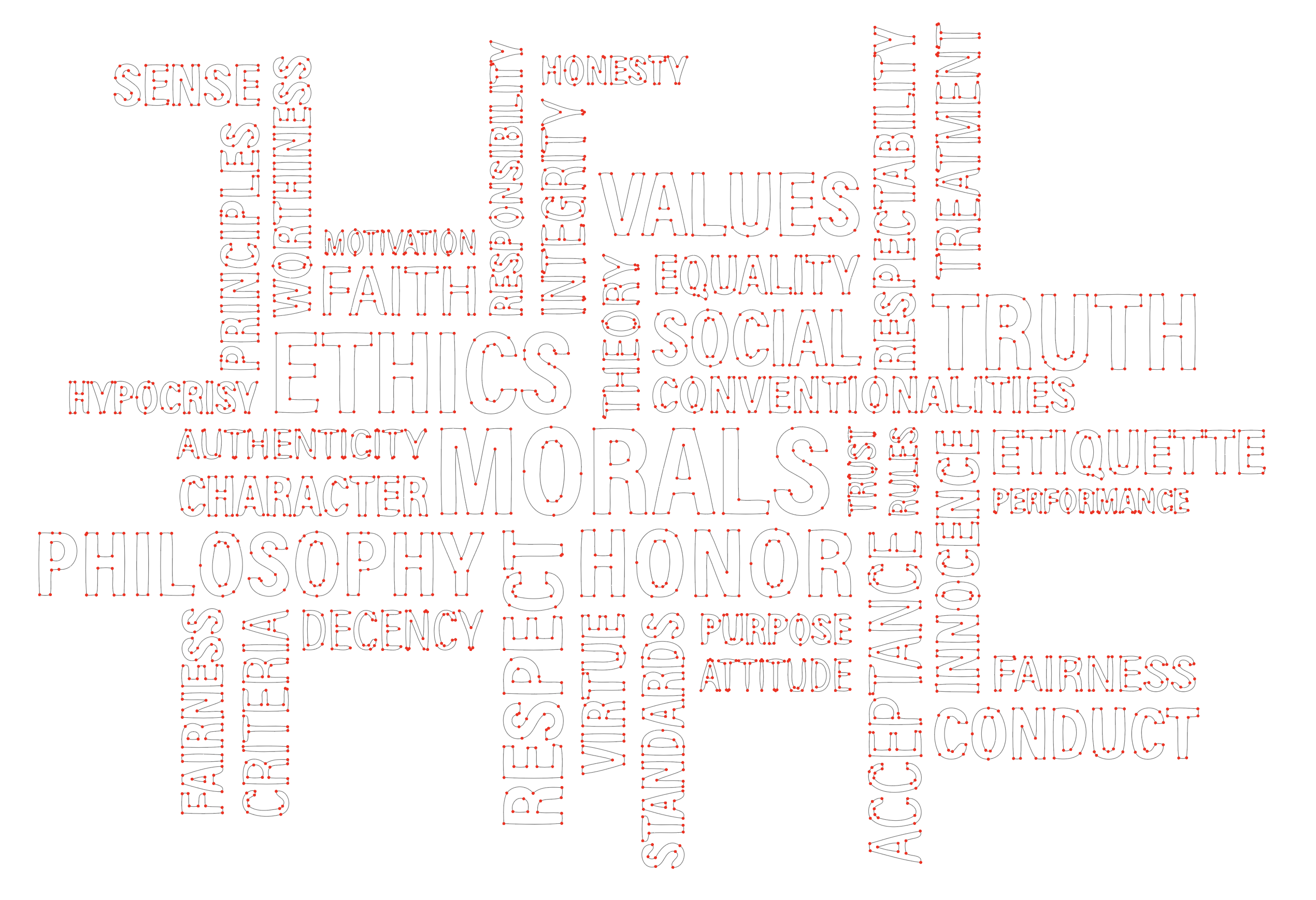}\\
			(g)&(h)\\
			\includegraphics[scale=0.15]{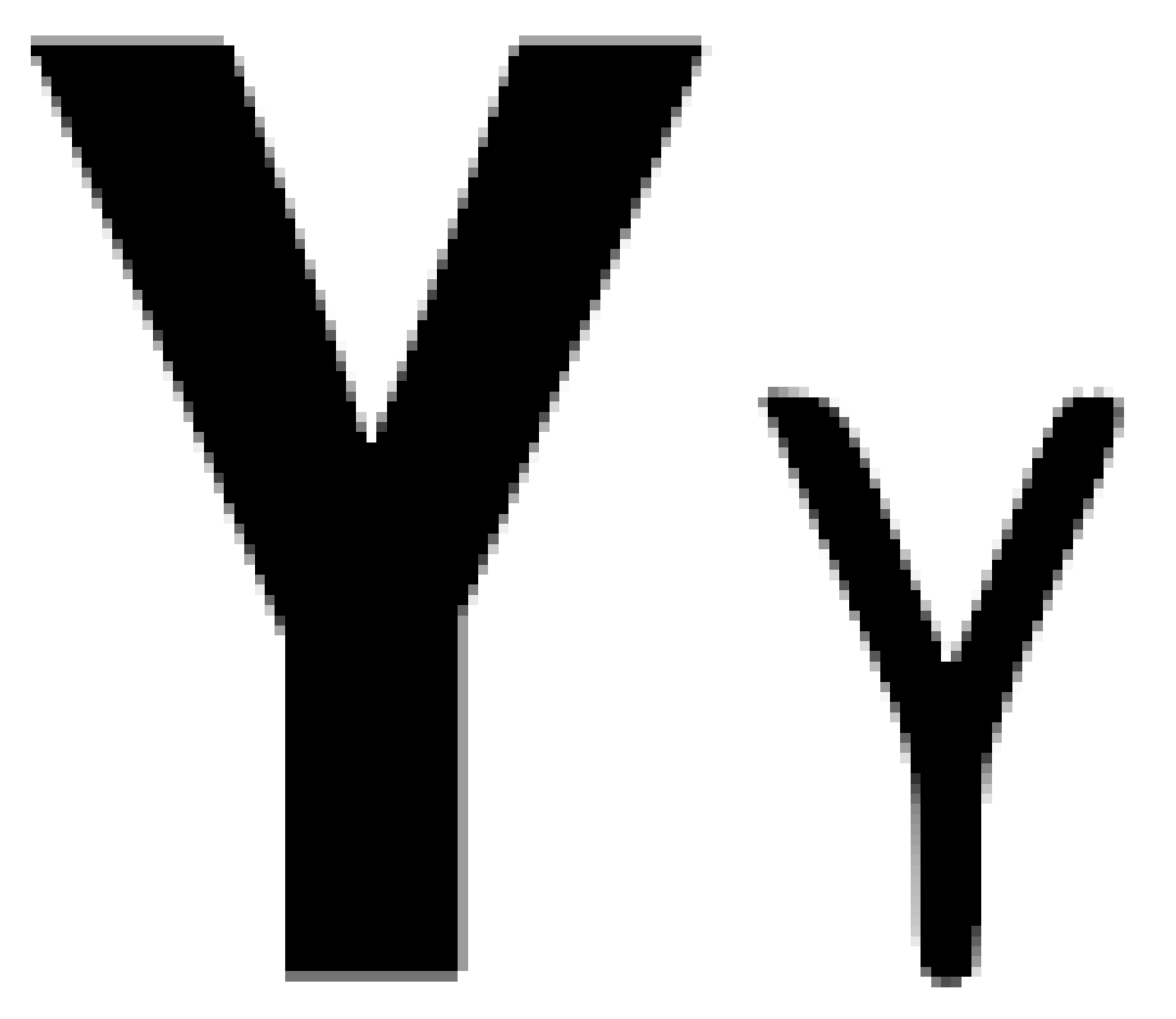}&
			\includegraphics[scale=0.15]{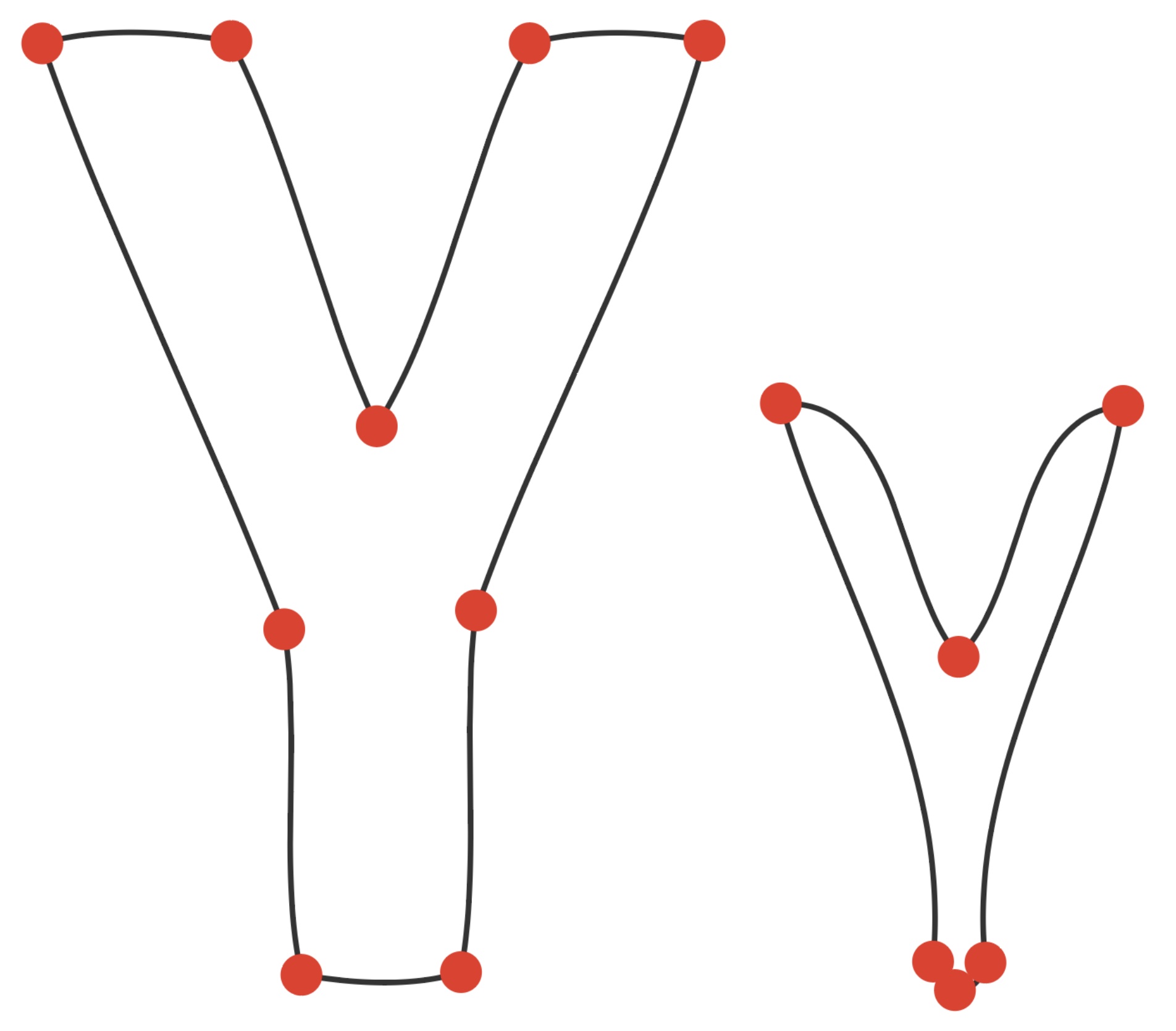}
		\end{tabular}
		\caption{Examples of results of the proposed algorithm for silhouette vectorization. (a) Cat and (b) its vectorized outline ($42$ control points). (c) Butterfly and (d) its vectorized outline ($158$ control points). (e) Text design and its vectorized outline ($2683$ control points). Each red dot signifies the location of a control point. (g) Two letters exerted from (e) scaled up with the same magnitude. (h) Zoom-in of the vectorization (f) on the two letters corresponding to (g). }\label{fig_example}
	\end{figure}

	We present some results of the our proposed algorithm in Figure~\ref{fig_example}. In (a), we have a silhouette of a cat. It has a single outline curve which contains multiple sharp corners on the tail, near the neck and around the paws, etc. These features provide informative visual cues for silhouette recognition, and  our algorithm identifies them as control points for the silhouette vectorization shown as the red dots in (b). The outline of  a butterfly in (c) has multiple connected components. In addition to the control points corresponding to corners, we observe in (d) some others on smooth segments of the outline. They are inserted during the refinement step of our algorithm, where a single B\'{e}zier cubic is inadequate to guarantee the accuracy specified by the error threshold $\tau_e=1$. In (e), we show a tessellation of words and (f) presents the  vectorized result. The input is a PNG image of dimension $1934\times 1332$ and takes $346$ KB in the storage. In contrast, its silhouette vectorization, saved as a SVG file,  has in total $2683$ control points and takes $68$ KB  if the coordinates are stored in float, $36$ Kb if stored in integers. In this example, our algorithm provides a compresion ratio of about $80.35\%$  for float type, and a $89.60\%$  compression ratio for the integer type.  Moreover, the total computational time for this case only takes $0.83$ seconds. Similar statistics for the other two examples are summarized in Table~\ref{tab}.  Our algorithm is both effective and efficient. 
	
	\begin{table}
		\centering
		\begin{tabular}{|c|c|c|c|c|c|}
			\hline
			Shape& Image Dim. & Size& Result Size (float) &  Result Size (int) & Proc. Time\\\hline
			Cat (a) &  $700\times537$ &$5$ KB&$2$ KB ($60\%$)&$1$ KB ($80\%$)&$0.10$ Sec.\\
			Butterfly (c) &  $732\times 596$ &$178$ KB &$5$ KB ($97.19\%$)&$3$ KB ($98.31\%$)&$0.15$ Sec.\\
			Text (e) &  $1934\times1332$ &$346$ KB&$68$ KB ($80.35\%$)&$36$ KB ($89.60\%$)&$0.83$ Sec.\\\hline
		\end{tabular}
		\caption{Performance statistics of the proposed algorithm applied to examples in Figure~\ref{fig_example}. The float type result stores the control point coordinates as float type, and the int type result stores them as integer type.}~\label{tab}
	\end{table}
	
	\subsection*{Degenerate Cases}
\begin{figure}
		\centering
		\begin{tabular}{cc}
			(a)&(b)\\
			\includegraphics[scale=0.5]{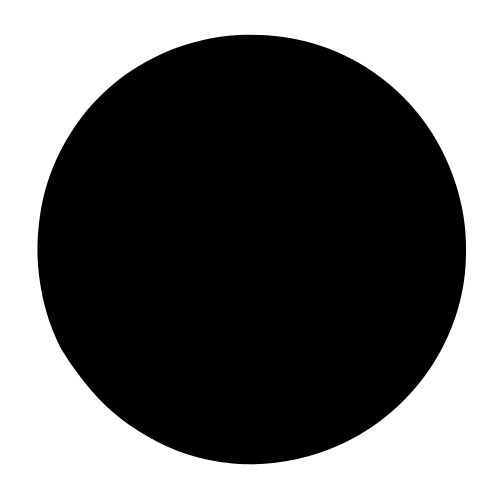}&
			\includegraphics[scale=0.28]{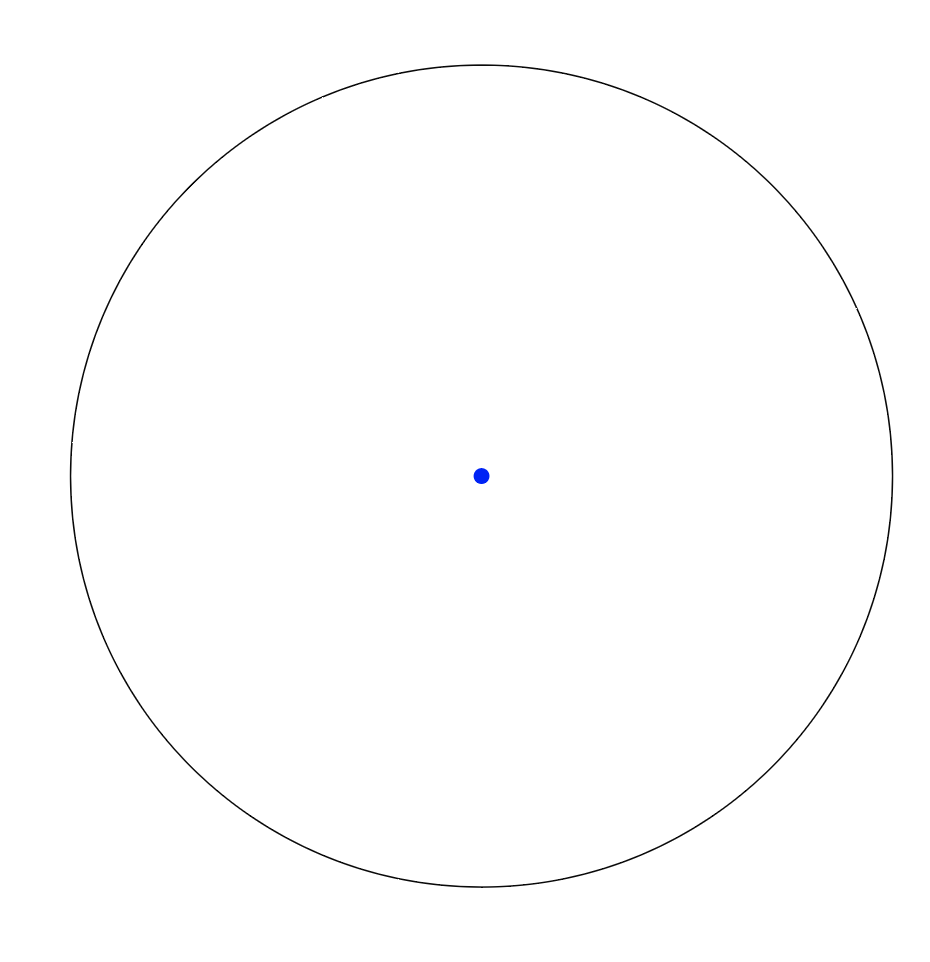}\\
			(c)&(d)\\
			\includegraphics[scale=0.5]{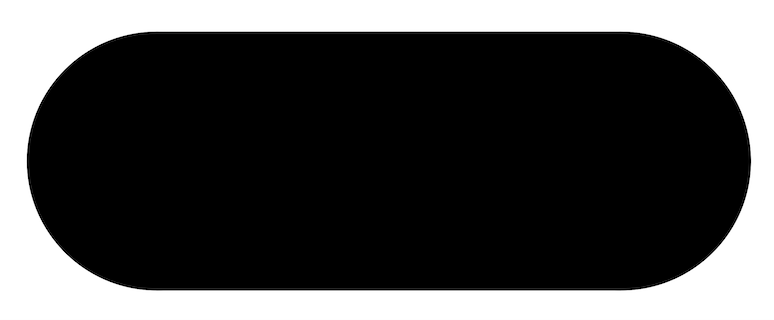}&
			\includegraphics[scale=0.28]{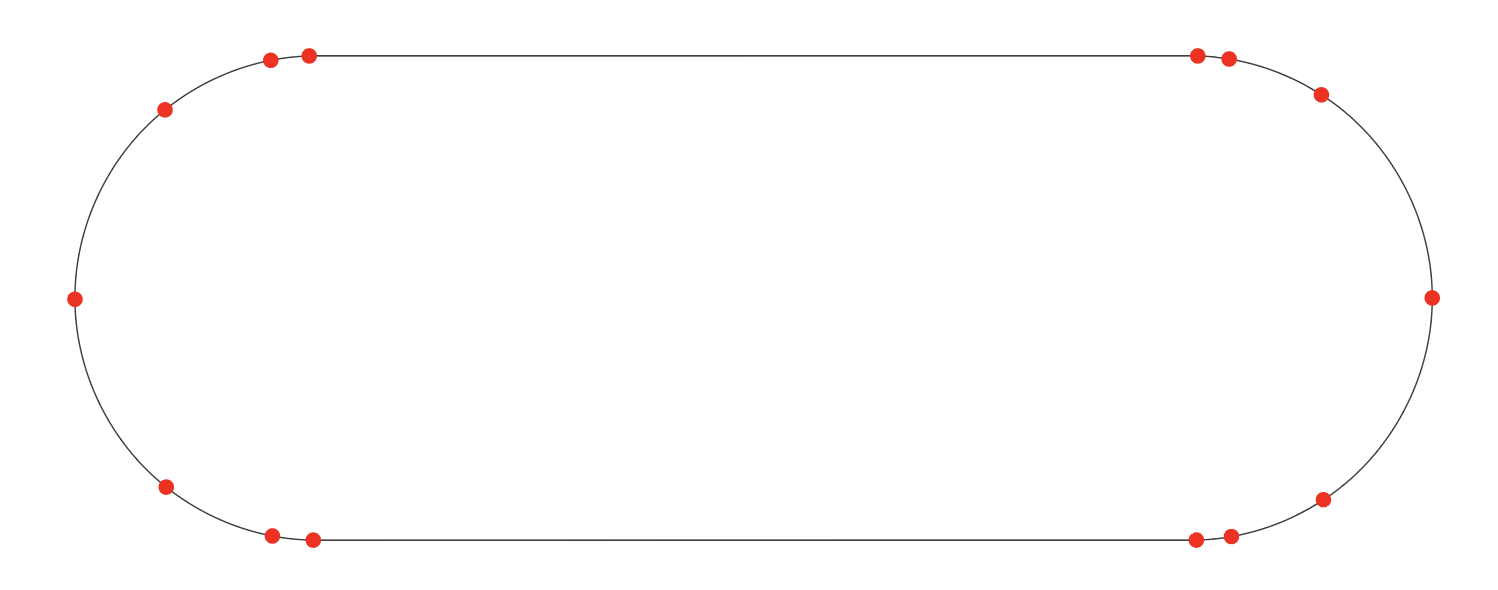}\\
			(e)&(f)\\
			\includegraphics[scale=0.8]{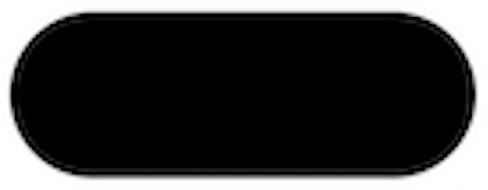}&
			\includegraphics[scale=0.7]{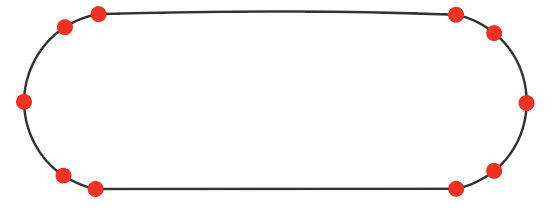}\\
			(g)&(h)\\
			\includegraphics[scale=0.8]{Figures/taiji}&
			\includegraphics[scale=0.42]{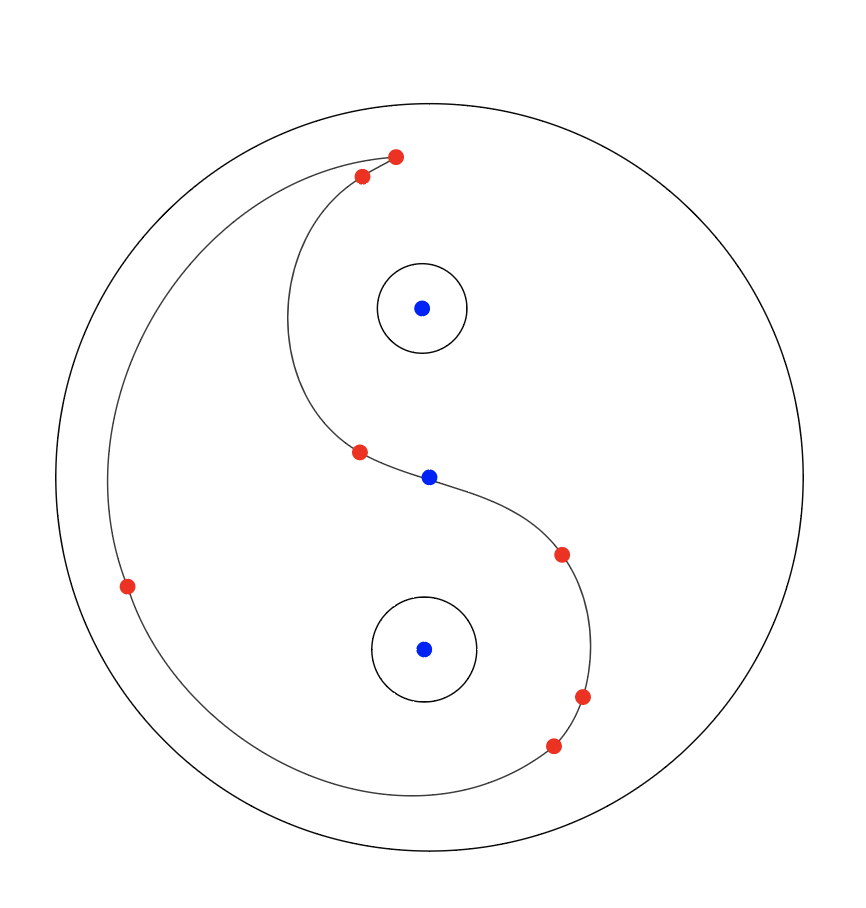}
		\end{tabular}
		\caption{Degenerate cases. In (a) and (c), no candidate control points were identified. Our algorithm handles such situations by checking if the outline is a circle. If it is, e.g. (a), the center and radius are computed and a circle is drawn without B\'{e}zier fitting; hence, there is no control point (red dots) on the vectorized outline (b). The blue dot indicates the center of the circle. If it is not a circle, e.g., (c), a pair of most distant points are inserted to initiate the B\'{e}zier fitting, such as in (d). (e) shows the low resolution version of (c) and (f) displays its vectorization. When the resolution is reduced, all the control points are identified curvature extrema.  In (g), three of the outline curves are identified as circles and the others are fitted by B\'{e}zier polygons. (h) shows the vectorized result.}\label{fig_degenerate}
	\end{figure}

	An important feature of our algorithm is that it offers flexibility in face of degenerate cases, where the silhouette does not have identifiable curvature extrema on its outline. A disk, as shown in Figure~\ref{fig_degenerate} (a), is the most common example. Once our algorithm classifies the outline as a circle, instead of fitting B\'{e}zier cubics, it directly approximates the center and radius of the circle and orders the SVG output to draw a perfect circle. See Figure~\ref{fig_degenerate} (b). 
	
	Figure~\ref{fig_degenerate} (c) shows another degenerate case. It consists of a rectangle in the middle and two half disks attached on its opposite sides, whose diameters are equal to the height of the rectangle. This particular silhouette has no strict curvature extrema on its outline. By computation,  its area is $172644$ and perimeter is $1742.07$; since $4\pi\text{Area}/\text{Perimeter}^2 = 4\times \pi \times 172644/ (1742.07)^2=0.7149 <1$, the outline is not a circle.  Hence the algorithm inserts a pair of most distant points on the outline, the left-most and the right-most points in this case, and conducts the B\'{e}zier fitting routine as in the non-degenerate cases.
	
	The design of this special procedure for degenerate cases is important for two reasons. First, it makes the algorithm adaptive to image resolutions. If we reduce the resolution of (c) from $774\times 320$ to $144\times 58$, whose magnified version is shown in (e), due to strong effects of pixellization, all the control points are identified as local curvature extrema. (f) shows the magnified vectorization of the low resolution image. Second, it improves the compression ratio. To fit a circle using B\'{e}zier polygon requires at least two pieces of cubics, hence we need to store the coordinates of at least $6$ points. With our algorithm, only the coordinate of the center and the value of the radius are required, which saves the space for $9$ float or int type data. Figure~\ref{fig_degenerate} (g) shows  mixture of degenerate and non-degenerate outline curves. The vectorization in (h) shows that the circles are represented as perfect circles, and the others are represented as B\'{e}zier polygons.

		\subsection*{Importance of the Control Point Update}
\begin{figure}
		\centering
		\begin{tabular}{ccc}
			(a)&(b)&(c)\\
			\includegraphics[scale=0.425]{Figures/knot}&
			\includegraphics[scale=0.26]{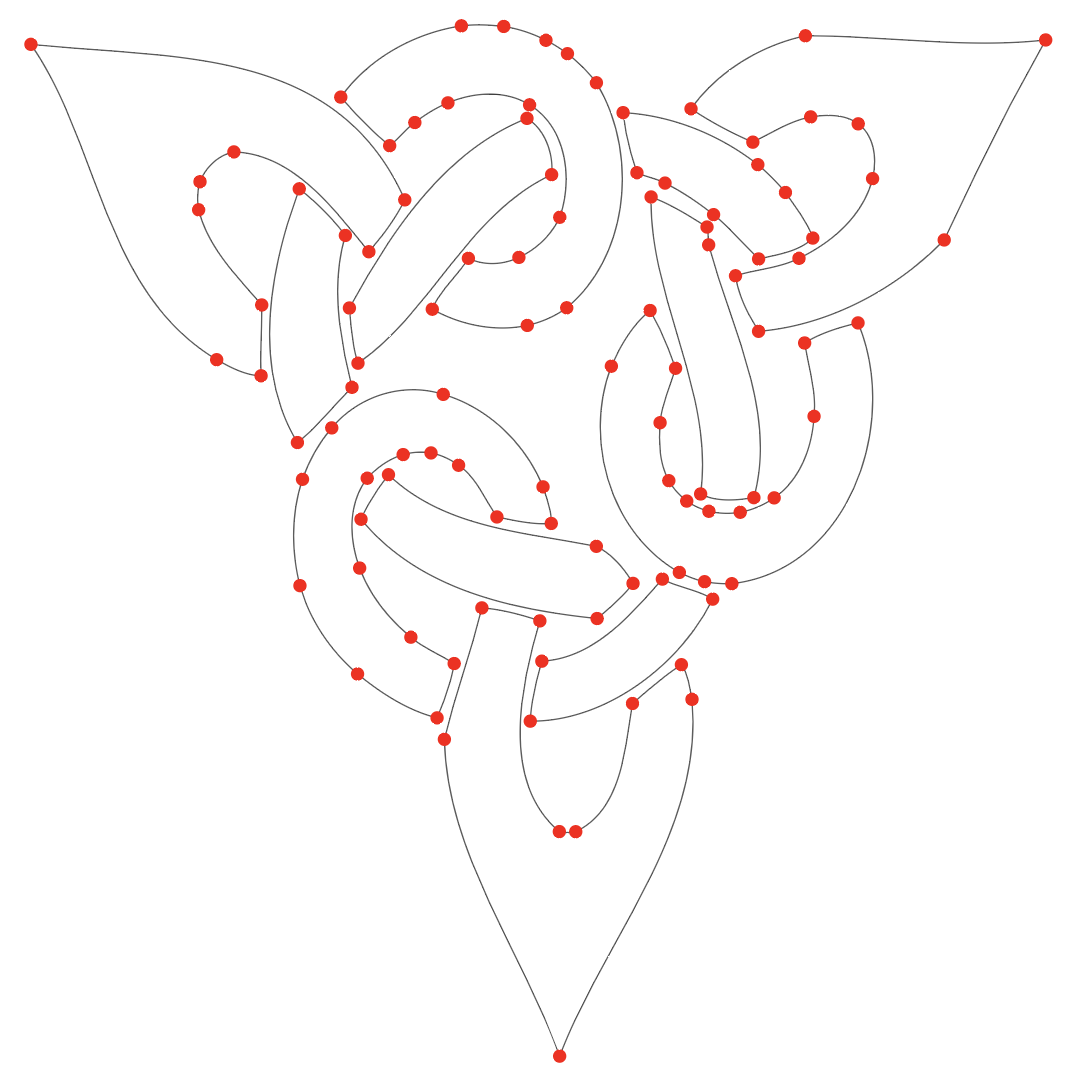}&
			\includegraphics[scale=0.26]{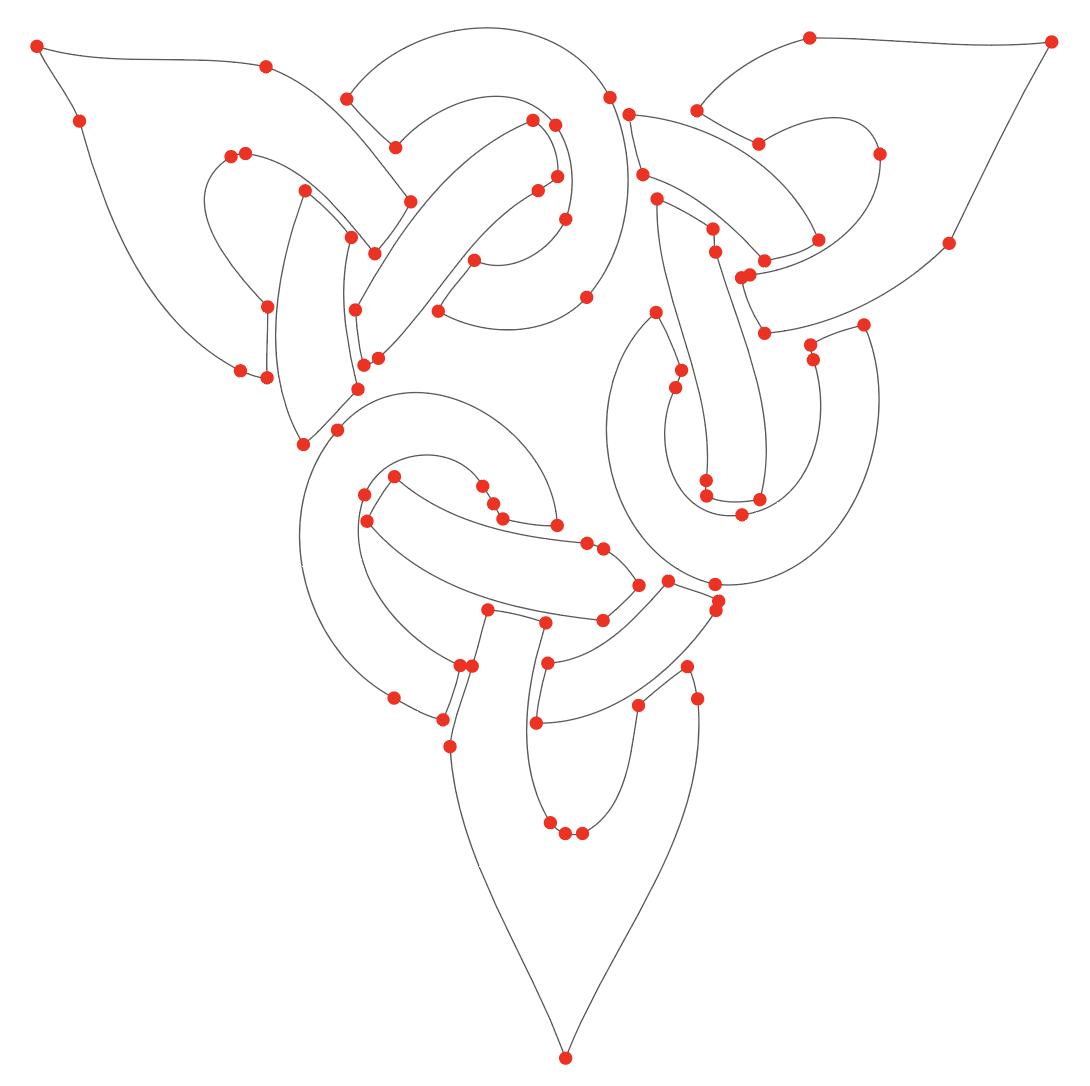}
		\end{tabular}
		\caption{Importance of refinement. (a) A silhouette of a knot. (b) Result without refinement (106 control points). (c) Result with refinement (86 control points). Compared to (b),  (c) has fewer control points distributed on the smooth curve segments, and some new control points are introduced to enhance accuracy.}\label{fig_refinement}
	\end{figure}

	Figure~\ref{fig_refinement} compares the vectorization using control points before and after the control point update, which are described in Section~\ref{sec_deletion} and \ref{sec_insertion}. The  outline of the knot silhouette in (a) shows curvature variations of multiple scales. If no refinement is applied, as shown in (b), the corners of larger scale are captured, while some inflection points are missed on the top-left component. Moreover,  many unnecessary control points appear on several arcs. In contrast, with the refinement, the result in (c) has fewer control points, but with a better approximation accuracy.  Notice that on the top-left component of the knot, the newly inserted control points are close to the inflection points, and on the arcs, only  $1$ or $2$ control points are generally needed.
	
	By refinement, we remove  extrema that do not represent salient corners and inset new control points to meet the accuracy requirement. Although the total number of control points may or may not decrease after refinement,  the distribution of the refined control points shows more correlation with the geometric features of the outline, and the vectorized result is improved.

	\subsection*{Effect of the Error Threshold $\tau_e$}
	The error threshold $\tau_e$ controls the accuracy of the B\'{e}zier polygon approximating the  outline.  When the value of $\tau_e$ is reduced, the user requires higher accuracy of the B\'{e}zier fitting.  Since any B\'{e}zier cubic  contains at most one inflection point, a single cubic only allows a limited amount of variations. Hence, by adding more control points to split the outline into shorter segments, the specified accuracy is achieved. 
	
\begin{figure}
		\centering
		\begin{tabular}{cc}
			(a) & (b)\\
		\includegraphics[scale=0.2]{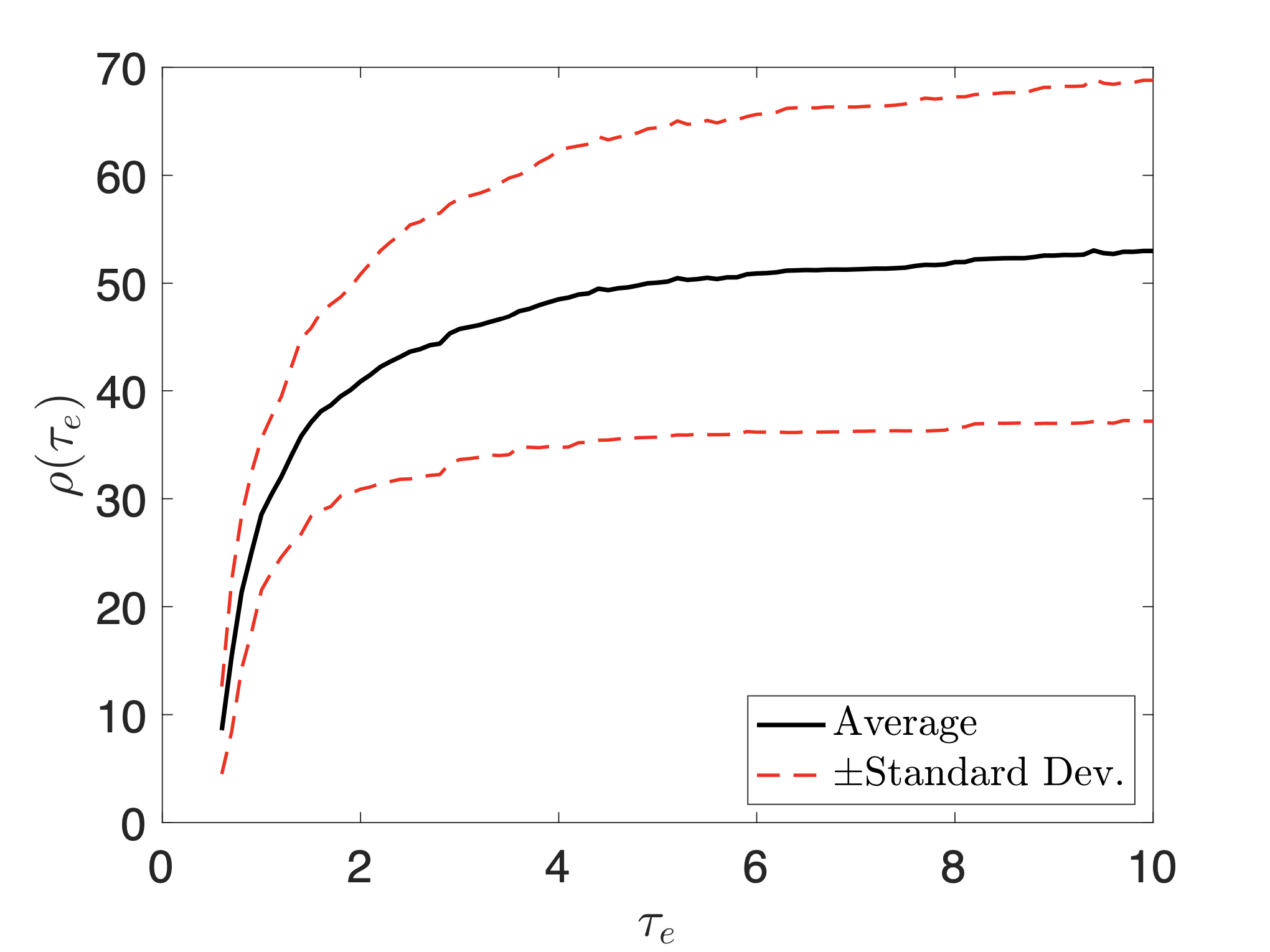}&
		\includegraphics[scale=0.2]{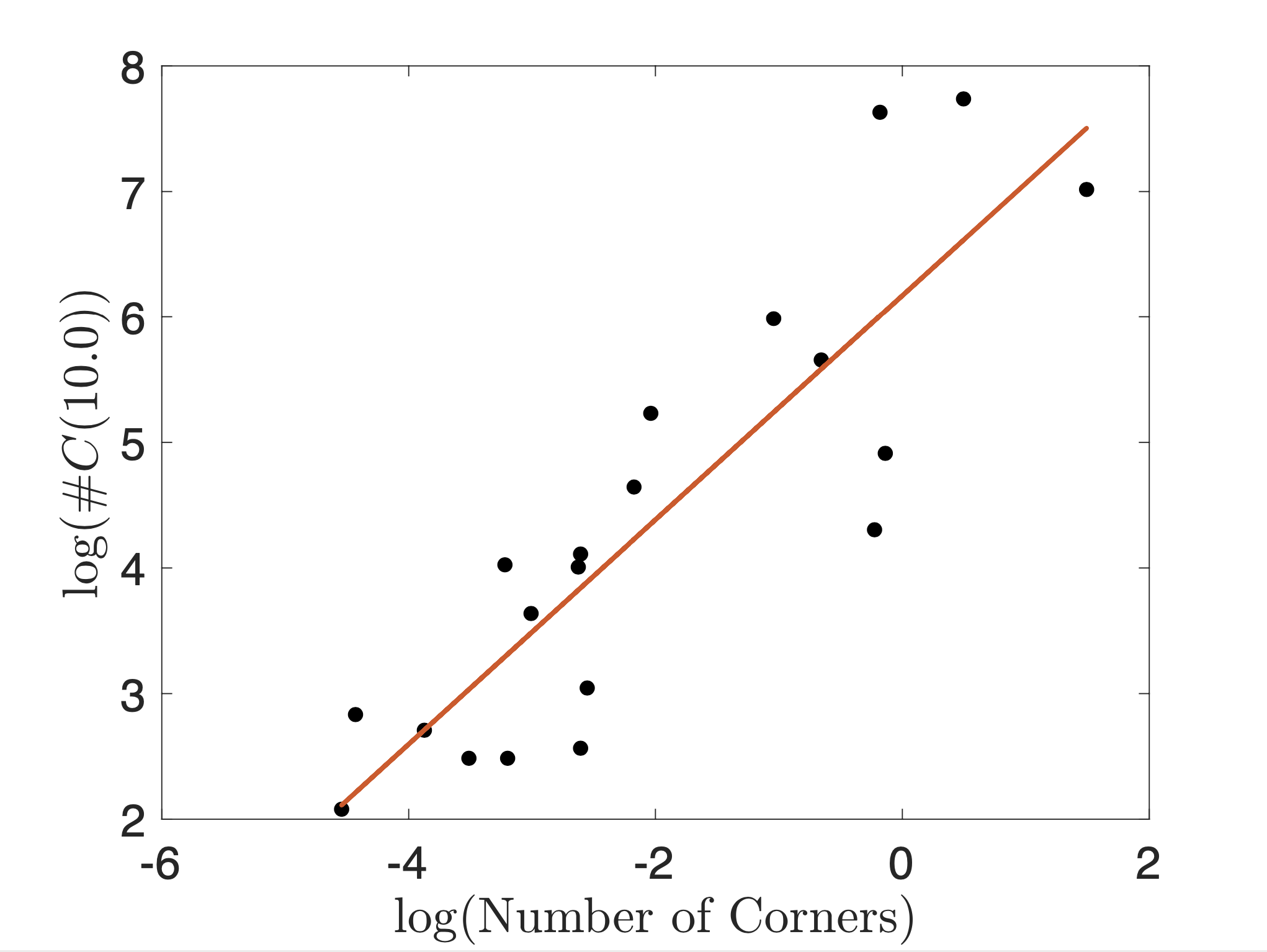}
		\end{tabular}
		\caption{(a) For the $20$ silhouettes in our data set (Table~\ref{tab_dataset}), the solid curve shows the average relative reduction of the number of control points $\rho(\tau_e)$~\eqref{eq_percentage}, and the dashed curves indicate the standard deviations.  (b) The positive relation between the number of control points when $\tau_e=10.0$ is large and the number of corners of a silhouette. Each dot represents a sample in our data set. The red curve is computed by linear regression with a goodness of fit $R^2=0.75592$. }\label{fig_threshold2}
	\end{figure}

	To better illustrate the effect of varying the threshold $\tau_e$, we computed in percentage the reduction of the number of control points when the threshold is $\tau_e>0.5$ compared to that when the threshold is $0.5$:
	\begin{align}
	\rho(\tau_e) = \frac{\# C(\tau_e)-\#C(0.5)}{\# C(0.5)}\times 100\%\;,\quad \tau_e>0.5\;.\label{eq_percentage}
	\end{align}
	Here $\#C(\tau_e)$ denotes the number of control points when the threshold is $\tau_e$.  Figure~\ref{fig_threshold2} (a) shows the average values and the standard deviations of~\eqref{eq_percentage} when we apply the proposed method to the $20$ silhouettes in our data set. We observe that when $\tau_e<1$, the effect of increasing $\tau_e$ is the strongest: the number of control points reduces exponentially. On average, the percentage curves show inflection points around $\tau_e=1$, that is, when the fitted B\'{e}zier polygon has distance to the sub-pixel outline less than 1 pixel. After passing $\tau_e=1$, increasing $\tau_e$ has less impact on the variation of the number of control points. For even larger values of $\tau_e$,  there is almost no need of inserting new control points,  and the corresponding control points are closely related to the corners of the outline. This is justified by the regression in  Figure~\ref{fig_threshold2} (b), where each point represents a silhouette in our data set. It shows that there is a positive relation between the number of corners computed by the Harris-Stephens corner detector~\cite{harris1988combined} and the number of control points when $\tau_e=10.0$, which is relatively large.
	
	With large values of $\tau_e\gg 1$, the silhouette representation is more compact yet less accurate. With small values of $\tau_e<1$, we have a more accurate representation yet less efficient. From this point of view,  we would recommend $\tau_e=1$.
	
	\subsection*{Effect of the Smoothness Parameter $\sigma_0$}
	The smoothness parameter $\sigma_0$ adjusts the regularity of the smooth bilinear outline which approximates $\partial\mathcal{S}$. With larger values of  $\sigma_0$, oscillatory features of the given outline are suppressed, while with smaller values of $\sigma_0$, the vectorized silhouette preserves sharp corners.
	
\begin{figure}
	\centering
	\begin{tabular}{cccc}
		(a)&(b)&(c)&(d)\\
		\includegraphics[scale=0.2]{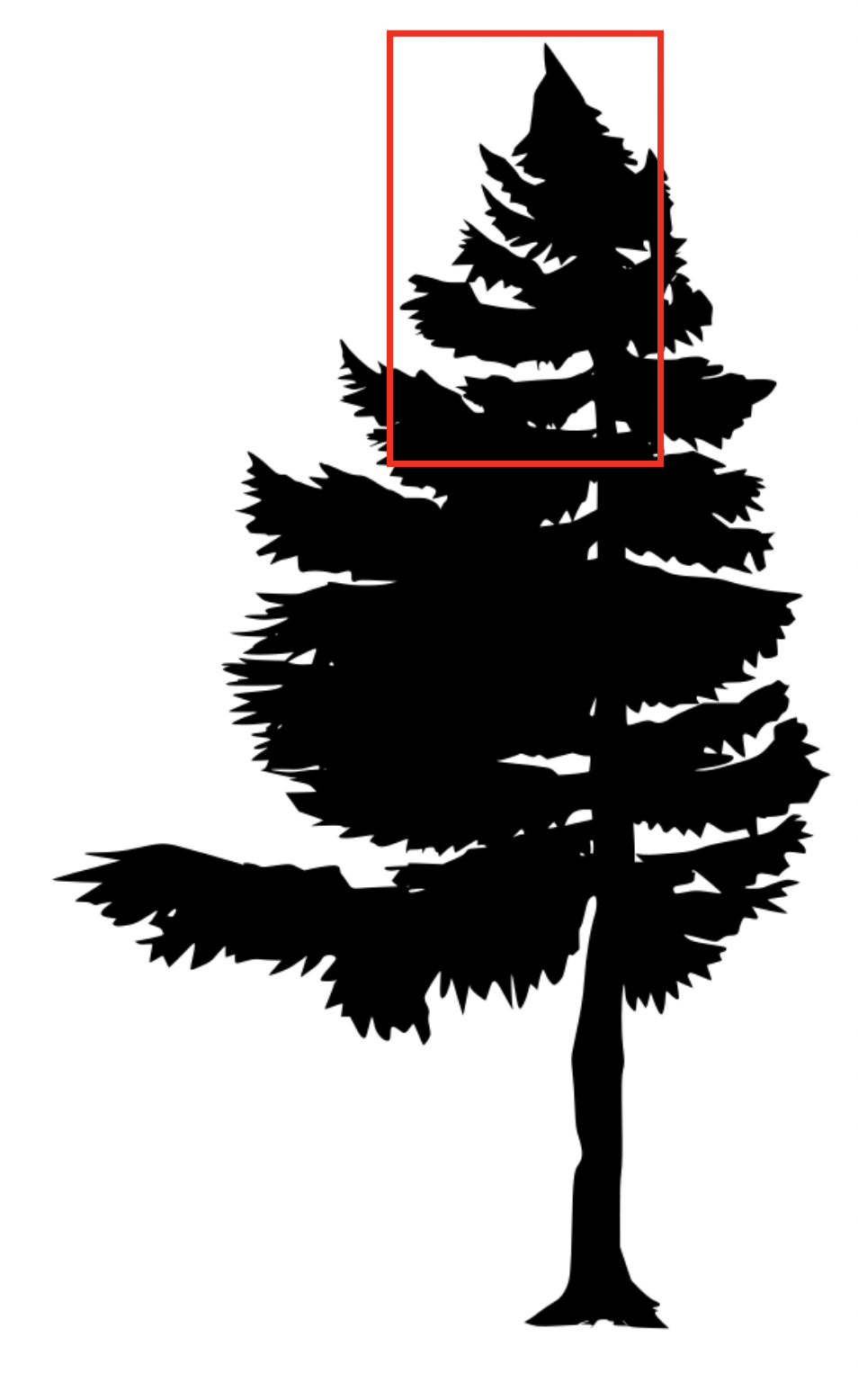}&
		\includegraphics[scale=0.24]{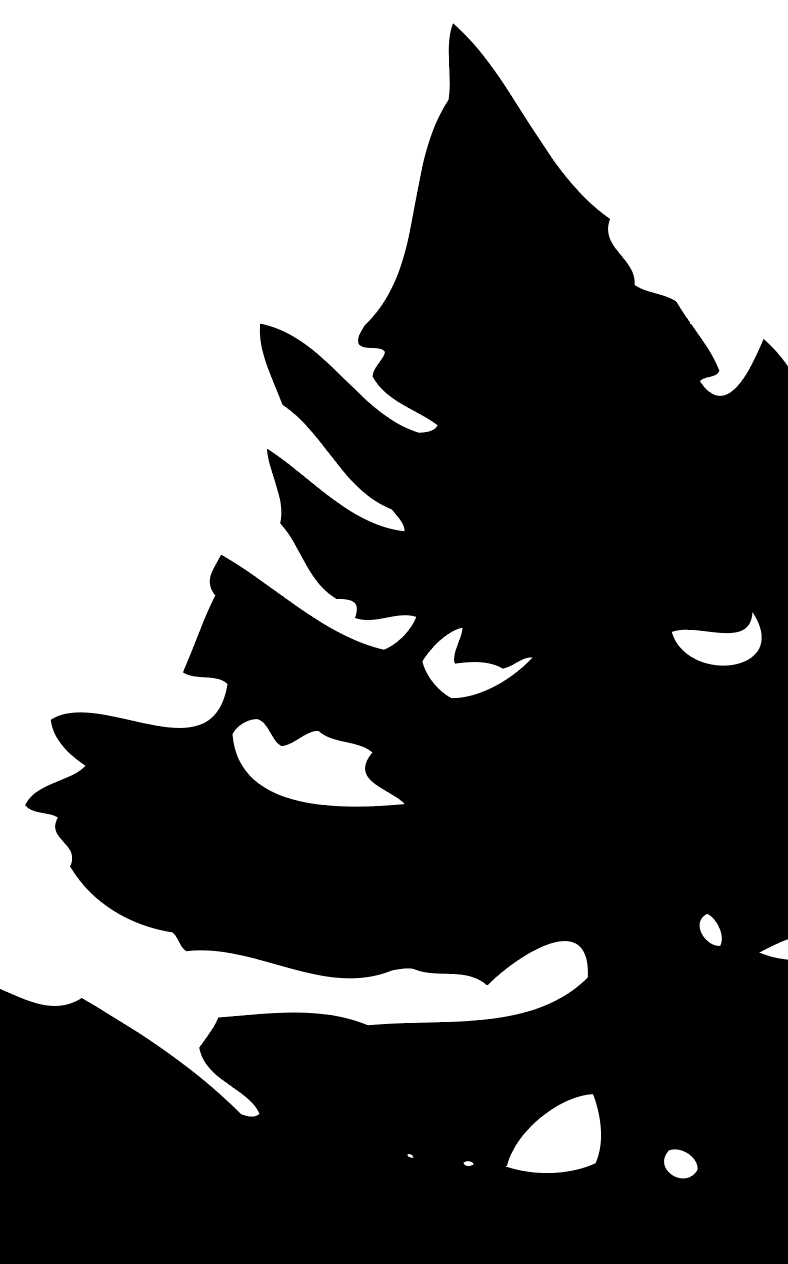}&
		\includegraphics[scale=0.24]{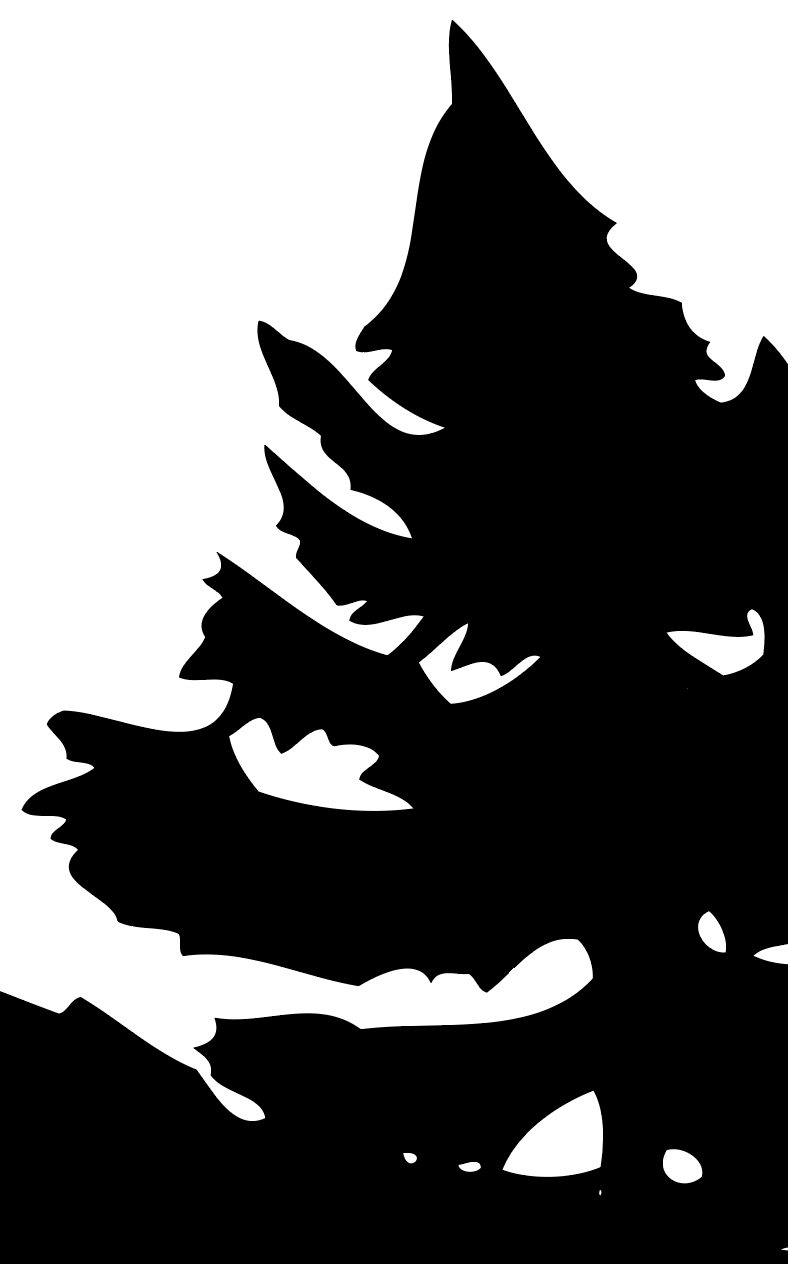}&
		\includegraphics[scale=0.24]{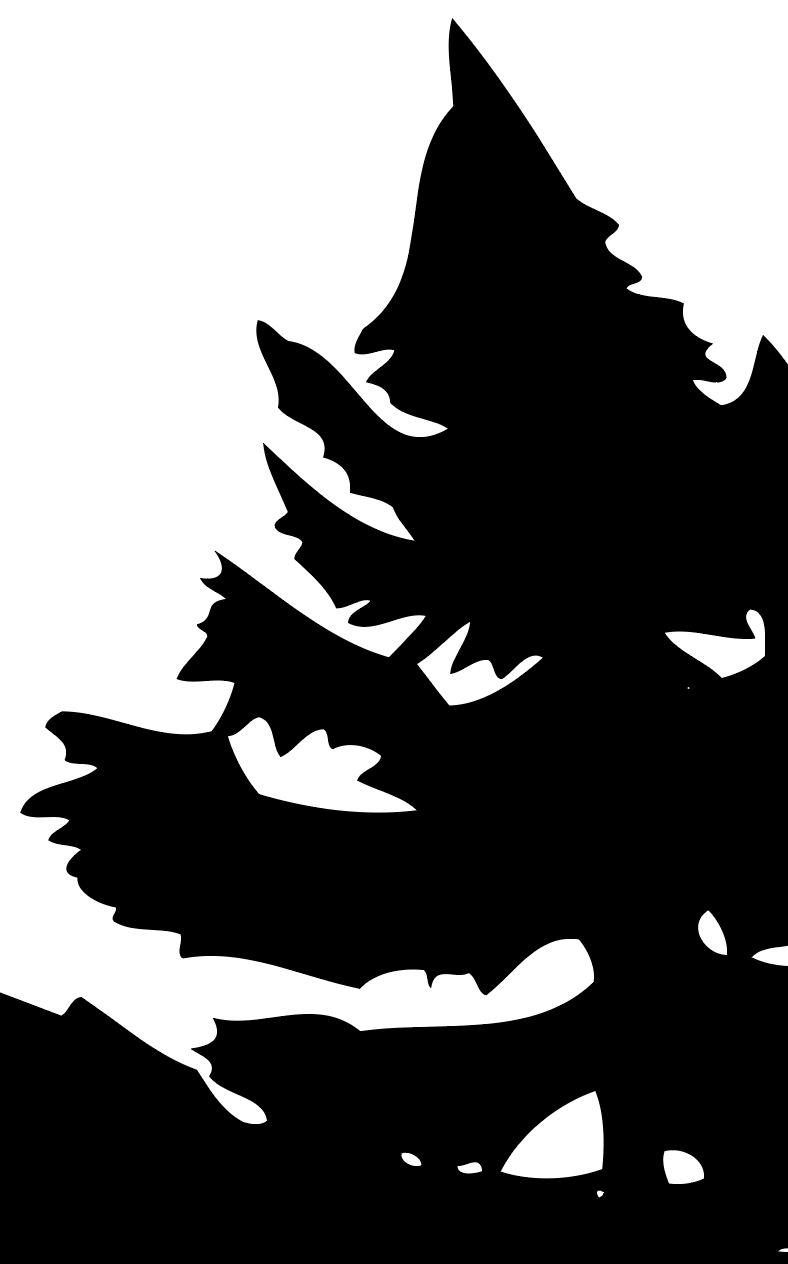}
	\end{tabular}
	\caption{Effect of the smoothing parameter $\sigma_0$. (a) A silhouette of a tree where the boxed region is examined in detail. Vectorization using (b) $\sigma_0=2.0$  ($362$ control points) (c) $\sigma_0=1.0$ ($448$ control points),  and  (d) $\sigma_0=0.5$ ($500$ control points). With smaller values of $\sigma_0$, the vectorized outline is sharper, and the number of control points increases.}\label{fig_smooth}
\end{figure}

	Figure~\ref{fig_smooth} demonstrates this effect of $\sigma_0$. We apply the proposed method using $\sigma_0=2.0$, $1.0$ and $0.5$ on the silhouette of a tree~(a), and the zoom-ins of vectorization results within the boxed region of (a) are presented in (b), (c), and (d), respectively.  Observe that the zig-zag feature around the tree's silhouette is better preserved by reducing $\sigma_0$. As a trade-off, this introduces  more control points to recover the sharpness of the outline.
	
	\subsection*{Stability Under Affine Transformations}
	
\begin{figure}
		\centering
		\begin{tabular}{ccc}
			(a)&(b)&(c)\\
			\includegraphics[scale=0.3]{Figures/cat2_1}&
			\includegraphics[scale=0.3]{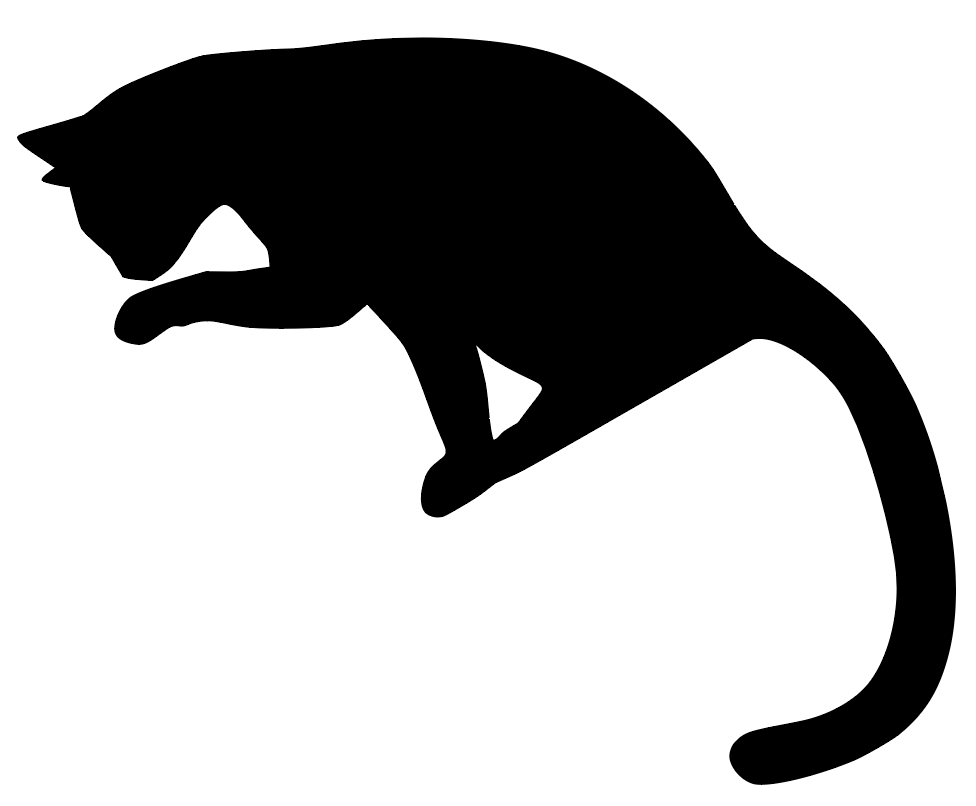}&
			\includegraphics[scale=0.3]{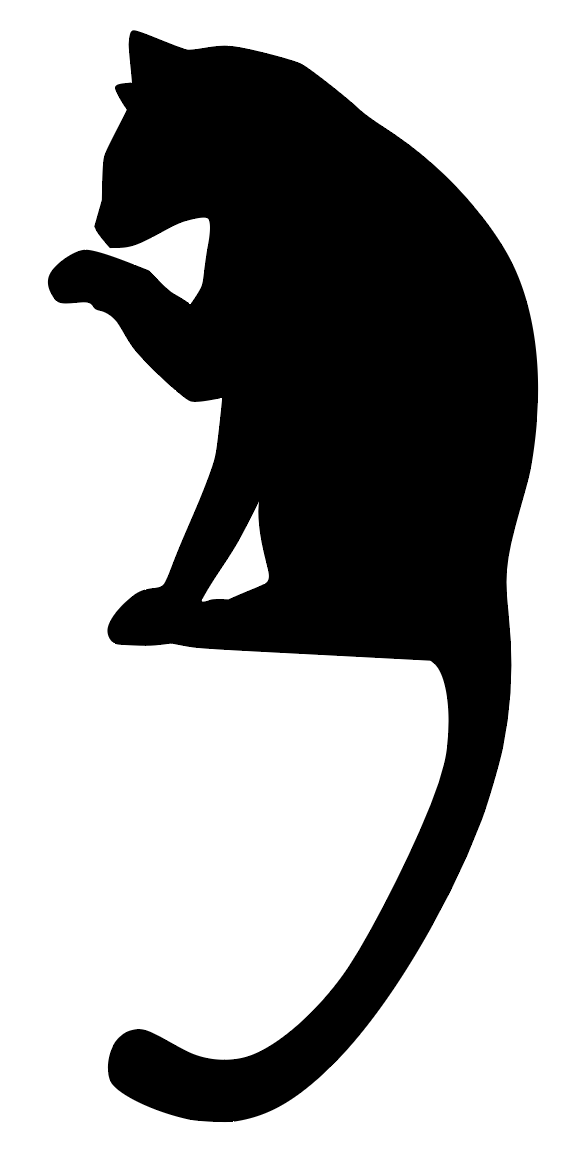}\\
			(d)&(e)&(f)\\
			\includegraphics[scale=0.3]{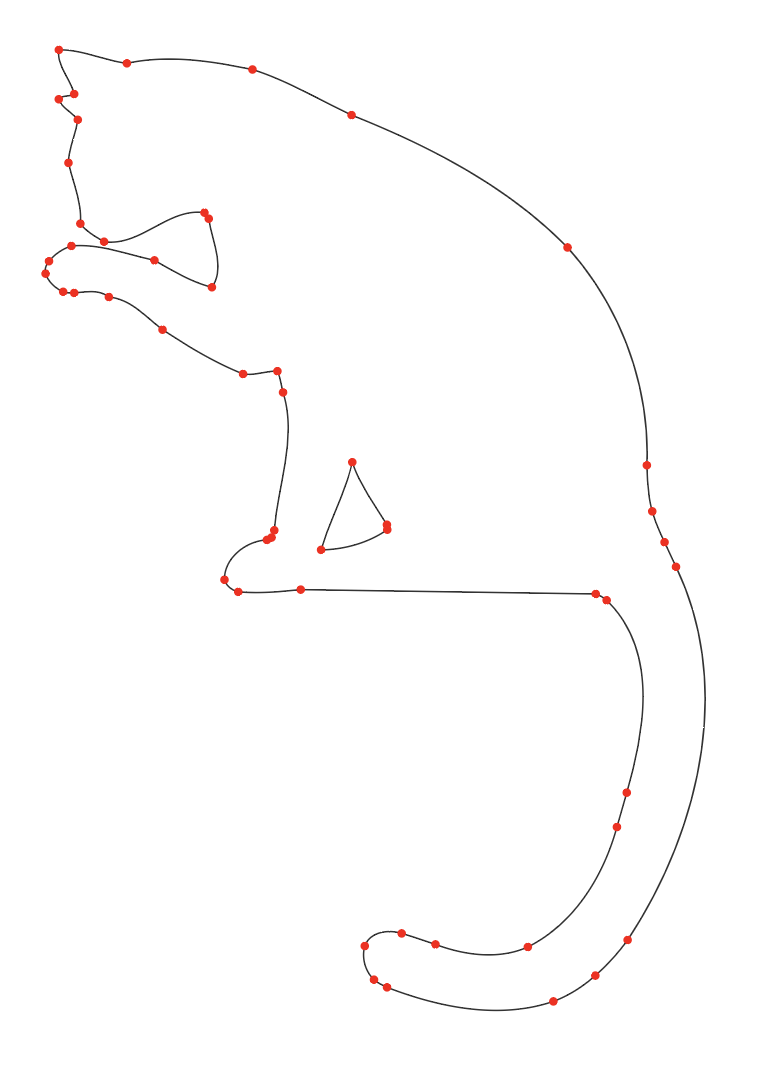}&
			\includegraphics[scale=0.3]{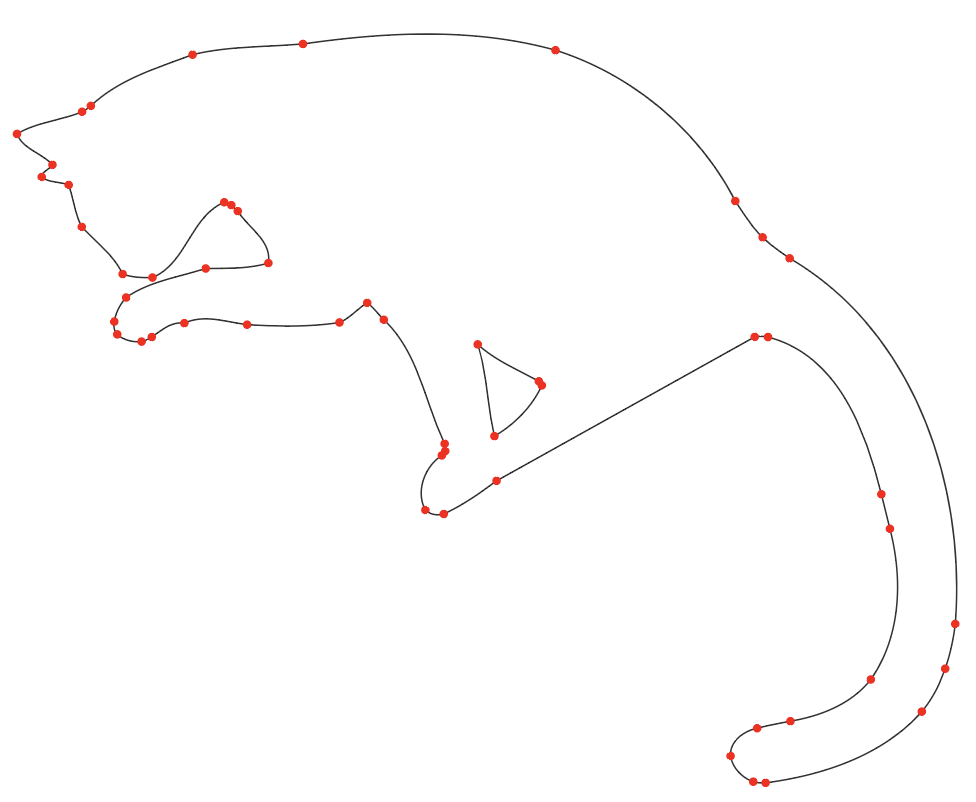}&
			\includegraphics[scale=0.3]{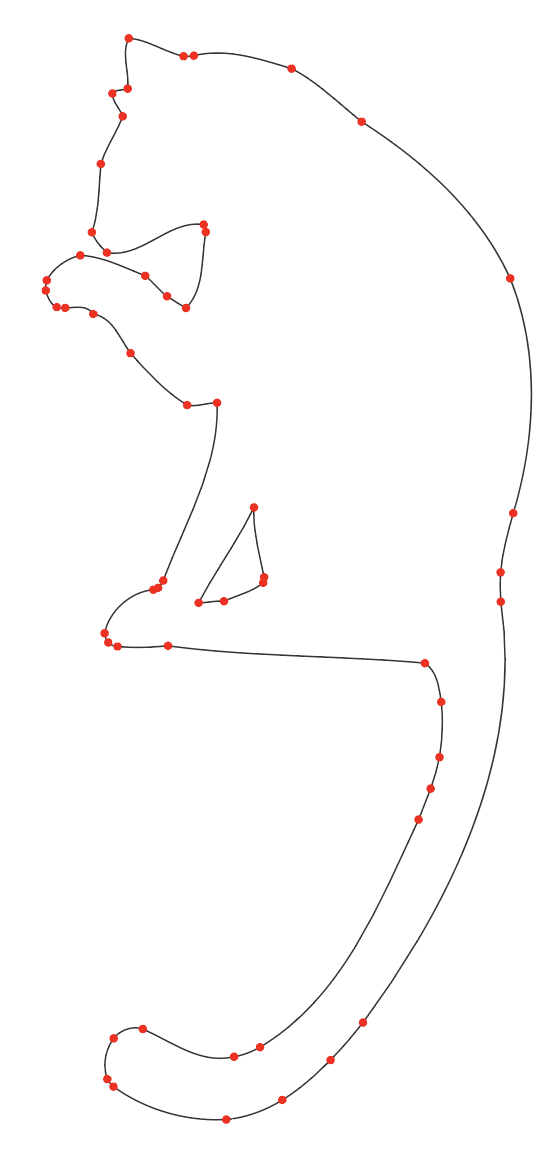}\\
			(g)&(h)&(i)\\
			\includegraphics[scale=0.3]{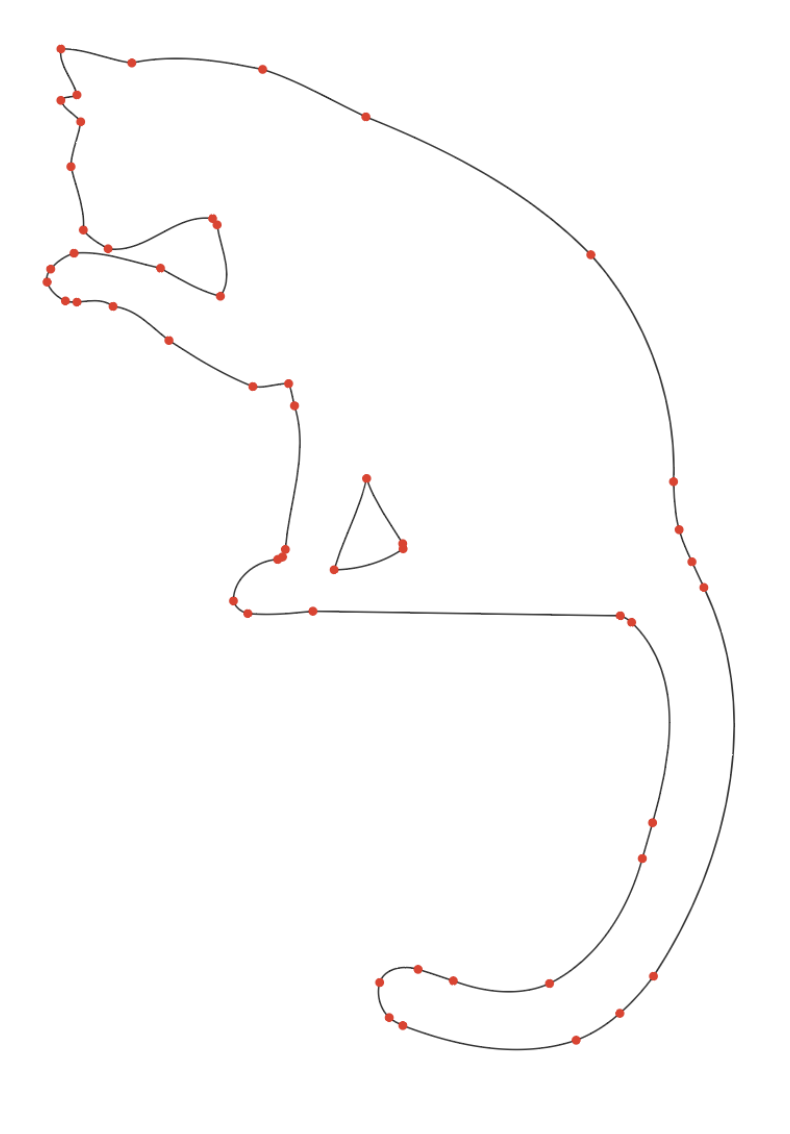}&
			\includegraphics[scale=0.3]{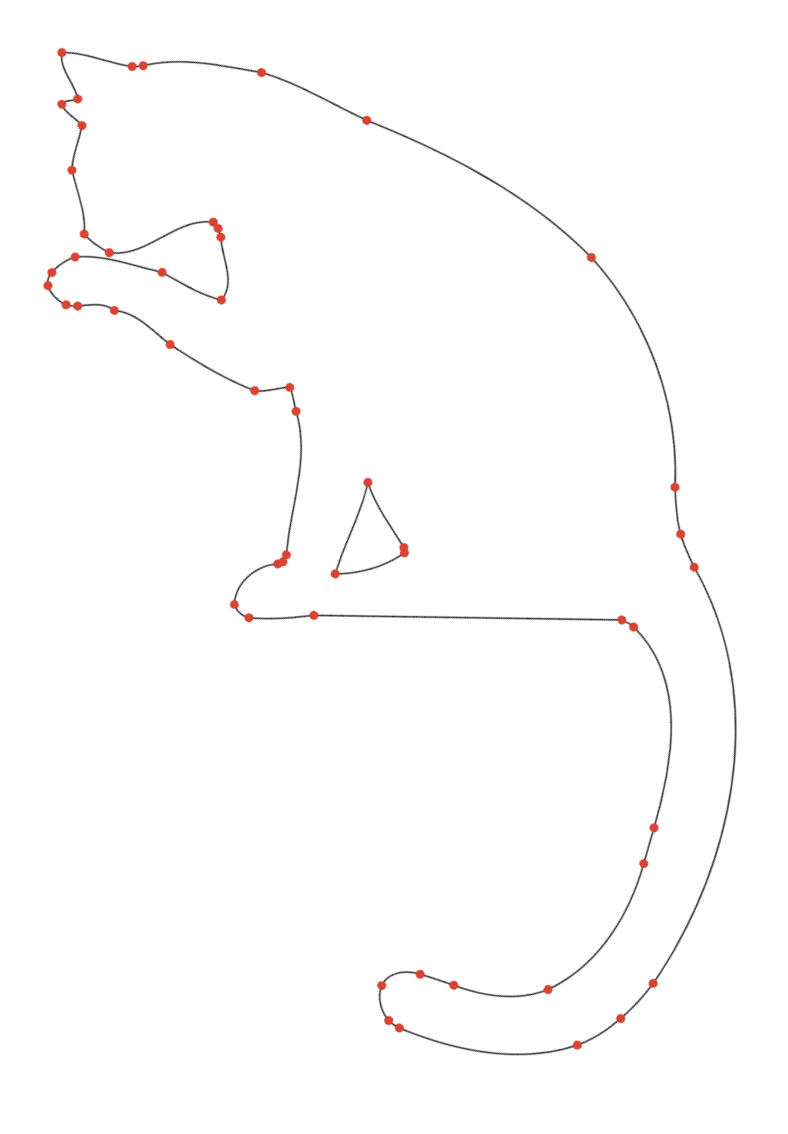}&
			\includegraphics[scale=0.3]{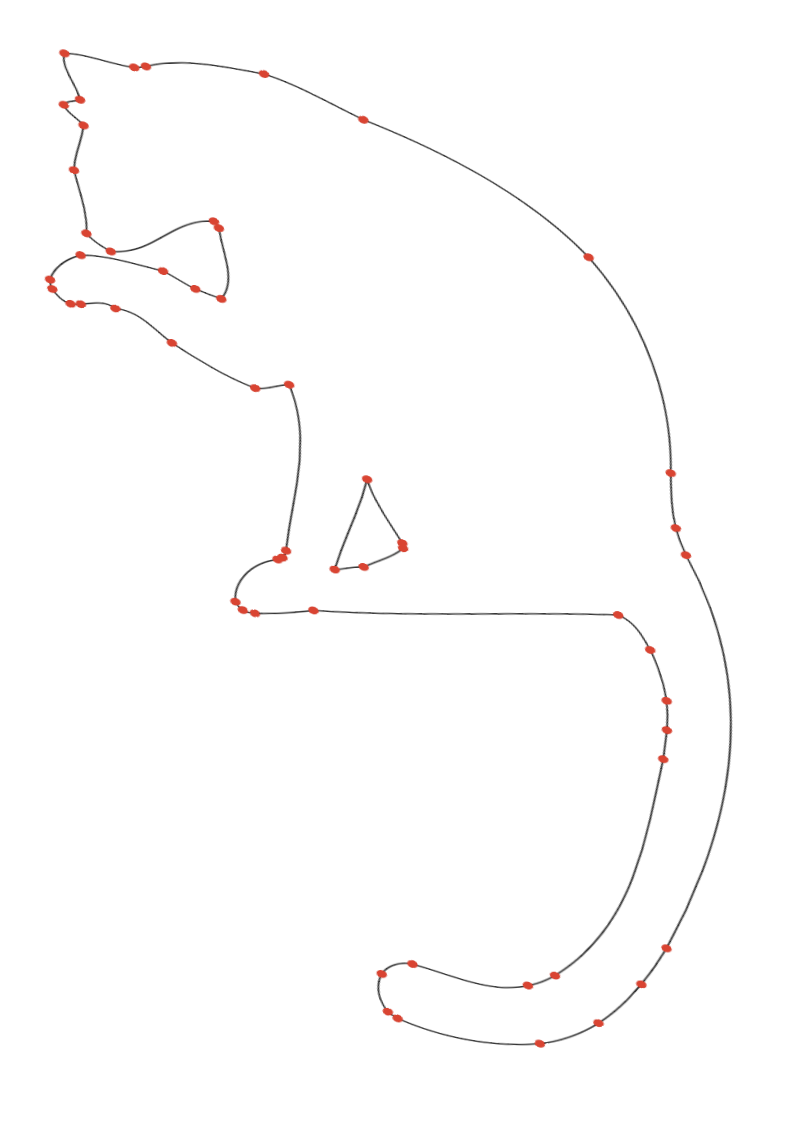}\\
		\end{tabular}
		\caption{Stability of control points under affine transformations. (a) Silhouette of a cat. (b) Rotation of (a). (c) Shear of (a). (d) Vectorized outline of (a). (e) Vectorized outline of (b). (f) Vectorized outline of (c). (g) is identical as (d) for comparison with (h) the inverse affine transform of (e), and (i) the inverse affine transform of (f). The control points in (g)--(i) are similarly distributed along the outline.  }\label{fig_affine}
	\end{figure}

	 We qualitatively explore the geometric stability of the proposed control points under affine transformations. Figure~\ref{fig_affine}~(a) shows a silhouette of a cat. (b) is a rotation of (a), and (c) is a sheared (a). The vectorized results of these silhouettes are presented in (d), (e), and (f), respectively. To better compare the distributions of the control points  on these vectorized outlines, we applied the corresponding inverse affine transformations to (d)--(f) and show the results  in (g)--(i). The numbers of control points are similar: (g) has $52$ control points ($38$ before refinement), (h) has $53$ control points ($37$ before refinement), and (i) has $56$ control points ($41$ before refinement).  The distributions of control points between (g) and (h) are almost identical, while locations of the control points in (i) are slightly shifted, especially those on the tail. This is because B\'{e}zier fitting is not affine invariant,  and these shifted points are inserted to guarantee the accuracy of approximating the transformed outline using a B\'{e}zier polygon.

	\subsection*{Qualitative Comparison with  Feature Point Detectors}
\begin{figure}
		\centering
		\begin{tabular}{cccc}
			(a)&(b) & (c) &(d)\\
			\includegraphics[width=1.5in]{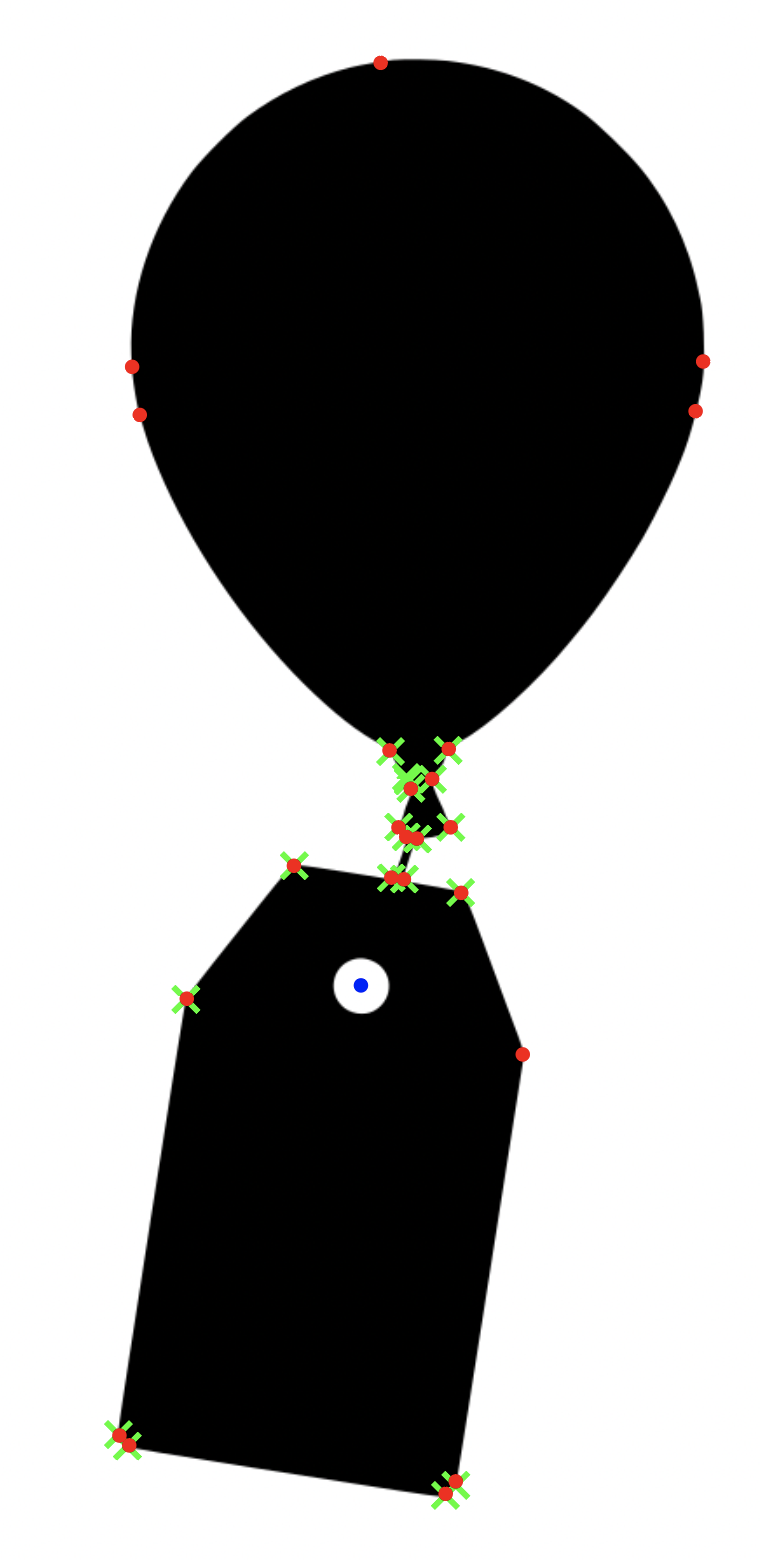}&
			\includegraphics[width=1.5in]{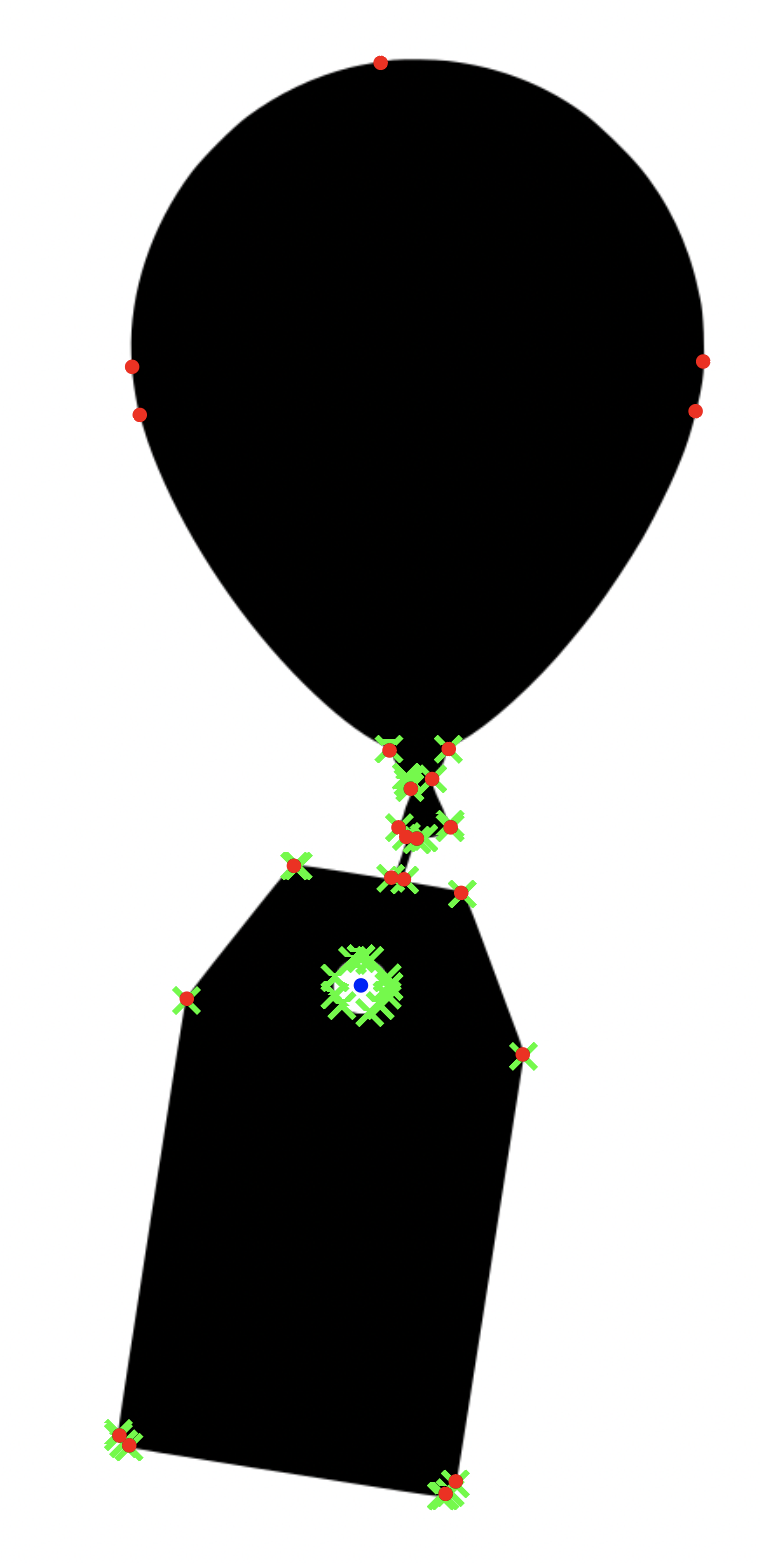}&
			\includegraphics[width=1.5in]{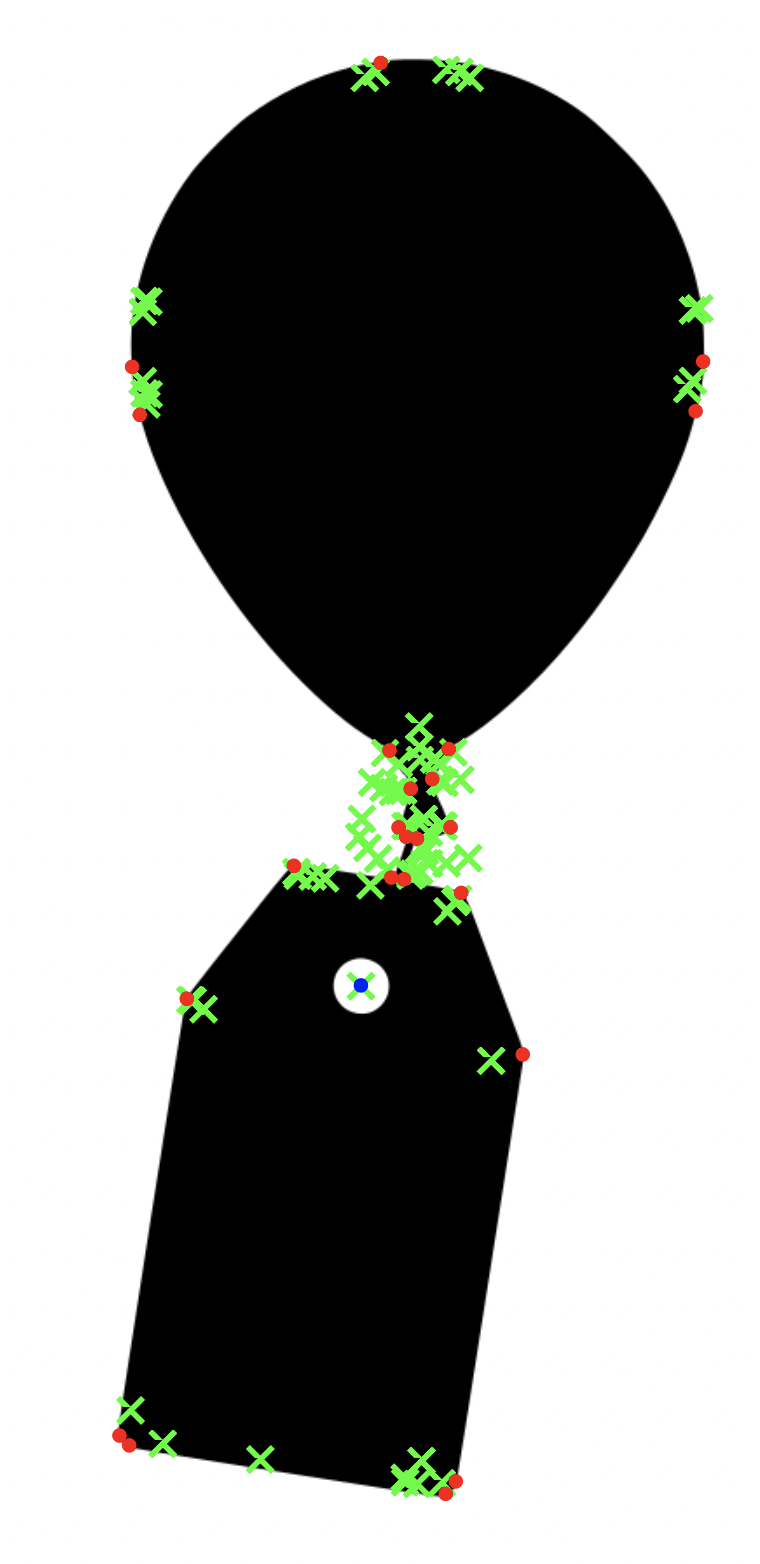}&
			\includegraphics[width=1.5in]{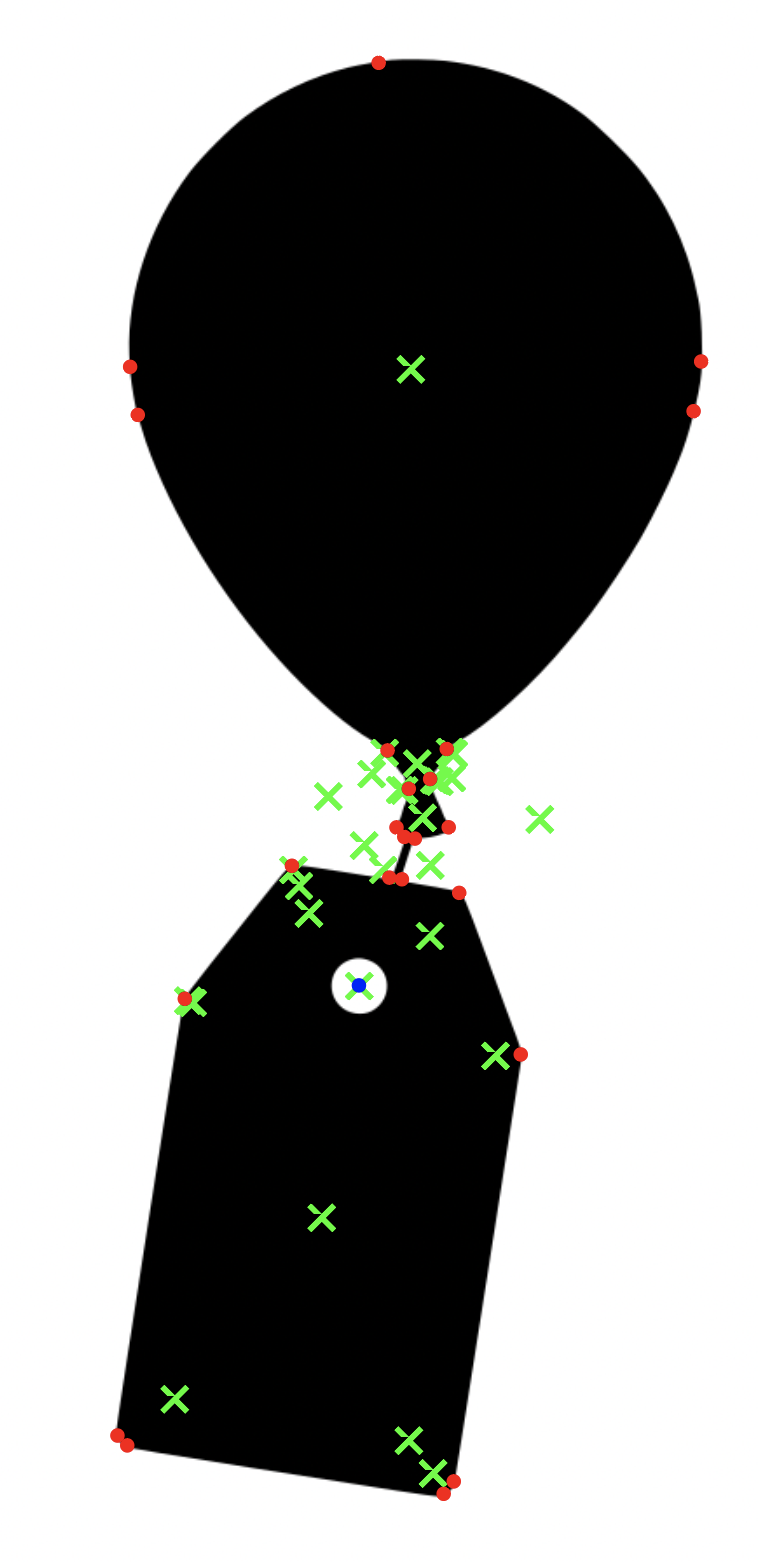}
		\end{tabular}
		\caption{Comparison between the control points (red dots) plus the centers of circles (blue dots) produced by the proposed algorithm and other point feature detectors (green crosses). (a) Compared with the Harris corner detector~\cite{harris1988combined}. (b) Compared with the FAST feature detector~\cite{rosten2005fusing}. (c) Compared with the SURF  detector~\cite{bay2006surf}. (d) Compared with the SIFT detector~\cite{lowe1999object} }\label{fig_detector}
	\end{figure}

	Our algorithm produces a set of informative point features of the outline. This includes the control points which separate the outline curves into segments for cubic B\'{e}zier fitting and the centers of circles. In Figure~\ref{fig_detector}, we compare the distribution of these points with the results of some extensively applied feature point detectors: the Harris-Stephens corner detector~\cite{harris1988combined}, the features from Accelerated Segment Test (FAST) detector~\cite{rosten2005fusing}, the Speeded Up Robust Features (SURF) detector~\cite{bay2006surf}, and the Scale-Invariant Feature Transform (SIFT)~\cite{lowe1999object}.
	
	The Harris-Stephens corner detector is a local auto-correlation based method. It locally filters the image with spatial difference operators and identifies corners based on the  response. In (a), the Harris-Stephens corner detector identifies all the corners except for the one on the right side of the label. The set of control points produced by our algorithm contains all the corners found by the Harris-Stephens detector plus the missed one.  
	
	The FAST detector only considers the local configurations of pixel intensities, hence it is widely applied in real-time applications. From (b), we see that FAST identifies all the prominent corners same as our method does. Similarly to (a), there are no FAST points identified around the balloon. On the circular outline at the center instead, FAST detects multiple false corners; this illustrates how our algorithm is more robust against pixellization

	The SURF detector combines a fast Hessian measure computed via integral images and the distribution of local Haar-wavelet responses to identify feature points that are scale- and  rotation-invariant.  There is a similarity that it utilizes the Gaussian scale-space and scale-space interpolation to localize the  points of interest. The SURF points are marked over scales, hence we see most of the green crosses in (c) form sequences converging toward the outline. These limit points correspond exactly to our control points (red dots) distributed over the outline, including those around the balloon. Moreover, there is a SURF point at the center of the circular hole in the label, which overlaps with our identified center of circle (blue dot). Rather than showing feature points over scales, our method locates them directly on the original outline. In~(c), notice that our identified points are much simpler compared to SURF points.
	
	The SIFT detects scale-invariant features of a given image. As shown in (d), SIFT successfully indicates the presence of corners and marks the centers of the balloon as well as the label, which are visually robust features of the silhouette Our method focuses on the outline instead of the interior points and provides interesting boundary points' locations exactly. Around the balloon, the symmetric distribution is compatible with the SIFT point at the center.
	
	The set of control points plus the center of circles produced by our algorithm is comparable to some of the frequently used feature point detectors in the literature.  Hence,in addition to being an effective silhouette vectorization method, the identified control points can be used for other applications where feature point detectors are needed.
	
	\subsection*{Quantitative Comparison with  Feature Point Detectors}
	 To further justify that our method can be applied as a  stable point feature detector for silhouettes, we compare the techniques discussed above with ours by quantitatively evaluating their performances via the repeatability ratio~\cite{schmid2000evaluation}. It measures the geometric stability of the detected feature points under various transformation. 
	 
\begin{figure}
		\centering
		\begin{tabular}{c|c|c}
			Silhouette&  Under Rotation& Under Scaling\\\hline
			\includegraphics[scale=0.12]{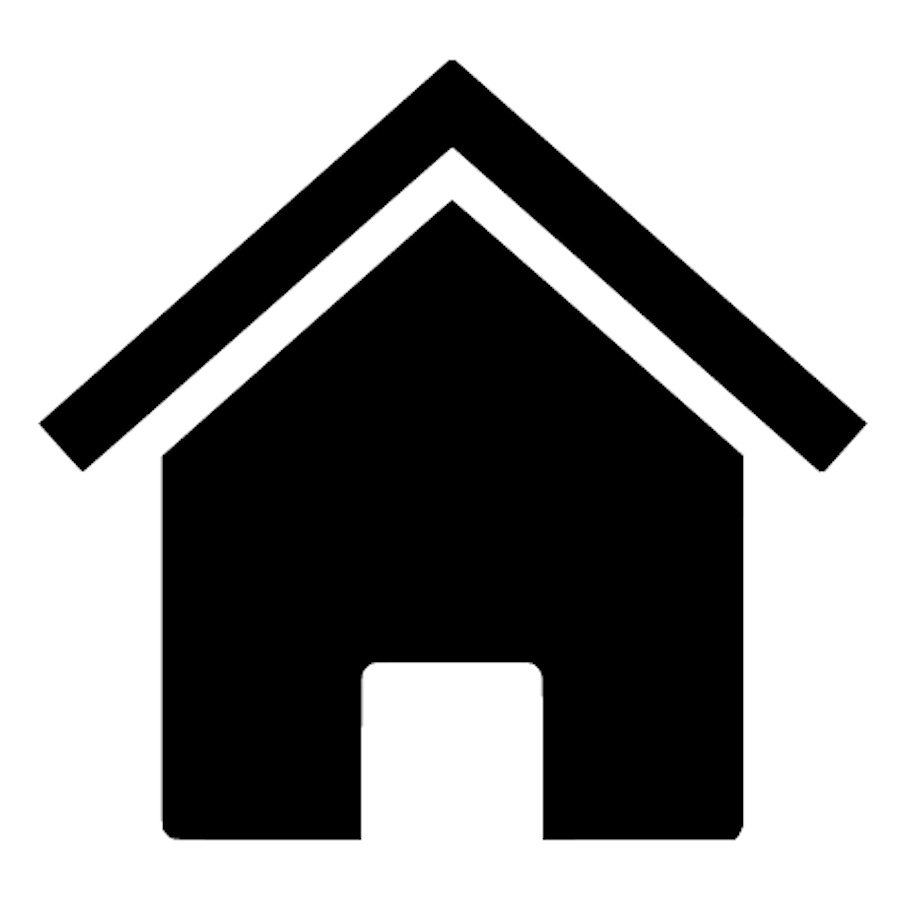}&
			\includegraphics[scale=0.2]{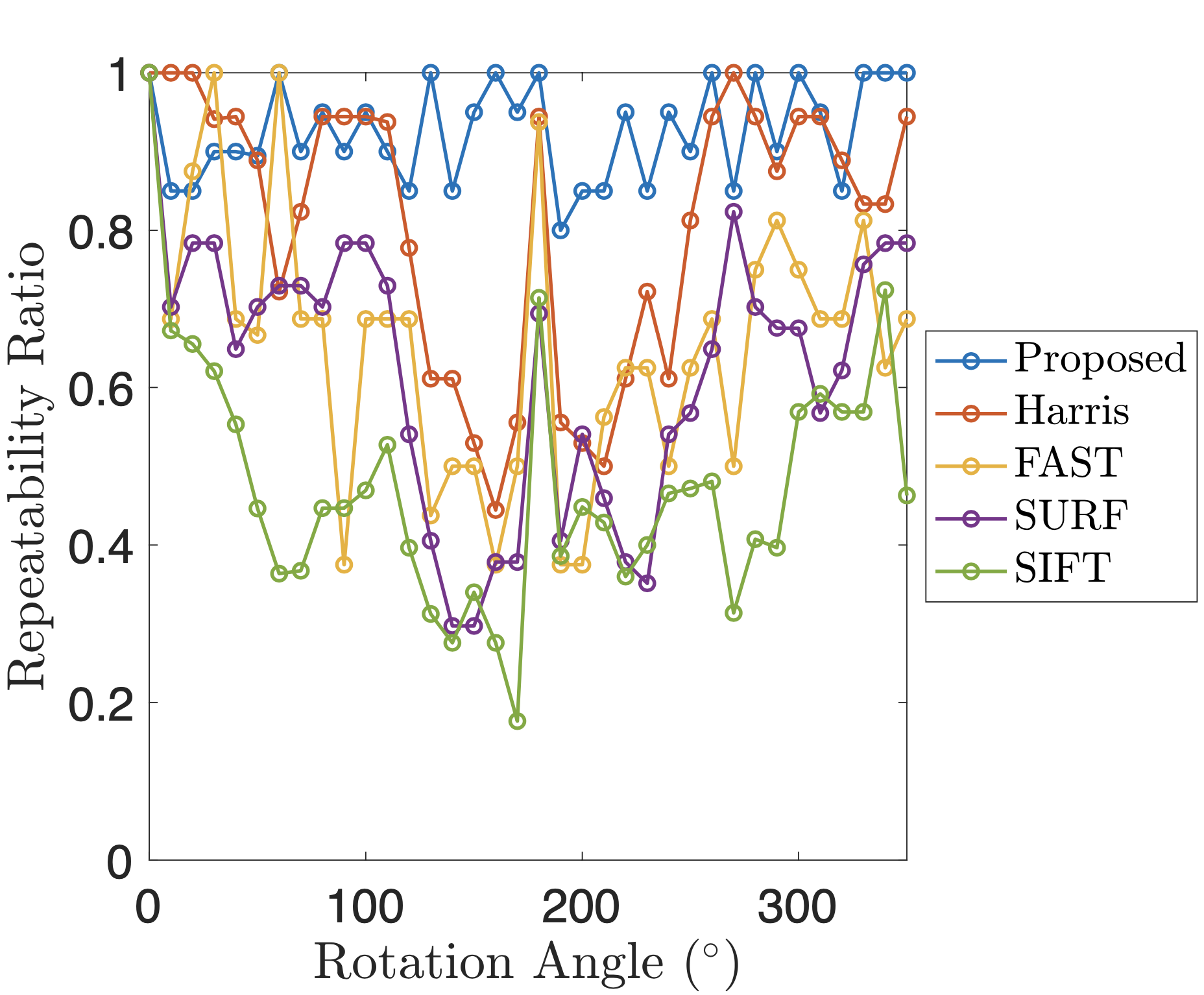}&
			\includegraphics[scale=0.2]{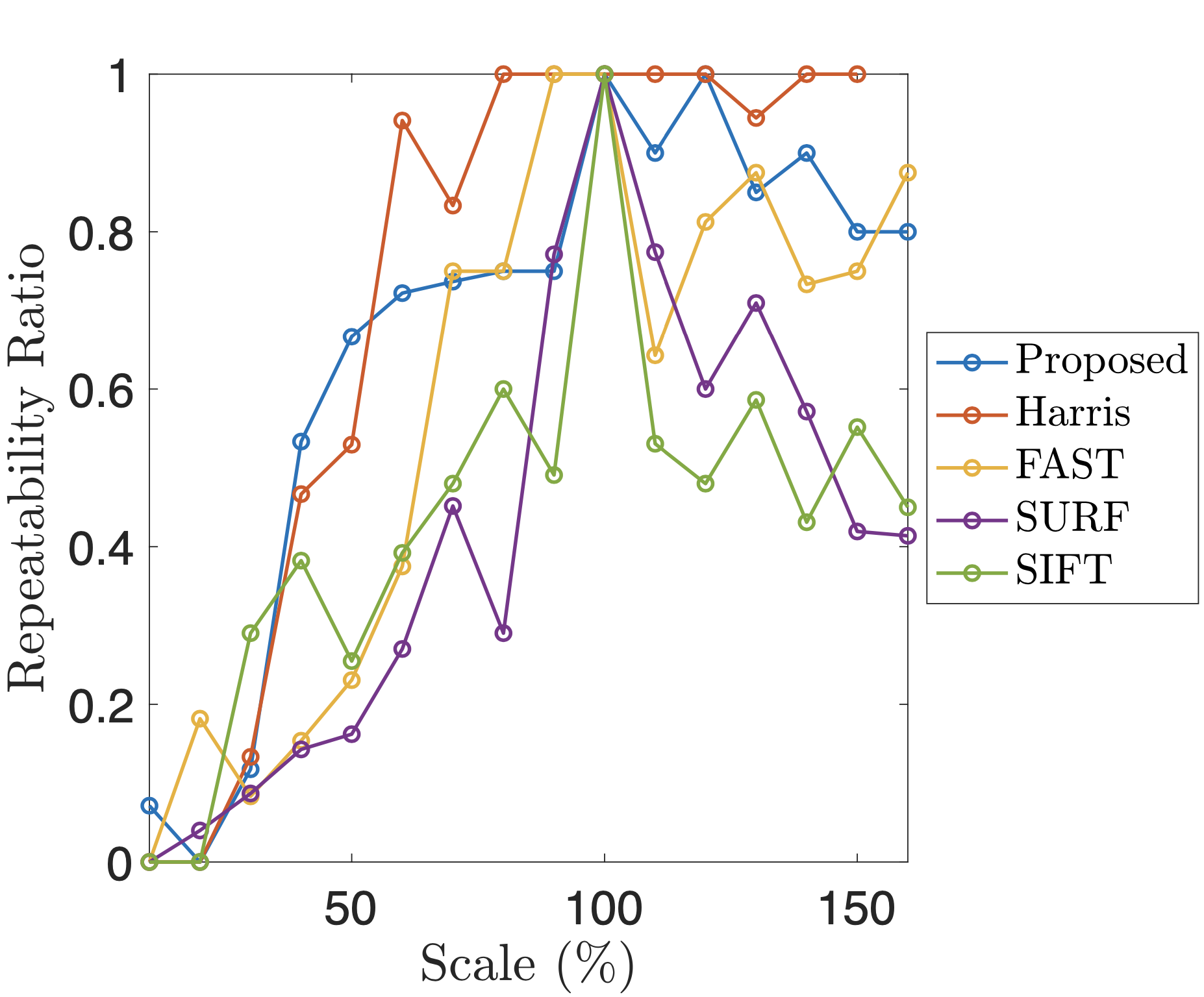}\\
			\includegraphics[scale=0.15]{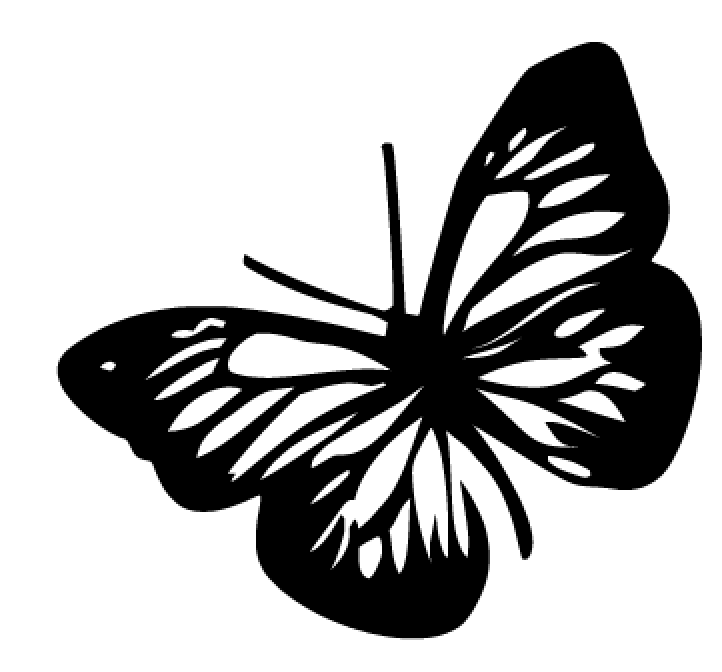}&
			\includegraphics[scale=0.2]{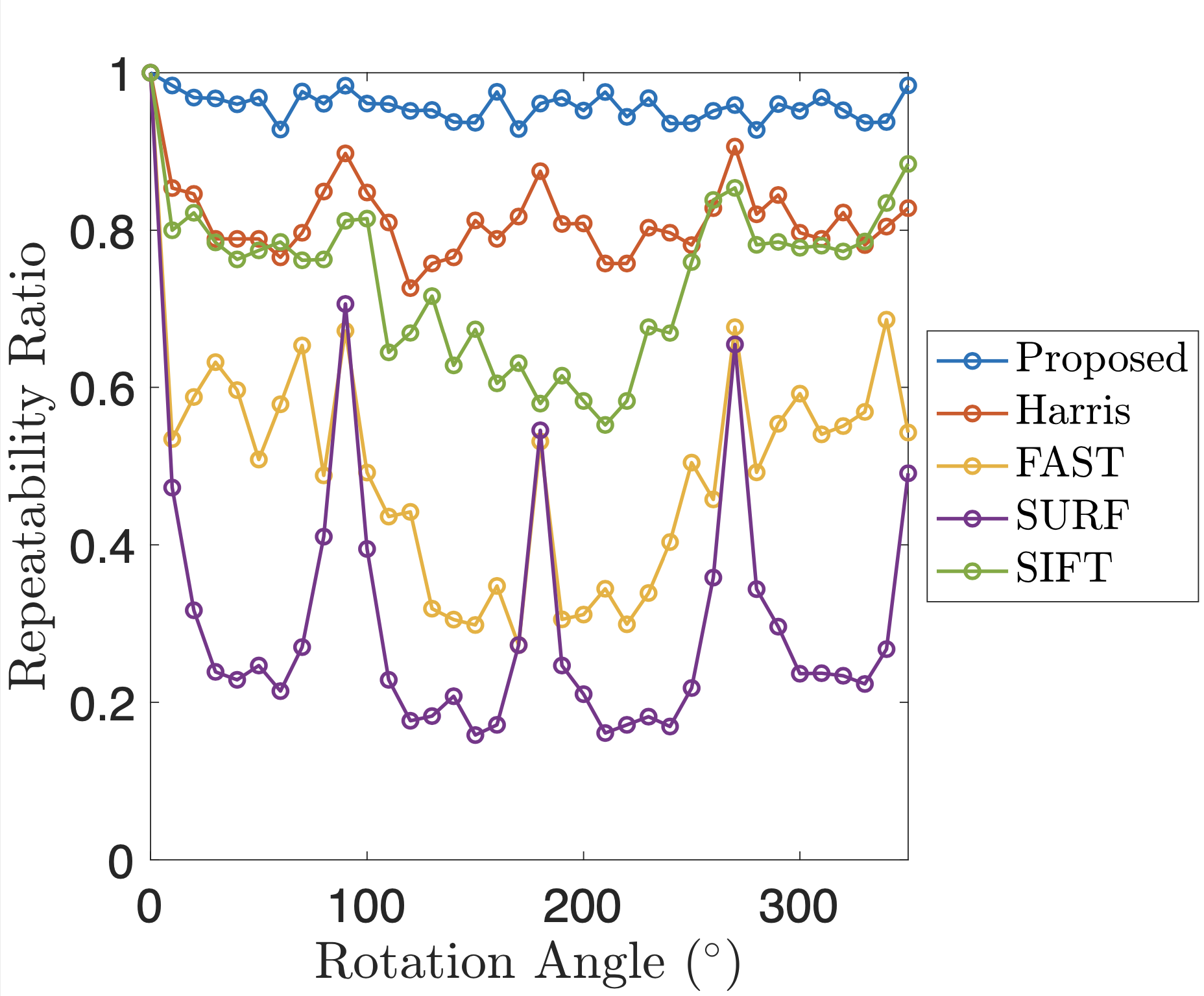}&
			\includegraphics[scale=0.2]{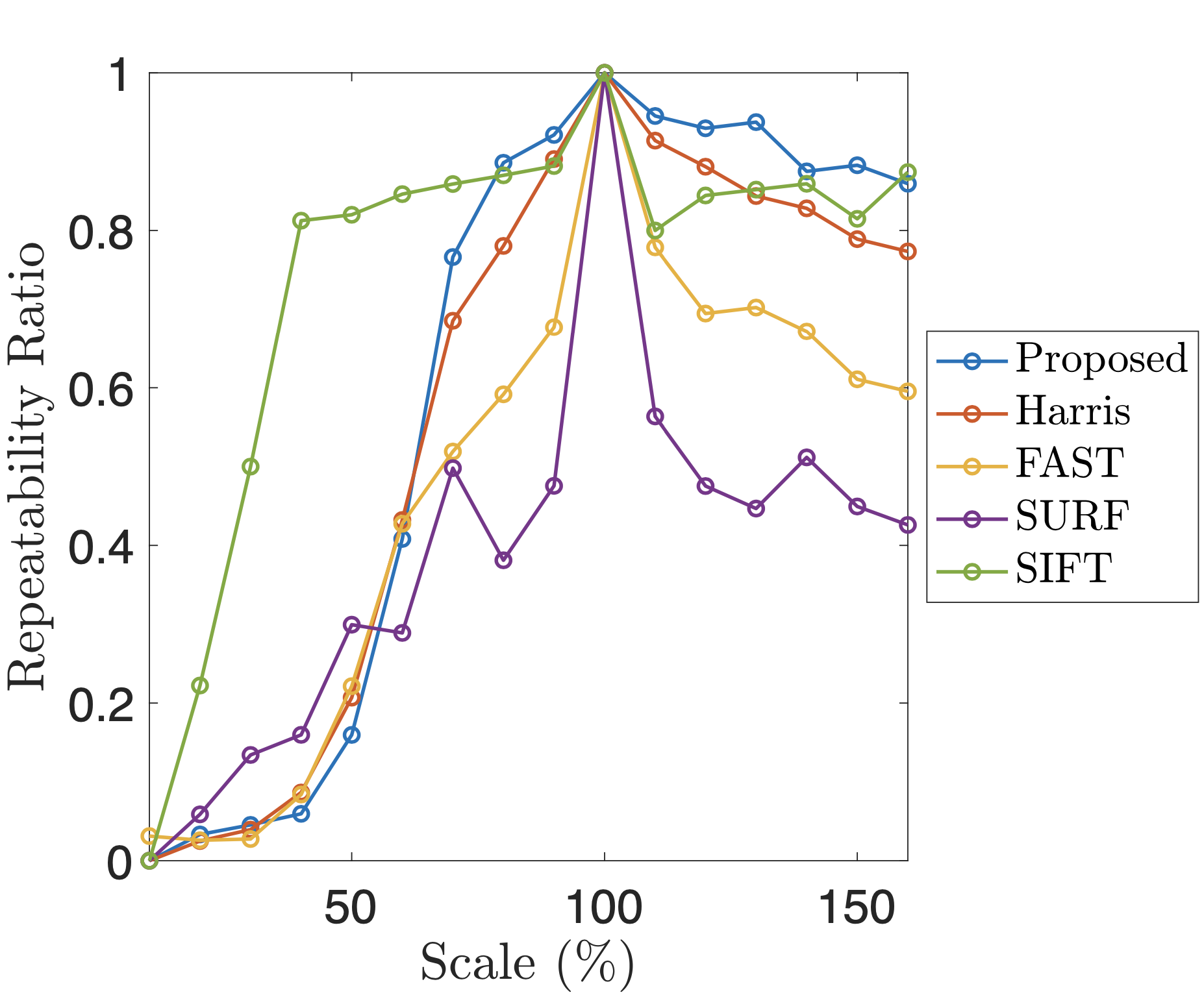}\\
			\includegraphics[scale=0.15]{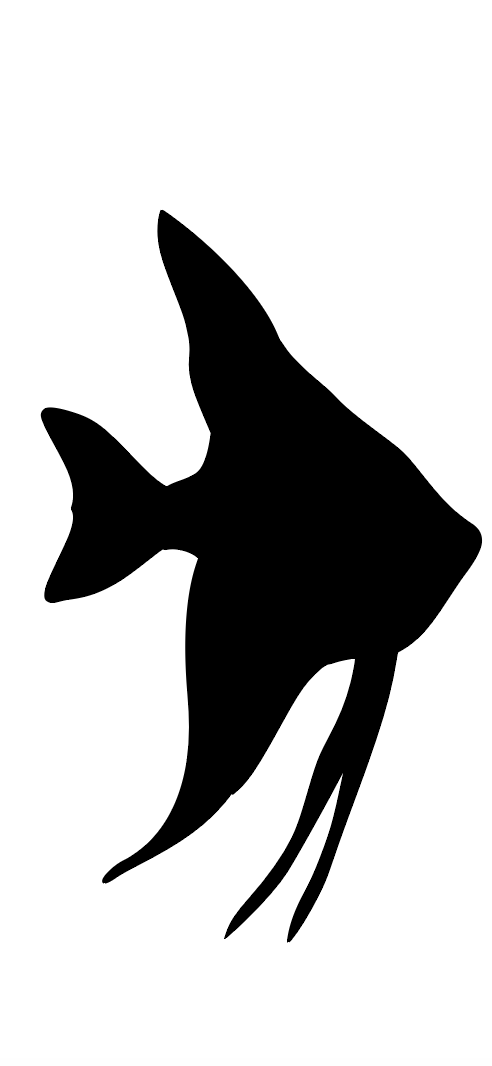}&
			\includegraphics[scale=0.2]{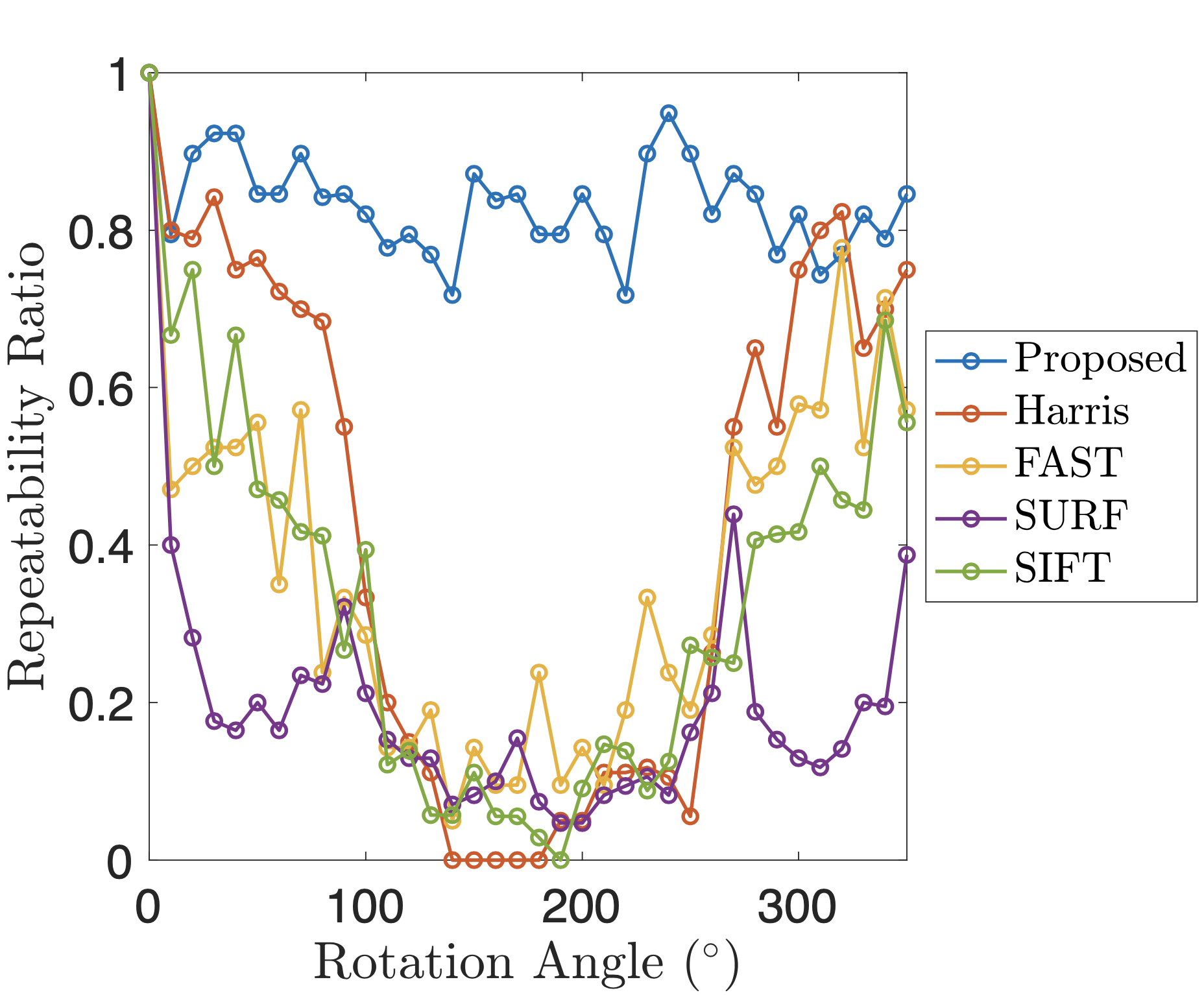}&
			\includegraphics[scale=0.2]{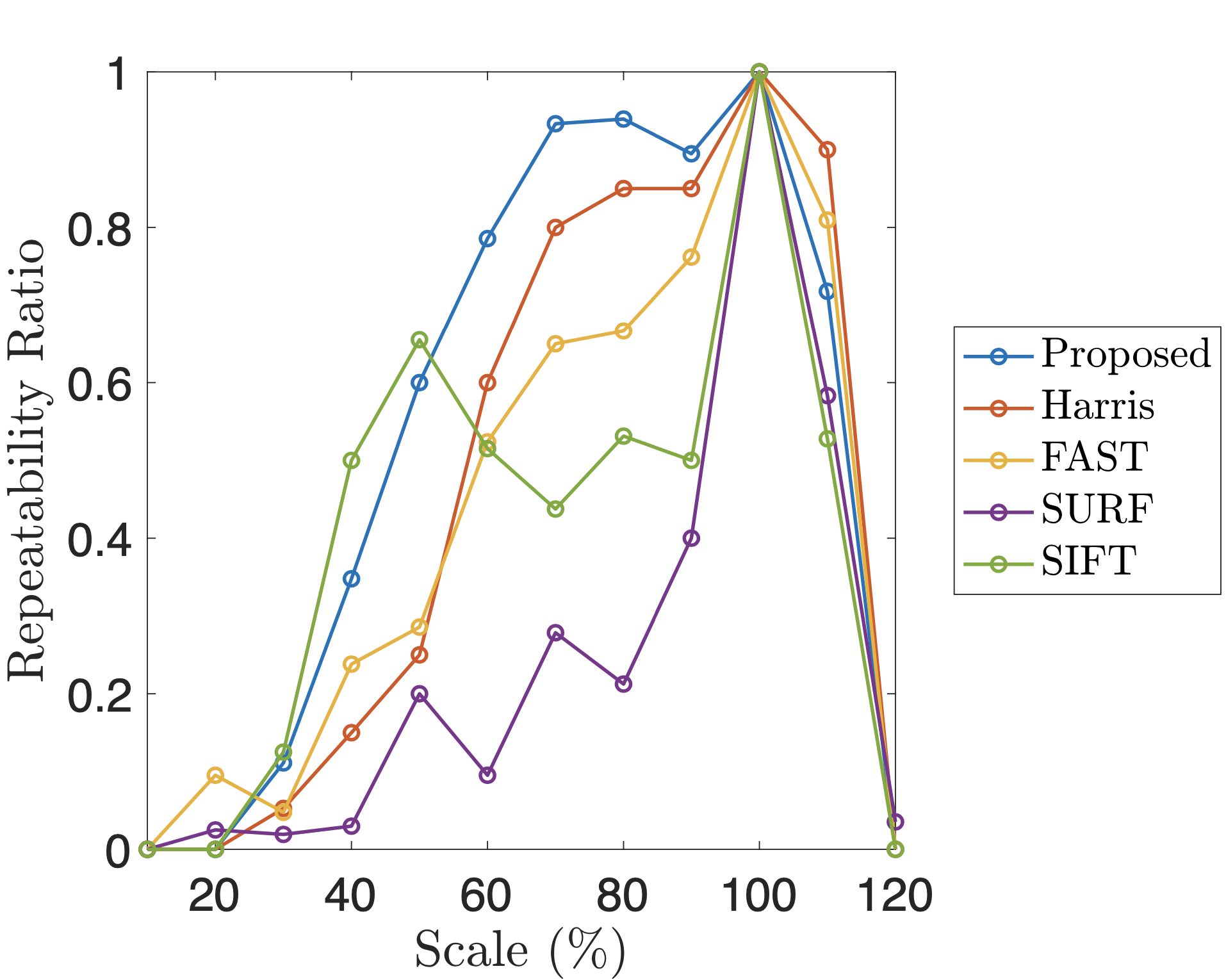}\\\hline
		\end{tabular}
		\caption{Repeatability ratios of the methods in comparison when the  silhouettes in the first column are rotated or scaled. Notice that the blue lines (proposed method) are near $1$. The performance of our method is the most consistent across these silhouettes with different geometric features.}
		\label{fig_quant_eval}
	\end{figure}

	 In particular, for each method, given any angle $\alpha$, $0^\circ<\alpha<360^\circ$, we rotate the silhouettes in the first column of Figure~\ref{fig_quant_eval} with respect to their centers by $\alpha$ respectively,  record the detected feature points,  apply the inverse transform on these points by rotating them by $-\alpha$, then compare their positions with the feature points detected on the original silhouette. Let $n_{\text{repeat}}=0$. For any rotated feature point, within its $\epsilon$-neighborhood, if we find at least one feature point on the original silhouette, we increase  $n_{\text{repeat}}$ by $1$. The $\epsilon$-repeatability ratio is computed by
	 \begin{align}
	 \frac{n_{\text{repeat}}}{\min\{n_0,n_\text{transform}\}}
	 \end{align}
	where $n_0$ denotes the number of feature points detected on the original silhouette, and $n_\text{transform}$ is the number of feature points detected on the transformed one. During the angle (or scale) changes, this value staying near $1$ indicates that the applied method is invariant under rotation (or scale). We fix $\epsilon=1.5$. 
	
	The second column of Figure~\ref{fig_quant_eval} shows the repeatability ratios under rotations. The set of feature points produced by our method has superior stability when the silhouette is rotated by arbitrary angles. In contrast, the other detectors have low repeatability ratios especially when the silhouette is turned almost upside-down. Moreover, our method performs consistently well for silhouettes with different geometric features. The house silhouette has straight outlines and sharp corners; the butterfly silhouette is defined by smooth curves; and the fish silhouette has prominent curvature extrema which are not perfect corners.
	
	For the third column of Figure~\ref{fig_quant_eval}, we compute the repeatability ratios when the transformation is replaced by scaling. Observe that our method is comparable with other detectors, and it is the most consistent one across these different silhouettes.

	\subsection*{Comparison with State-of-the-art Software}
	There are many software available for image vectorization, e.g., Vector Magic~\cite{VM}, Inkspace~\cite{IS}, and Adobe Illustrator 2020 (AI)~\cite{AI}. In the following set of experiments, we compare our method with these software using  the number of control points generated for given silhouettes as a criteria. This quantity is equal to the number of curve segments used for approximating the  outline, and a smaller value indicates a more compact silhouette representation.
	
	To perform this comparison, after acquiring  SVG files of various silhouettes, we rasterized them and used the PNG images as inputs. Table~\ref{tab_software} summarizes the results. For Vector Magic, we test three available settings: high, medium, and low for the vectorization quality. For AI, we choose the setting``Black and White Logo", as it is suitable for the style of our inputs. We also include the results when the automatic simplification is used, which are marked by daggers. For Inkspace, we use the default parameter settings as recommended. As shown by the values of the mean relative reduction on the number of control points in the last row, our method produces the most compact vectorization results.  
	
\begin{figure}
	\centering
	\includegraphics[scale=0.32]{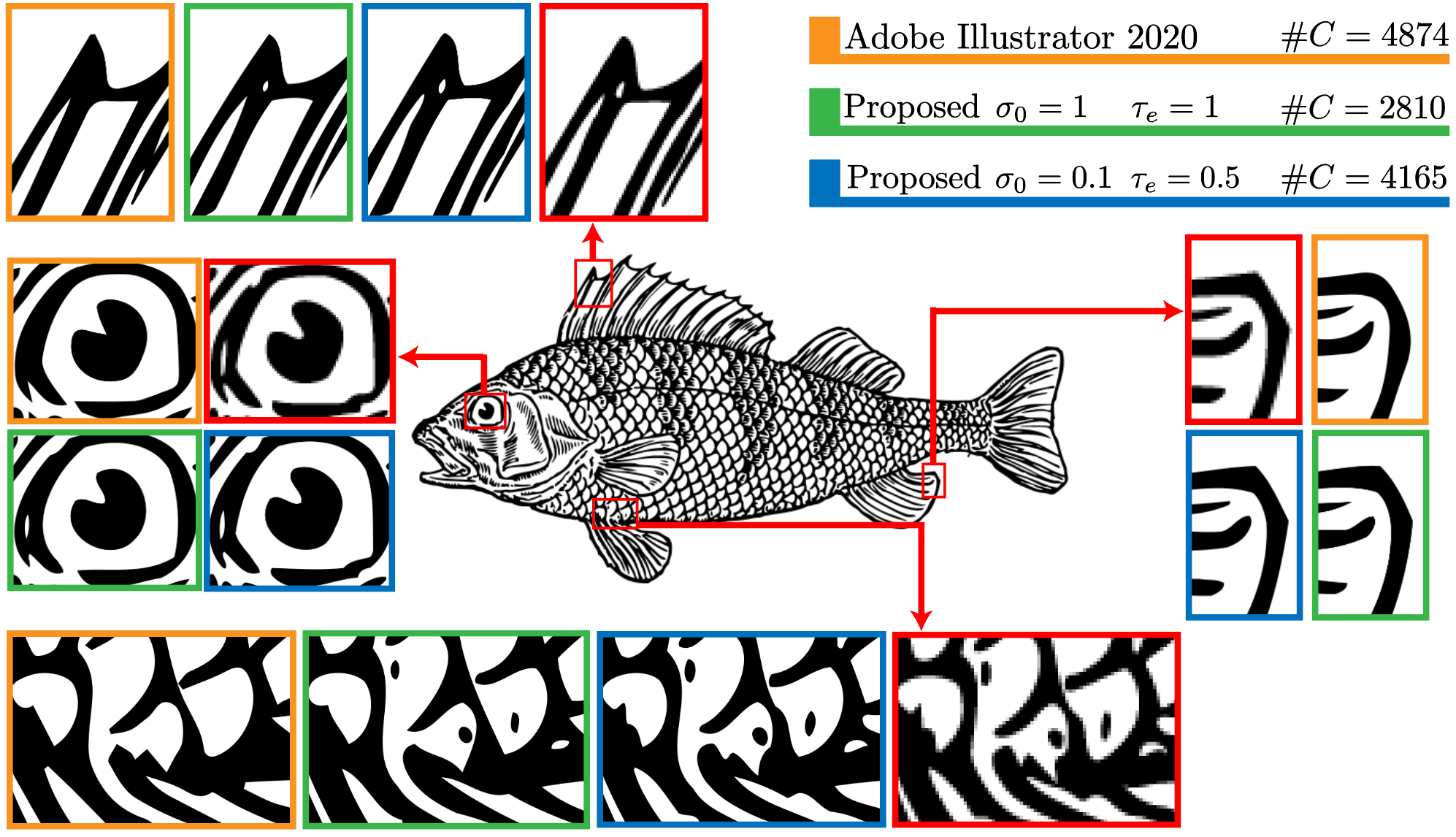}
	\caption{Comparison among the given raster image (red boxes), AI (orange boxes), the proposed with $\sigma_0=1$, $\tau_e=1$ (green boxes), and the proposed with $\sigma_0=0.1$, $\tau_e=0.5$ (blue boxes). With smaller numbers of control points ($\# C$), our method preserves better the geometric details  of the given silhouette. }\label{fig_detail_fish}
\end{figure}

	With such an effective reduction on the number of control points, we justify that our method does not over-simplify the representation.  We show a detailed comparison in Figure~\ref{fig_detail_fish} between our proposed method and AI. In particular, we use  AI without simplification and our method with two sets of parameters: $\sigma_0=1$, $\tau_e=1$ and $\sigma_0=0.1$, $\tau_e=0.5$. We note that $\sigma_0$ specifies the smoothness of the recovered outline, and $\tau_e$  controls the accuracy. Notice that under these settings, our method  gives less number of control points, yet our results  preserve more details of the given silhouettes, for example, the strokes on the scales at the bottom, and the sharp outlines on the rear fin.
	
\begin{figure}
	\centering
	\begin{tabular}{cc}
		(a) & (b)\\
		\includegraphics[scale=0.22]{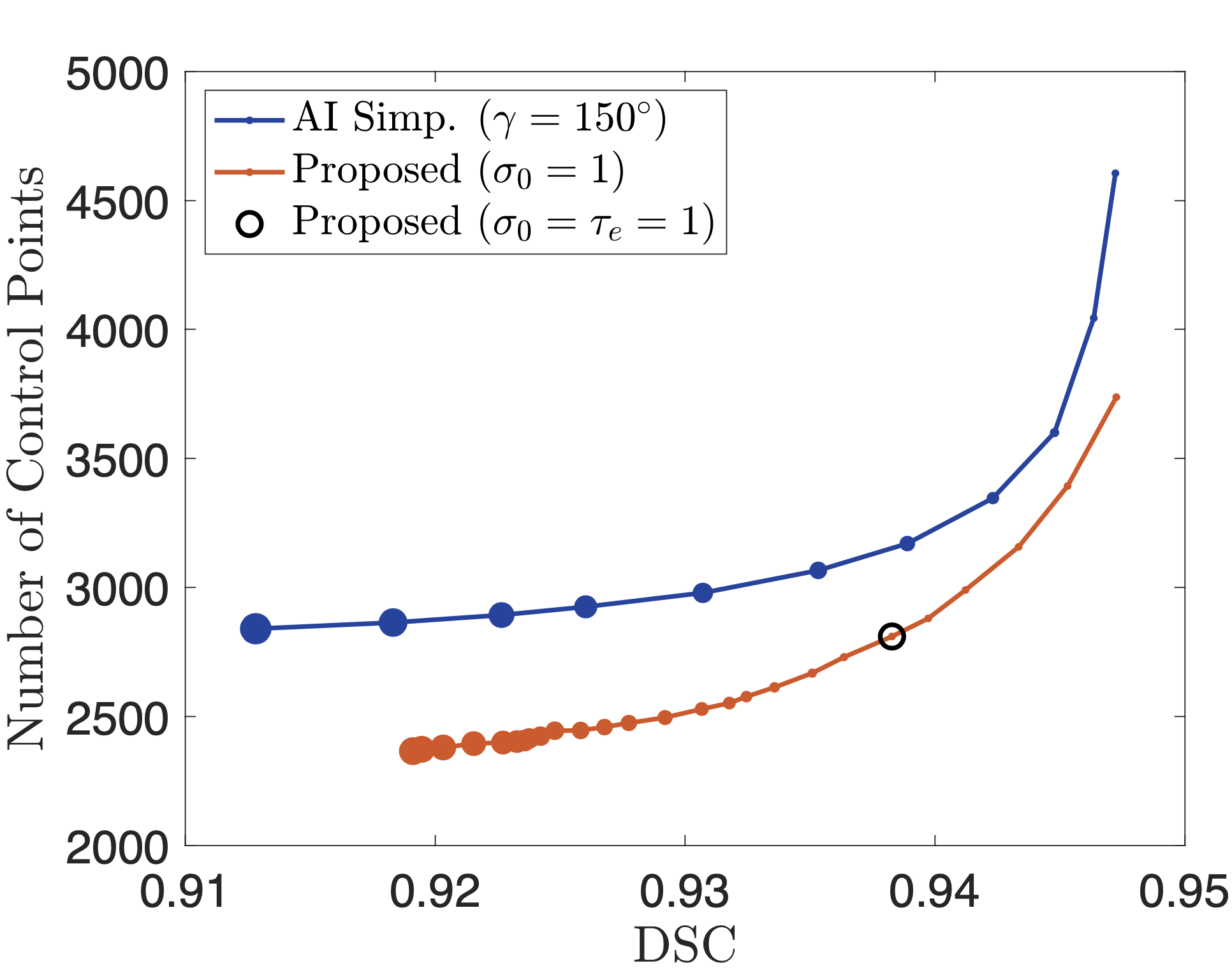}&
		\includegraphics[scale=0.22]{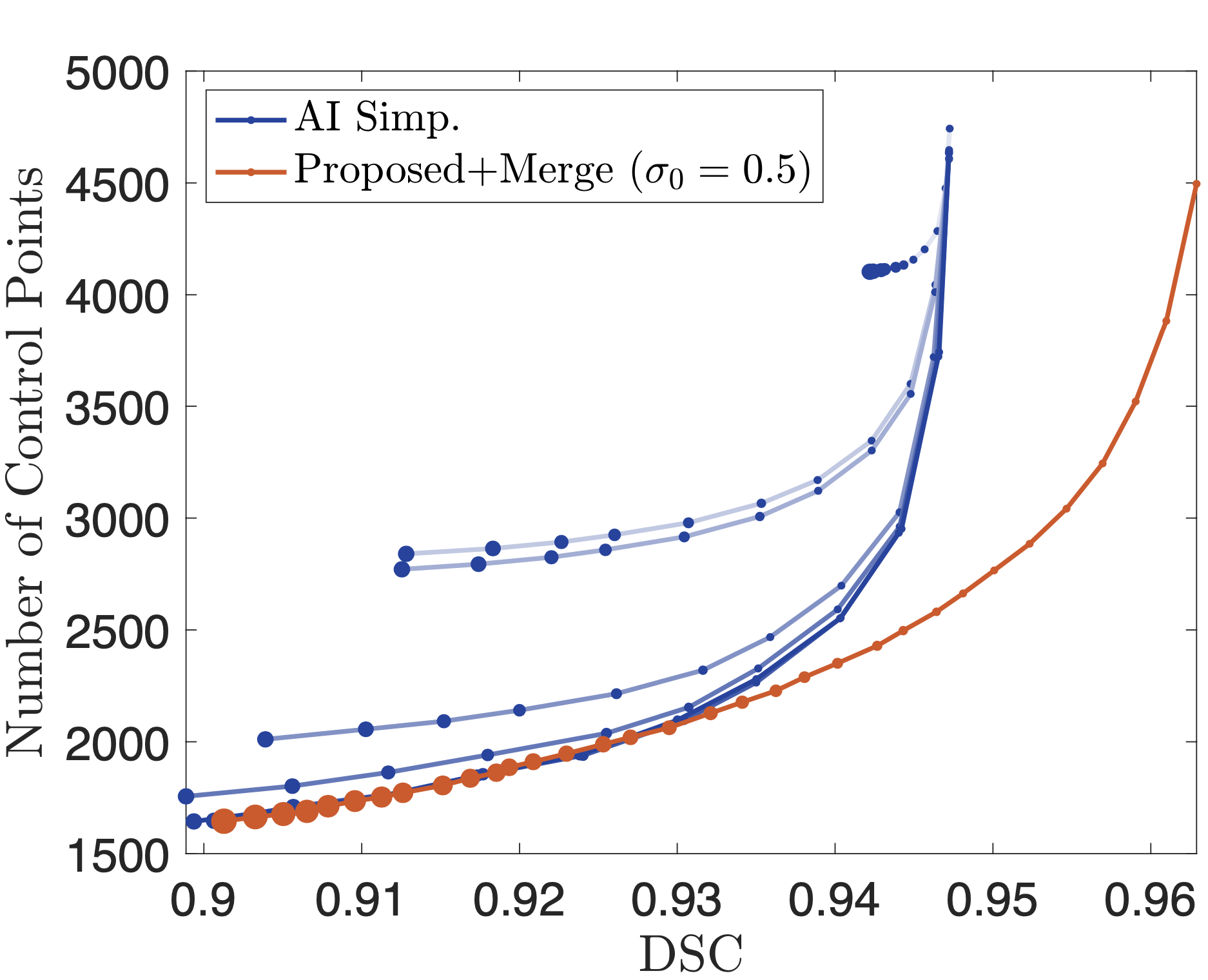}
	\end{tabular}
\caption{ (a) Comparison between AI ($\gamma=150^\circ$) and the proposed method ($\sigma_0=1$) when the complexity parameters ($\mu$ for AI, $\tau_e$ for ours) vary.  The circled dot corresponds to our default setting. (b) Comparison between AI with simplification specified by various combinations of $\mu$ and $\gamma$, and the proposed method using merging with fixed $\sigma_0=0.5$ and varying $\tau_e$. In both figures, smaller dots indicate  higher levels of complexity for AI  ($\mu$) and the proposed method ($\tau_e$), respectively. . }\label{fig_DSC_Bpn}
	
\end{figure}

	We quantify the performance of AI and our method by comparing the given image $I$ and the image $I'$ rasterized from the vectorization result. Denote $S_0=\{(x,y)\in\Omega\cap\mathbb{N}^2\mid I(x,y)<127.5\}$ and  $S_r=\{(x,y)\in\Omega\cap\mathbb{N}^2\mid I'(x,y)<127.5\}$ as the interior pixels of the given silhouette and the reconstructed one. We evaluate the similarity between $S_0$ and $S_r$ by
	\begin{align}
	\text{Dice similarity coefficient (DSC)~\cite{sorensen1948method}}\quad \text{DSC} = \frac{2|S_0\cap S_r|}{|S_0|+|S_r|}\;.
	\end{align}
	Higher values of DSC $(0\leq \text{DSC}\leq 1)$ imply a better matching between two silhouettes.   We evaluate the performance with wide ranges of parameters for both AI and the proposed method. For AI, we test various combinations of the curve simplification parameter $\mu$ ($0\%$--$100\%$) and the corner point angle threshold $\gamma$ ($0^\circ$--$180^\circ$). For our method, we use different combinations of $\tau_e$ and $\sigma_0$. Roughly speaking, $\mu$ in AI corresponds to $\tau_e$ in ours, which controls the approximating accuracy, and $\gamma$ in AI corresponds to $\sigma_0$ in ours, which adjusts the smoothness of the vectorized outline. Figure~\ref{fig_DSC_Bpn} plots the  number of control points against the corresponding DSC values for  various parameter settings in both methods. In (a), we fix the sharpness requirement, i.e., fixed $\gamma=150^\circ$ (default value for the automatic simplification used in AI) and fixed $\sigma_0=1$, and vary $\mu$ for AI (the blue curve) and $\tau_e$ for ours (the red curve). On the blue curve, larger dots correspond to smaller values of $\mu$; on the red curve, larger dots correspond to larger values of $\tau_e$. Moving from left to right along both curves indicates more accurate outline approximations.  Since the red curve stays below the blue one, compared to AI, our method produces less control points while achieving the same level of DSC values. In (b), we present the results of AI using simplification specified by a set of combinations of parameters ($\gamma=0^\circ,10^\circ,\dots,180^\circ$, $\mu=0\%,10\%,\dots,100\%$). They are organized such that each blue curve corresponds to a fixed value of $\gamma$; higher curves (lighter shades of blue) correspond to larger values of $\gamma$, while moving from left to right (smaller sizes of dots) along each of the curves corresponds to decreasing $\mu$. The red curve shows our results using different values of $\tau_e$ when the merging is applied and $\sigma_0$ is fixed at $0.5$. From left to right, the value of $\tau_e$ decreases. Observe that the red curve gives a close lower bound for the blue curves when DSC$<0.93$. For higher requirement on the accuracy (DSC$>0.93$), our method shows superior efficiency: it takes comparatively small number of control points to reach greater values of DSC. In contrast, for AI, the best DSC value it can achieve is around $0.95$, and adding more control points does not offer any improvement.
	\begin{table}
		\centering
		\begin{tabular}{c|ccccc}
			&\multicolumn{5}{c}{Number of Control Points ($\# C$)}\\\hline
			Test Image & Original& VM& IS & AI & Proposed \\\hline
				{\small
					\includegraphics[height=3.5em]{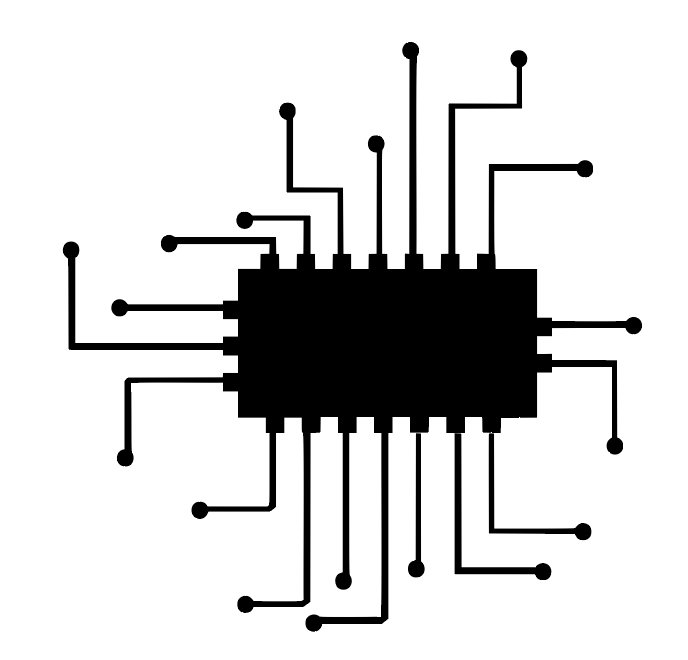}
				}
			&$405$& $248/256/245$& $330$& $280~(193^\dagger)$  & $168$\\\hline
			{\small
				\includegraphics[height=3.5em]{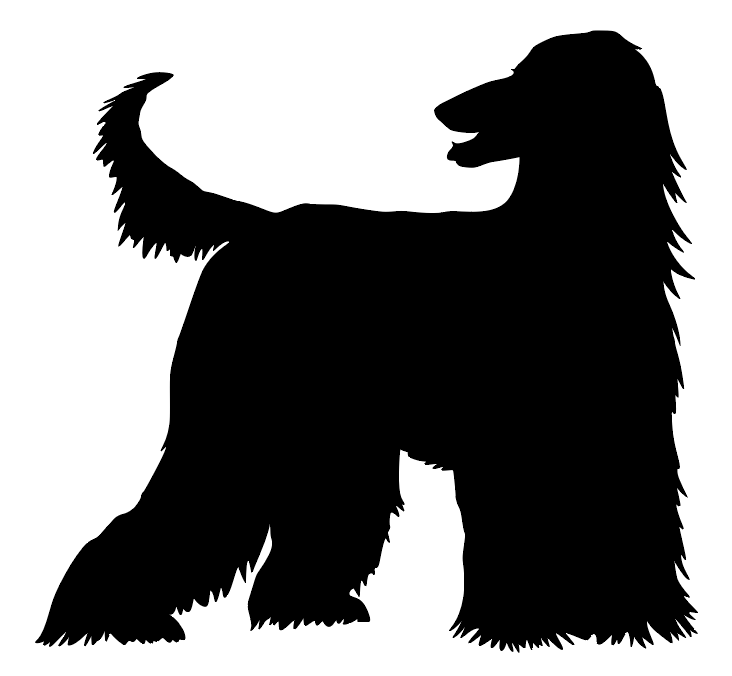}
			}&$611$& $359/343/325$& $383$& $340~(293^\dagger)$  & $222$\\\hline
			{\small
				\includegraphics[height=3.5em]{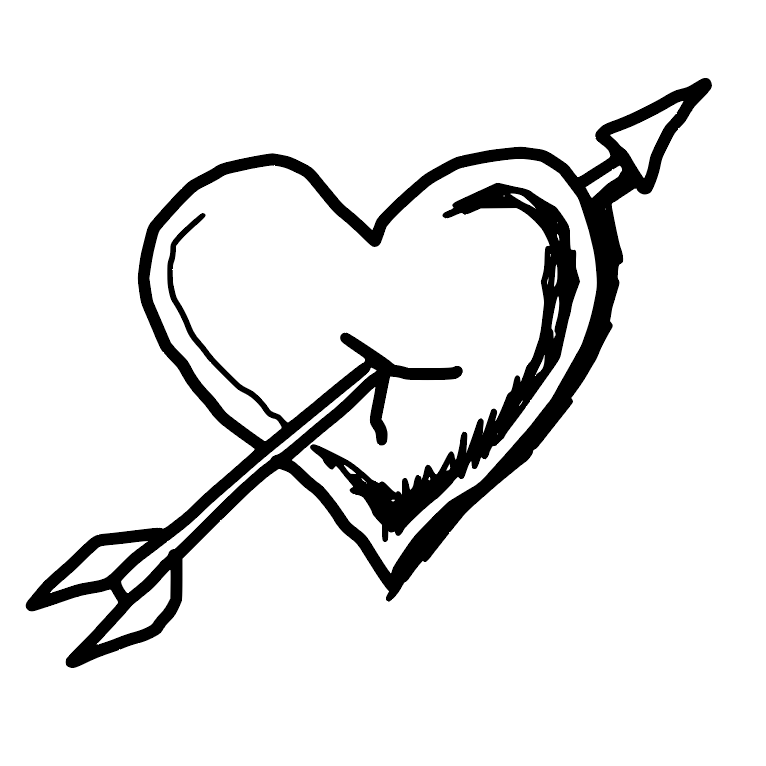}
			}&$682$& $296/294/263$& $272$& $211~(128^\dagger)$  & $120$\\\hline
			{\small
				\includegraphics[height=3.5em]{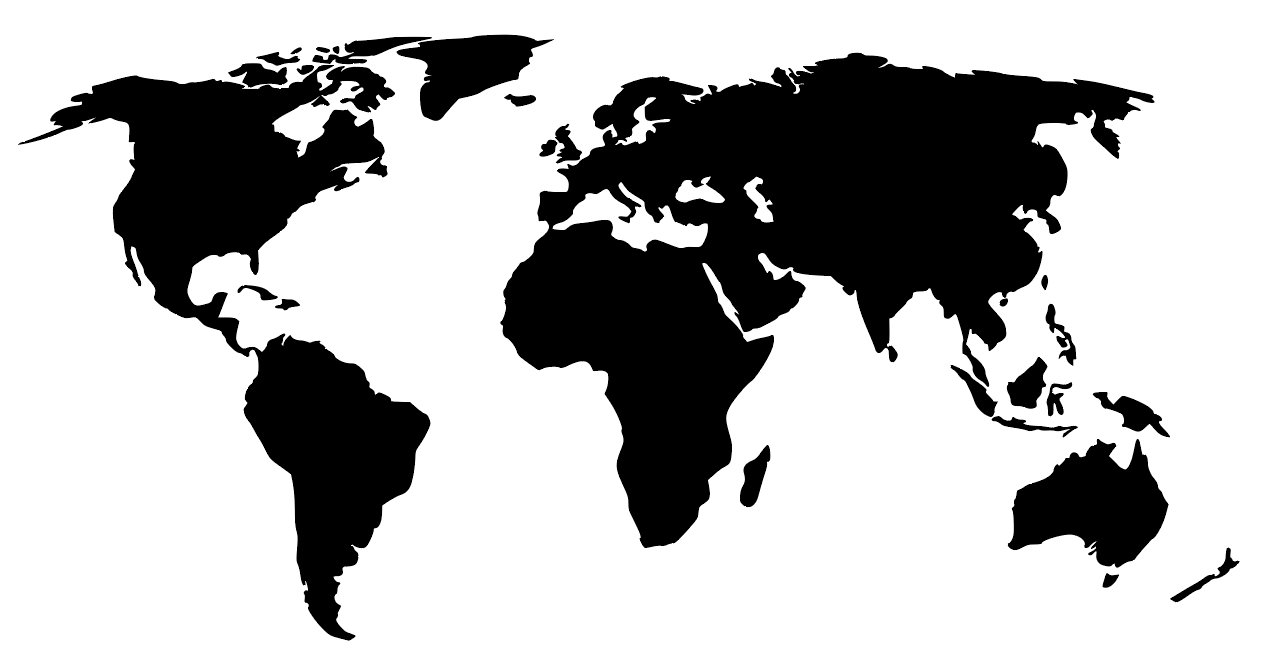}
			}&$1434$& $915/828/715$& $932$& $698~(462^\dagger)$  & $379$\\\hline
			{\small
				\includegraphics[height=3.5em]{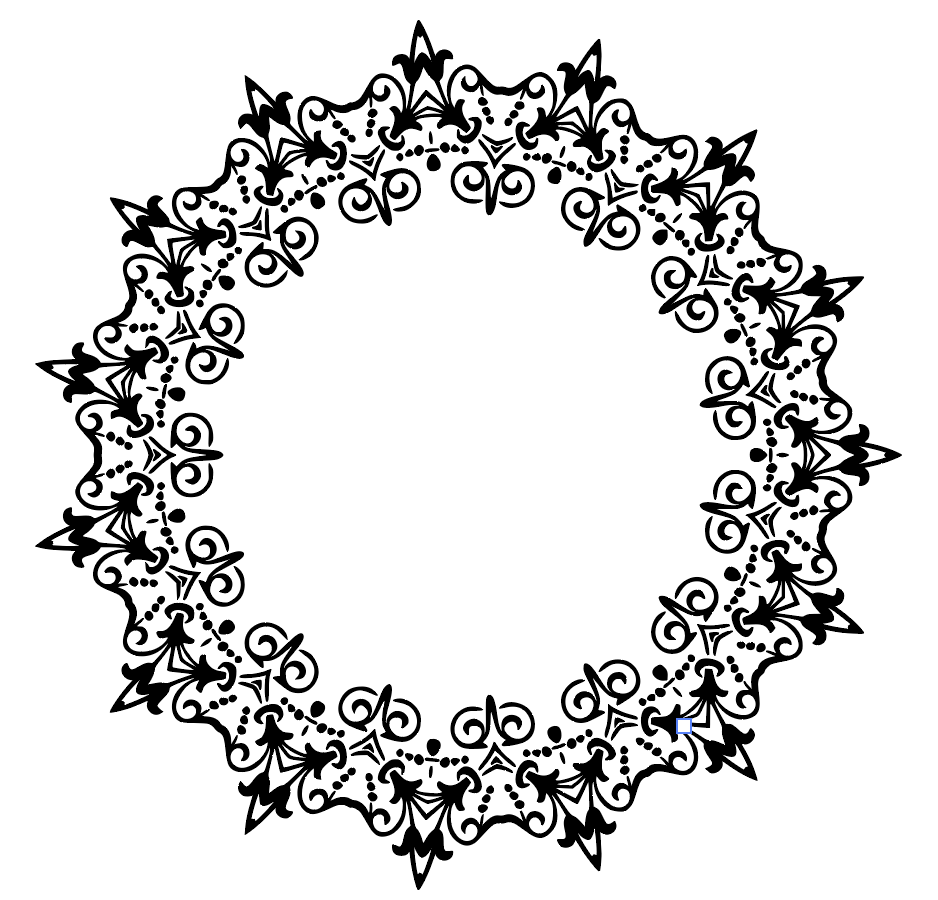}
			}&$4434$&$2789/2582/2370$& $3292$& $2120~(1431^\dagger)$  & $1407$\\\hline
			{\small
				\includegraphics[height=3.5em]{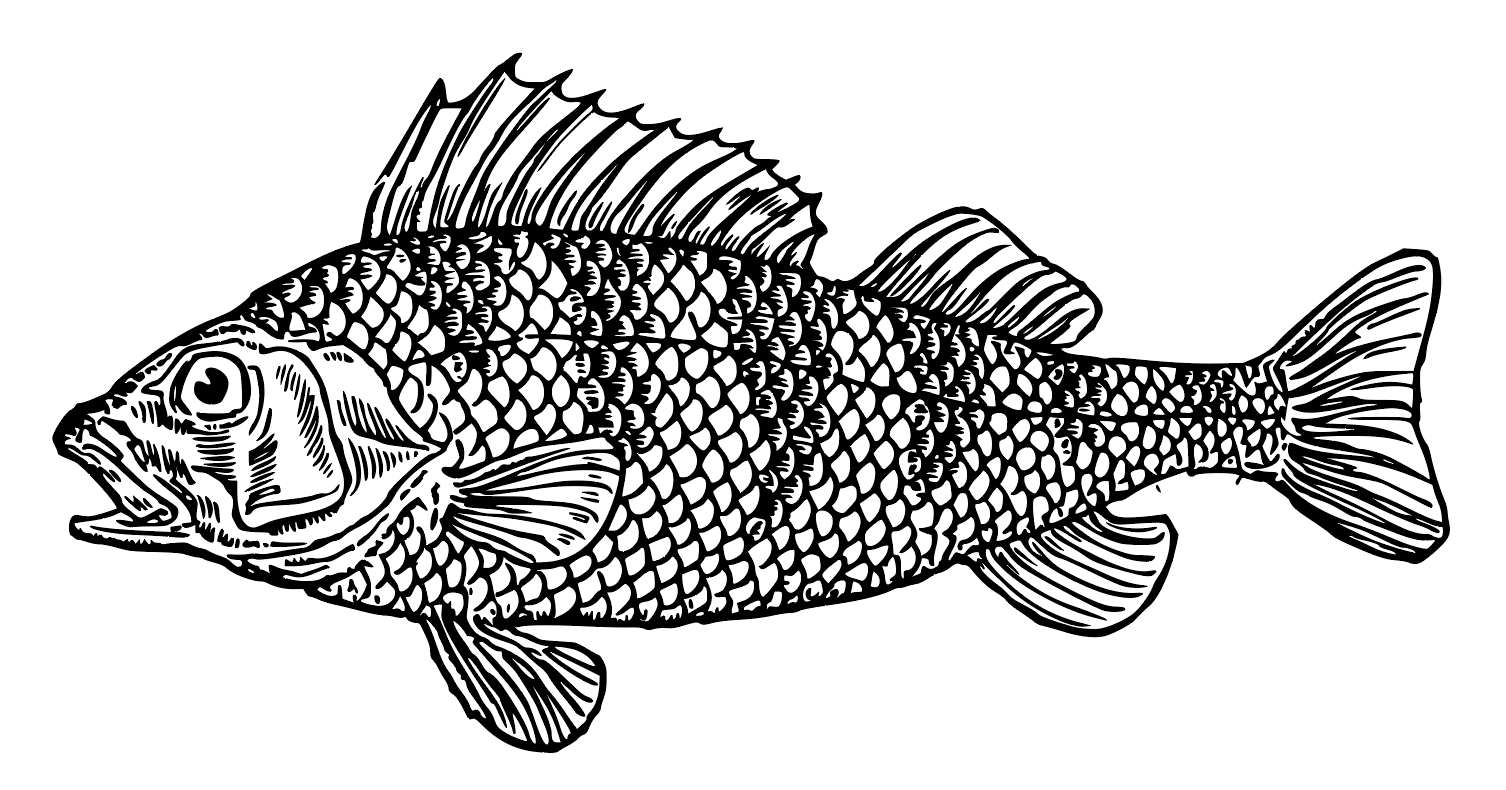}
			}&$6664$& $5470/5218/4955$& $6493$& $4870~(3441^\dagger)$  & $2810$\\\hline
			MRR& --- &$37..97\%/40.55\%/45.01\%$&$29.88\%$ &$45.79\%~(61.58\%^\dagger)$&$67.38\%$\\\hline
		\end{tabular}
		\caption{Comparison with image vectorization software in terms of the number of control points. We compared with Vector Magic (VM), Inkspace (IS), and Adobe Illustrator 2020 (AI). For VM, we report the number of control points using three optional settings: High/Medium/Low. For AI, the values with dagger$^\dagger$ indicate the numbers of control points produced by  the automatic simplification. The input image dimensions are $581\times 564$, $625\times 598$, $400\times 390$, $903\times 499$, $515\times 529$, and $1356\times 716$ from top to bottom. We also report the mean relative reduction (MRR) of the number of control points.}\label{tab_software}
	\end{table}
	
	\section{Conclusion}\label{sec_6}
	In this paper, we proposed an efficient and effective algorithm for silhouette  vectorization. The outline of the silhouette is interpolated bilinearly and uniformly sampled at sub-pixel level. To reduce the oscillation due to pixelization, we applied the affine shortening to the bilinear outline. By tracing the curvature extrema across different scales along  the well-defined inverse affine shortening flow, we identified a set of candidate control points. This set is then refined by deleting sub-pixel extrema that do not reflect salient corners, and inserting new points to guarantee any user-specified accuracy. We also designed special procedures to address the degenerate cases, such as disks, so that our algorithm adapts to arbitrary resolutions and offers better compression of information. Our method provides a superior compression ratio by vectorizing the outlines. When the given  silhouette undergoes affine transformations,  the distribution of control points generated by our method remains relatively stable. These properties are quantitatively justified by the repeatability ratio when compared with popular feature point detectors. Our method is competitive compared to some well-established image vectorization software in terms of producing results that  have less number of control points while achieving high accuracy.

	\appendix

	\section{Silhouette Data Set}
	In Table~\ref{tab_dataset}, we collectively display the $20$ silhouettes used in this paper. They are all downloadable from \url{https://svgsilh.com}, which are released under Creative Commons CC0.
		\begin{table}
		\centering
		\begin{tabular}{ccccc}
			\multicolumn{5}{c}{Silhouette Data Set}\\\hline
			{\small
				\includegraphics[height=3.5em]{Figures/cat.png}
			}&
			{\small
				\includegraphics[height=3.5em]{Figures/butterfly.png}
			}&
			{\small
				\includegraphics[height=3.5em]{Figures/text.png}
			}&
			{\small
				\includegraphics[height=3.5em]{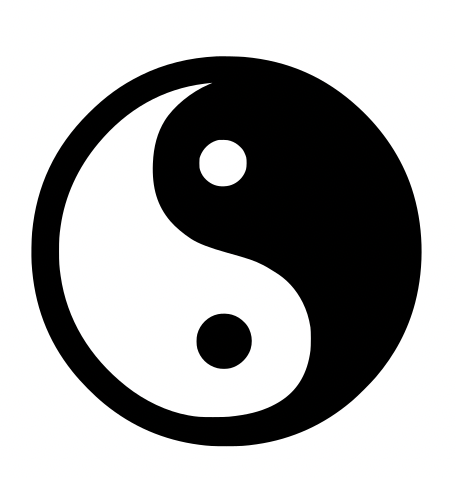}
			}&
			{\small
				\includegraphics[height=3.5em]{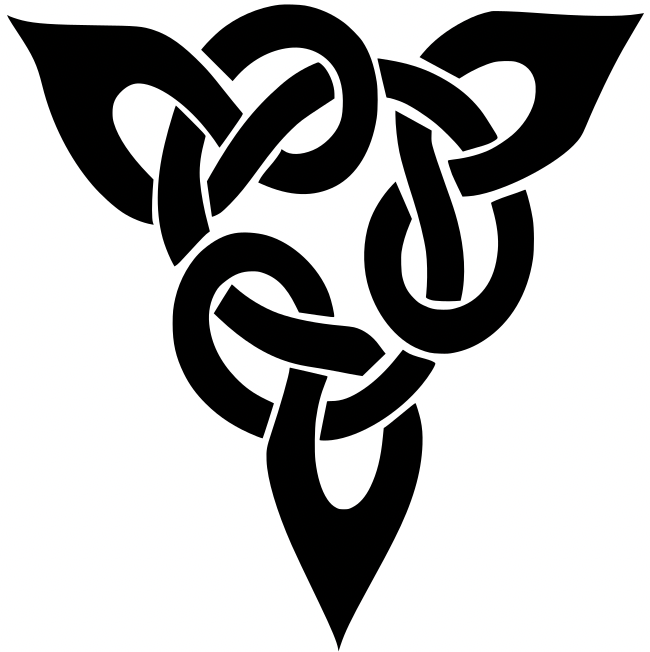}
			}\\
			{\small
				\includegraphics[height=3.5em]{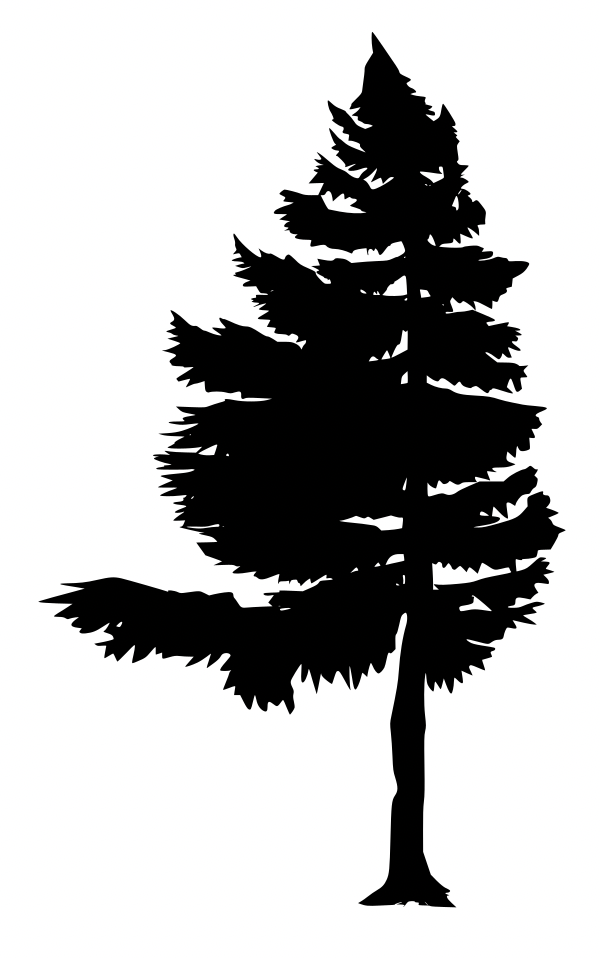}
			}&
			{\small
				\includegraphics[height=3.5em]{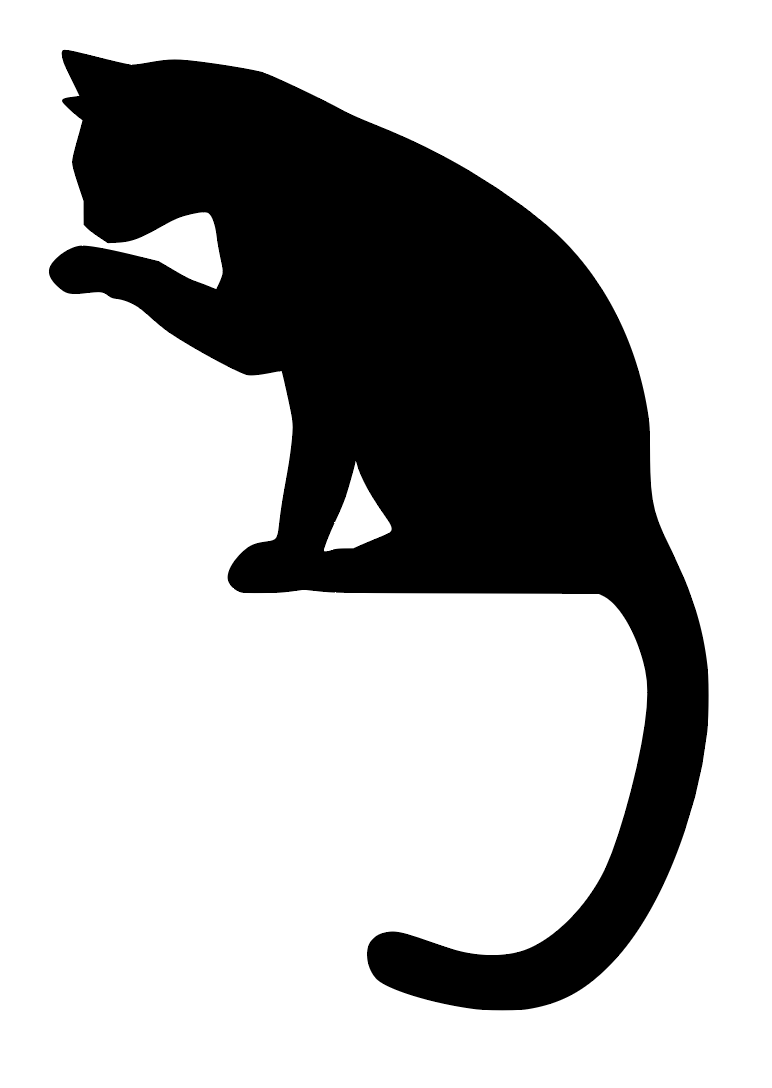}
			}&
			{\small
				\includegraphics[height=3.5em]{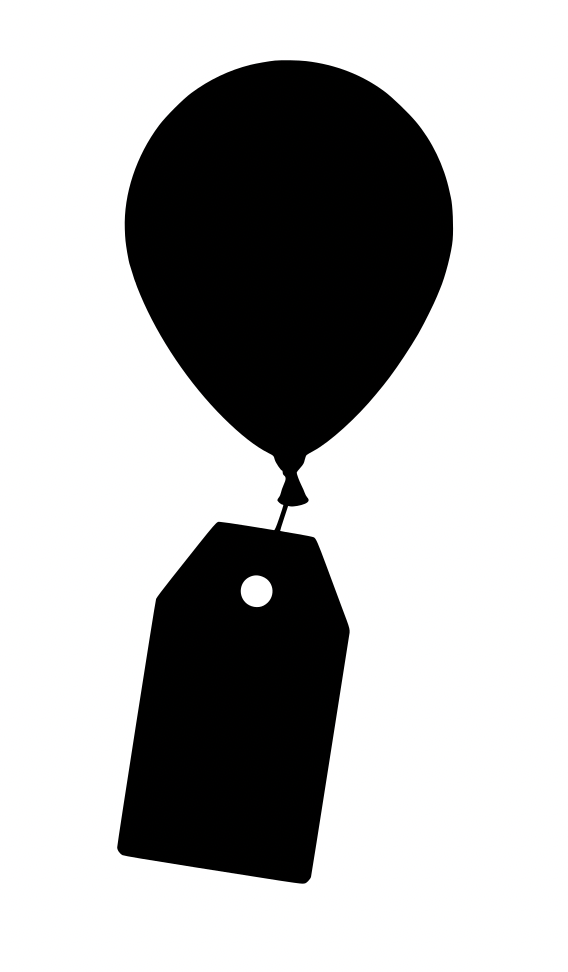}
			}&
			{\small
				\includegraphics[height=3.5em]{Figures/house_texample.png}
			}&
			{\small
				\includegraphics[height=3.5em]{Figures/fish_texample.png}
			}\\
			{\small
				\includegraphics[height=3.5em]{Figures/fish.png}
			}&
			{\small
				\includegraphics[height=3.5em]{Figures/panel.png}
			}&
			{\small
				\includegraphics[height=3.5em]{Figures/dog.png}
			}&
			{\small
				\includegraphics[height=3.5em]{Figures/heart.png}
			}&
			{\small
				\includegraphics[height=3.5em]{Figures/map.png}
			}\\
			{\small
				\includegraphics[height=3.5em]{Figures/circle.png}
			}&
			{\small
				\includegraphics[height=3.5em]{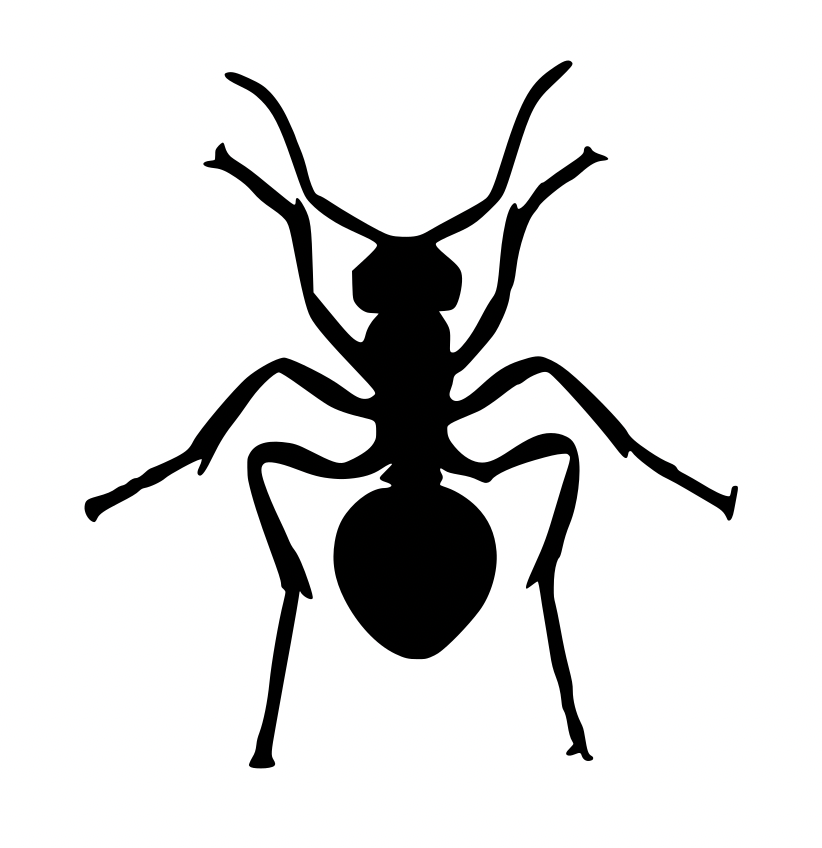}
			}&
			{\small
				\includegraphics[height=3.5em]{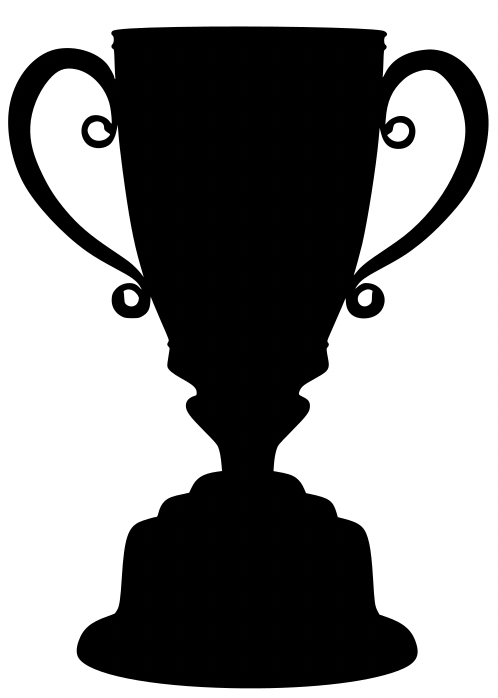}
			}&
			{\small
				\includegraphics[height=3.5em]{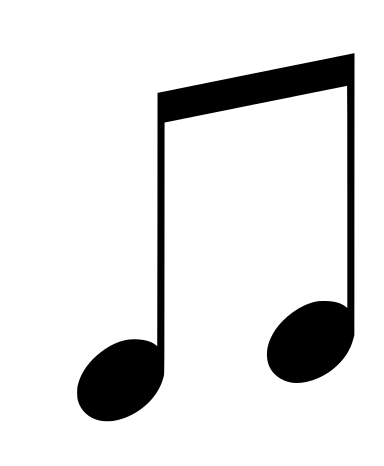}
			}&
			{\small
				\includegraphics[height=3.5em]{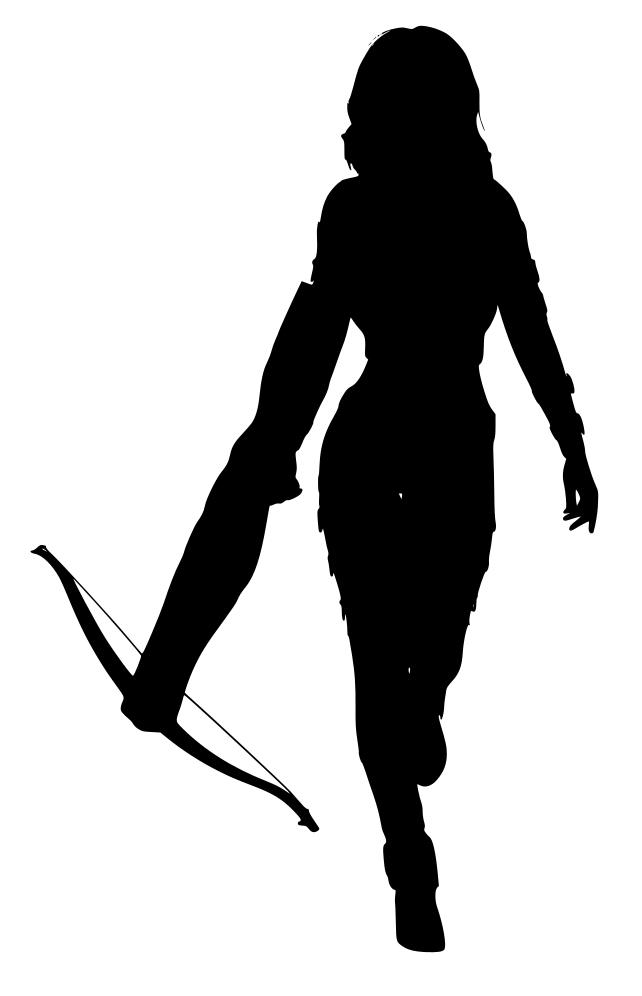}
			}\\\hline
		\end{tabular}
		\caption{Silhouette dataset used in the experiments. The last four are used in Figure~\ref{fig_threshold2} for computing the average $\rho(\tau_e)$. These silhouettes are chosen from~\cite{SVG}, which are released under Creative Commons CC0.}\label{tab_dataset}
	\end{table}
	
	\bibliographystyle{abbrv}
	\bibliography{ShapeVectorization}
	
\end{document}